\documentclass[showkeys,nofootinbib]{revtex4-2}

\usepackage[utf8]{inputenc} 
\usepackage[T1]{fontenc} 
\usepackage[english]{babel} 
\usepackage{xcolor} 
\usepackage{amsmath,amsfonts,amssymb,amsthm,braket,dsfont,comment,leftindex,bbm} 
\usepackage{graphicx} 
\usepackage{tikz} 
\usepackage{multirow,afterpage,tabularray} 
\usepackage{fancyhdr,orcidlink} 
\usepackage[capitalise]{cleveref} 
\usepackage{enumitem} 
\usepackage{titlesec} 
\usepackage{lipsum}
\usepackage{listings} 

\theoremstyle{definition}

\theoremstyle{plain}

\usetikzlibrary{arrows.meta}
\definecolor{asparagus}{rgb}{0.53, 0.66, 0.42}
\definecolor{burntorange}{rgb}{0.8, 0.33, 0.0}
\definecolor{armygreen}{rgb}{0.29, 0.33, 0.13}
\definecolor{atomictangerine}{rgb}{1.0, 0.6, 0.4}
\definecolor{bluebell}{rgb}{0.64, 0.64, 0.82}
\definecolor{blue(pigment)}{rgb}{0.2, 0.2, 0.6}

\newcommand{\diff}{\ensuremath{\mathrm{d}}}
\newcommand{\bal}{\begin{aligned}}
\newcommand{\eal}{\end{aligned}}

\renewcommand{\bm}[1]{\boldsymbol{#1}}

\titleformat{\subsubsection}
   {\normalfont\fontsize{9pt}{11pt}\selectfont\bfseries}
   {\thesubsubsection.}
   {1em}
   {\centering}
\titleformat*{\subsubsection}{\normalsize\itshape}

\begin{document}

\title{Two- and Three-gluon Glueballs within the Helicity Formalism}

\author{\surname{Cyrille} Chevalier \orcidlink{0000-0002-4509-4309}}
\email[E-mail: ]{cyrille.chevalier@umons.ac.be}
\affiliation{Service de Physique Nucl\'{e}aire et Subnucl\'{e}aire,
Universit\'{e} de Mons,
UMONS Research Institute for Complex Systems,
Place du Parc 20, 7000 Mons, Belgium}

\author{\surname{Vincent} Mathieu \orcidlink{0000-0003-4955-3311}}
\email[E-mail: ]{vmathieu@ub.edu}
\affiliation{Departament de F\'isica Qu\`antica i Astrof\'isica and Institut de Ci\`encies del Cosmos, Universitat de Barcelona, E08028, Spain.}

\date{\today}

\begin{abstract}
Both positive and negative charge conjugation glueball spectra are computed with a constituent gluon approach. We first compute the spectrum of the Hamiltionian describing two-gluon bound states having $C=+$, before tackling the three-gluon bound states having $C=-$. We review the construction of two and three particle helicity states in order to build totally symmetric wave functions for our glueball states. In the literature, two different couplings schemes are used to build three particles states. We derive for the first time the relation between these two three-body state definitions. The glueball spectra are compared with those obtained from quantum chromodynamics simulations on a lattice. The two-gluon glueball spectra show a good agreement with the constituent approach and the numerical lattice simulations. The three-gluon spectrum is in overall agreement with the masses observed on the lattice but also predicts additional low-lying states.
\end{abstract}
\keywords{Glueballs, Constituent approaches, Helicity formalism, Three-body quantum problem}

\maketitle


\section{Introduction}
\label{sec:intro}

One of the earliest predictions of Quantum Chromodynamics (QCD) is the existence of color-singlet pure-gauge states, known as glueballs \cite{frit72}. Despite this theoretical prediction, a consensus on their properties and definitive experimental evidence remains elusive. Two-gluon glueball states have been extensively studied both theoretically \cite{math09,llan21} and experimentally \cite{cred09,llan21}. Theoretical results have been obtained using various phenomenological approaches, functional methods, and lattice QCD (LQCD). On the experimental side, notable experimental efforts by collaborations such as PANDA, Crystal Barrel, WA102 or BESIII continue the search for these states. In contrast, three-gluon glueballs have received relatively little attention due to the technical complexity involved. On the theory side, the LQCD spectrum is expected to include three-gluon states \cite{chen06,morn99,meye05,liu02}. Experimentally, the possible observation of odderon exchange at TOTEM is still debated \cite{petr22}.

This work aims to describe two- and three-gluon systems within the framework of constituent models. In these approaches, hadronic states are treated as colorless bound states of several constituent quarks, antiquarks and/or gluons. The requirement of color neutrality determines whether a given combination of constituent particles can form a hadron. Constituent gluons transform under $SU(3)_c$ with its adjoint representation, denoted $8$, leading to the following decompositions for two- and three-gluon systems \cite{boul08}
\begin{align}
&8 \otimes 8 = 1 \oplus ... &&8 \otimes 8 \otimes 8 = 1 \oplus ...
\end{align}
These decompositions indicate the possibility of forming bound states of two and three constituent gluons, so-called two- and three-gluon glueballs, in constituent approaches. The dynamics of these bound states are implemented using a phenomenological but QCD inspired Hamiltonian. Properties of constituent particles often differ from their QCD counterparts (for instance, constituent quarks are often treated as heavier than their QCD equivalents). For two-gluon glueballs, it has been shown in reference \cite{math08} that constituent gluons have to be considered as massless particles with helicity degrees-of-freedom in order to reproduce results from LQCD calculations. As a result, extending this approach to three-gluon glueballs seems to necessitate addressing the three-body helicity formalism.

The article is organised as follows. Section \ref{sec:1BS2BS} introduces the helicity formalism for one- and two-body systems. Given its important role in most of the subsequent calculations and because the management of three-gluon glueballs requires to demonstrate novel properties within this formalism, this section provides a relatively detailed overview. Additionally, for pedagogical purpose and for the sake of generality, the case of massive particles is included in the description. Section \ref{sec:2GB} applies these results to compute the spectrum of two-gluon glueballs. As already mentioned, a detailed study of two-gluon glueballs in constituent approach already exists in the literature \cite{szcz03,szcz96,math08}. However, the extension to three-gluon glueballs necessitates revisiting the two-gluon spectrum and introducing a methodology better suited for three-gluons systems. Section \ref{sec:3BS} develops the helicity formalism for three-body systems. Section \ref{sec:3GB} investigates three-gluon glueballs. This Section is decomposed into three parts: first, helicity states with proper symmetry properties are constructed; secondly, expressions to compute Hamiltonian matrix elements (ME) on these states are prepared; and, lastly, the spectrum is obtained and analysed. The manuscript closes in Section \ref{sec:conclu} with a recap of the key results and suggestions for prospects.

\begin{figure}
\centering
\includegraphics[scale=0.25]{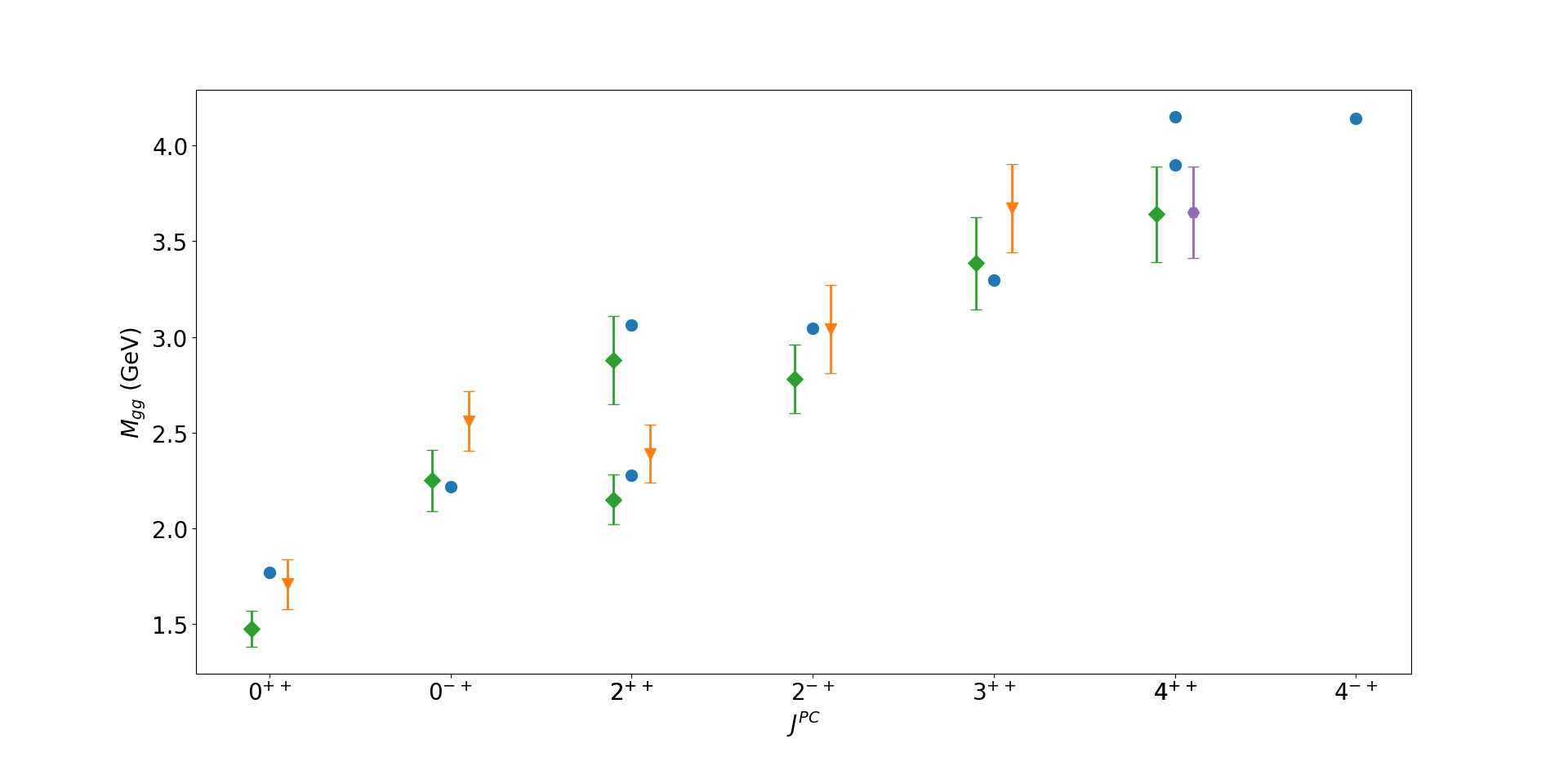}
\caption{Comparison of two-gluon glueball spectra. Upper bounds obtained with the current approach (blue circles) are compared to LQCD results from \cite{chen06} (orange triangles), \cite{meye05} (green diamonds) and \cite{liu02} (purple hexagon).}
\label{fig::intro_2GB}
\end{figure}

\begin{figure}
\centering
\includegraphics[scale=0.33]{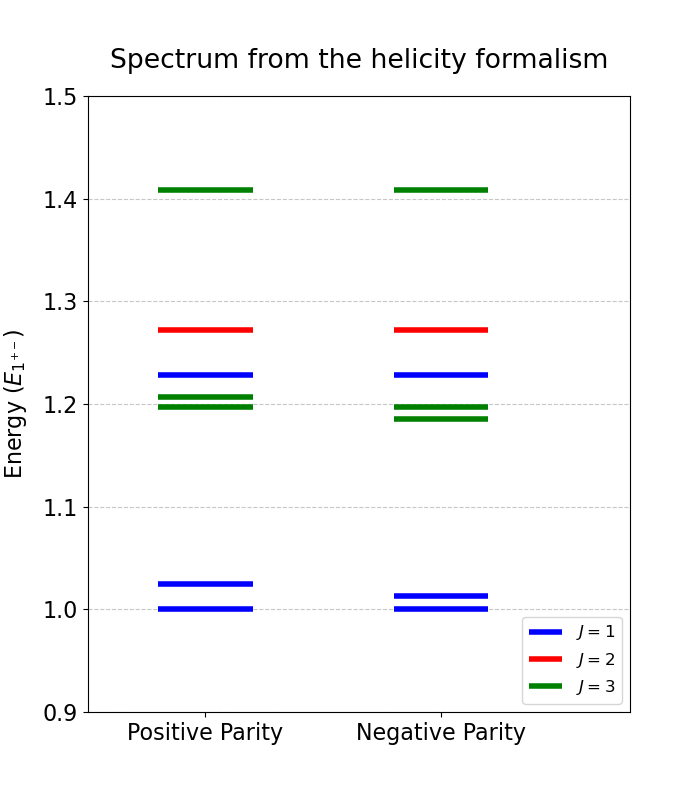}
\includegraphics[scale=0.33]{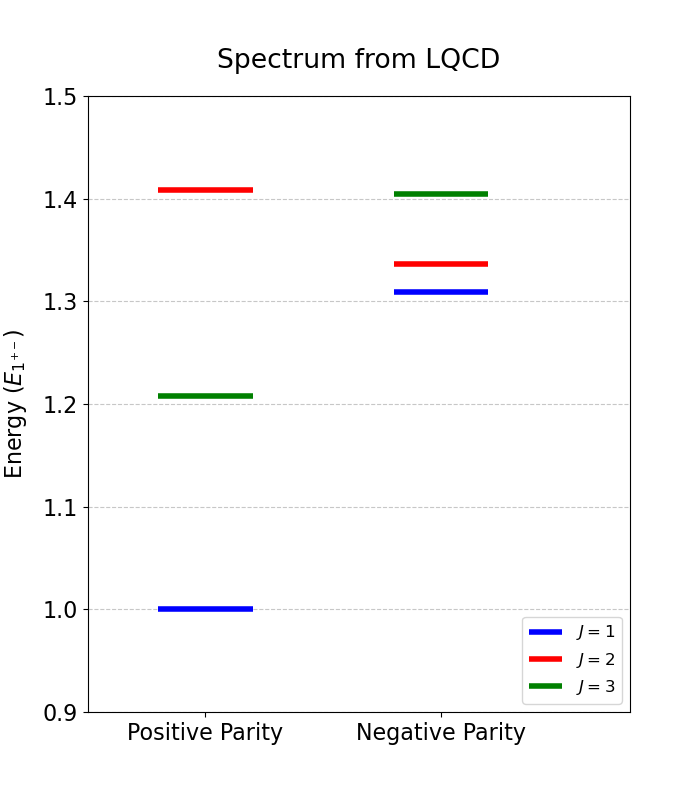}
\caption{Comparison of glueball spectra obtained using the helicity formalism (left, this work) and LQCD (right, references \cite{chen06,morn99}). In both cases, spectra are provided in unit of the lowest mass.}
\label{fig::intro_3GB}
\end{figure}

In a few words, the use of helicity degrees-of-freedom proves to allow for an accurate reproduction of the two-gluon glueball spectrum observed in LQCD, as illustrated in Fig.~\ref{fig::intro_2GB}. Extending this approach to three-gluon systems yields the spectrum shown in Fig.~\ref{fig::intro_3GB}. Overall, each state identified by LQCD has a corresponding counterpart in the spectrum obtained with the helicity formalism, but additional low-lying states are also predicted by the method. While definitive explanations for these discrepancies remain elusive, possible interpretations are suggested at the end of Section \ref{sec:3GB}.

\section{The Helicity Formalism for One- and Two-body Systems}
\label{sec:1BS2BS}

This Section reviews the main concepts underlying the helicity formalism for one- and two-body systems. A detailed recap is provided, as the analysis of three-body systems requires demonstrating new properties of helicity states. The discussion is kept general to enable application to a variety of systems. Glueball-specific considerations are deferred to the next section.


\subsection{One-body Helicity States}
\label{ssec:1BS}

To start with, let us focus on massive particles. Massive particles are classified according to their eigenvalue for both Casimir operators of the Poincaré group (PG), namely the mass operator $P^2$ and the spin operator $W^2$ (also known as the Pauli-Lubanski Casimir). Any state $\ket{\psi;m;s}$ such that\footnote{We are using natural units, $\hbar=c=1$, thorough this work.} 
\begin{align}
    &P^2 \ket{\psi;m;s} = m^2 \ket{\psi;m;s} \text{ and } W^2 \ket{\psi;m;s} = -m^2s(s+1) \ket{\psi;m;s}
\end{align}
is a possible state to describe a particle of mass $m$ and spin $s$. The current work uses the mostly minus convention for the Minkowski metric. To write the actual state of the particle, one has to resort to complete sets of states in which it can be decomposed. In the current work, two different complete sets will be used.

\subsubsection{The one-body canonical states}

\begin{figure}
\begin{center}
\begin{tikzpicture}
\draw[->] (0,0) - - (0,2);
\draw[->] (0,0) - - (0.9,0);
\draw[->] (0,0) - - (-0.3,-0.3);
\draw[-{Implies},double,very thick,teal] (0.2,0.3) - - (0.2,1.5);
\node[right, teal](s) at (0.2,0.9) {$\ket{sm_s}$};
\draw[->,very thick] (1.8,0.35) - - (3.2,0.35);
\node[above](L) at (2.6,0.5) {$R(\phi,\theta,0)^{-1}$};

\draw[->] (5,0) - - (5,2);
\draw[->] (5,0) - - (5.9,0);
\draw[->] (5,0) - - (4.7,-0.3);
\draw[dashed] (4.4,1.5) - - (4.4,0.3);
\draw[dashed] (4.4,0.3) - - (5,0);
\draw[-{Implies},double,very thick,teal] (4.9,0.3) - - (4.4,1.5);
\node[below](angles) at (4,0.3) {\scriptsize $(\theta,\phi)^{-1}$};
\draw[->,very thick] (6.8,0.35) - - (8.2,0.35);
\node[above](R) at (7.5,0.4) {$L_z(p)$};

\draw[->] (9,0) - - (9,2);
\draw[->] (9,0) - - (9.9,0);
\draw[->] (9,0) - - (8.7,-0.3);
\draw[-{Stealth},very thick,red] (9.2,0.3) - - (9.2,1.5);
\node[right, red](s) at (9.2,0.9) {$p\vec{e}_z$};
\draw[-{Implies},double,very thick,teal] (8.9,0.3) - - (8.4,1.5);
\draw[->,very thick] (10.8,0.35) - - (12.2,0.35);
\node[above](R) at (11.5,0.4) {$R(\phi,\theta,0)$};

\draw[->] (13,0) - - (13,2);
\draw[->] (13,0) - - (13.9,0);
\draw[->] (13,0) - - (12.7,-0.3);
\draw[-{Implies},double,very thick,teal] (13.2,0.3) - - (13.2,1.5);
\draw[-{Stealth},very thick,red] (13.15,0.15) - - (14.05,1.15);
\draw[dashed] (14,1.1) - - (14,-0.3);
\draw[dashed] (14,-0.3) - - (13,0);
\node[below](angles) at (14,-0.3) {$(\theta,\phi)$};
\end{tikzpicture}
\caption{Graphic illustration for the definition \eqref{eq::1BS_pcanonical} of one-body $p$-canonical states. \label{fig::1BS_pcanonical}}
\end{center}
\end{figure}
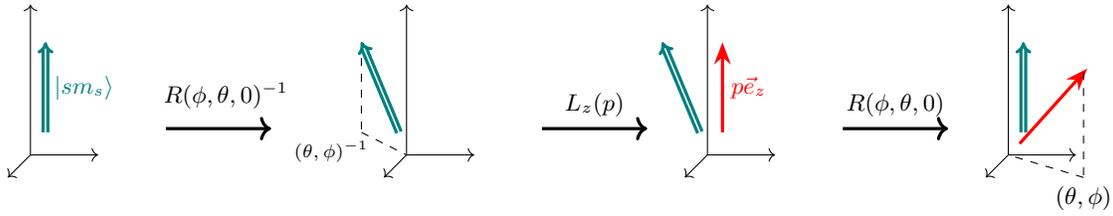

The first set is filled with the common eigenstates of the third component of the Pauli-Lubanski $W_3$ and of the four $P_\mu$ operators. In the following, such common eigenstates, referred to as \textit{one-body canonical states}, will be denoted $\ket{m;p\theta\phi;sm_s}$. The labels in the notation provides the eigenvalues of the aforementioned operators,
\begin{align}
    &P^2 \ket{m;p\theta\phi;sm_s} = m^2 \ket{m;p\theta\phi;sm_s}, &&W^2 \ket{m;p\theta\phi;sm_s} = -m^2s(s+1) \ket{m;p\theta\phi;sm_s},\label{eq::1BS_pcanMS}\\[5pt]
    &P_0 \ket{m;p\theta\phi;sm_s} = \sqrt{m^2+p^2} \ket{m;p\theta\phi;sm_s}, &&P_2 \ket{m;p\theta\phi;sm_s} = p\sin\phi\sin\theta \ket{m;p\theta\phi;sm_s},\label{eq::1BS_pcanP_1}\\[1pt]
    &P_1 \ket{m;p\theta\phi;sm_s} = p\cos\phi\sin\theta \ket{m;p\theta\phi;sm_s},
    &&P_3 \ket{m;p\theta\phi;sm_s} = p\cos\theta \ket{m;p\theta\phi;sm_s},\label{eq::1BS_pcanP_2}\\[8pt]
    &W_3 \ket{m;p\theta\phi;sm_s} = m\,m_s\ket{m;p\theta\phi;sm_s}. \label{eq::1BS_pcanW3}
\end{align}
Physically speaking, each $p$-canonical state has a definite mass, spin, four-momentum and spin projection along the $z$ axis. The associated lorentz-invariant orthonormality and completeness relations are written below,
\begin{equation}
\braket{m;\bar p\bar\theta\bar\phi;s\bar m_s|\,m; p\theta\phi;s m_s} = 2w\,
\delta(p-\bar p)\delta(\phi-\bar \phi) \delta(\cos\theta-\cos\bar\theta) \delta_{m_s\bar m_s},
\end{equation}
\begin{equation}
\sum_{m_s=-s}^s \int \ket{m;p\theta\phi;sm_s} \frac{\mathrm{d}^3p}{2w} \bra{m;p\theta\phi;sm_s} = \mathds{1},
\end{equation}
where $w=\sqrt{m^2+p^2}$. Normalisation conventions used in this document follow the ones from \cite{mart70}. In the literature, one-body canonical states are often decomposed as the application of a so-called canonical boost on a rest state $\ket{m;sm_s}$,
\begin{equation}
\begin{aligned}
    \ket{m;p\theta\phi;sm_s} &= U(L(m,p,\theta,\phi))\ket{m;sm_s} \\
    &= U(R(\phi,\theta,0) L_z(m,p) R(\phi,\theta,0)^{-1})\ket{m;sm_s}.
\end{aligned}\label{eq::1BS_pcanonical}
\end{equation}
Above, $R(\alpha,\beta,\gamma)$ refers to a rotation of Euler angles $(\alpha,\beta,\gamma)$ while $L_z(m,p)$ is a boost that provides a spatial momentum of modulus $p$ and along the $z$-axis to particle of mass $m$ initially at rest. Hereinabove, these transformations are combined to produce a so-called \textit{canonical boost} denoted $L(m,p,\theta,\phi)$. The notation $U$ reminds that whenever space-time transformations are applied on states, their action follows a given unitary representation of the PG. By construction, rest states $\ket{m;sm_s}$ transform under rotations using the spin $s$ irreducible representation of $SO(3)$. As a result, whenever applied on a rest state, rotations are concretely represented by the well-known Wigner $D$ matrices \cite{vars88},
\begin{equation}
    U(R(\alpha,\beta,\gamma))\ket{m;sm_s} = \sum_{m_s'=-s}^s D^s_{m_s'm_s}(\alpha,\beta,\gamma) \ket{m;sm_s'}. \label{eq::1BS_restrot}
\end{equation}
Formula \eqref{eq::1BS_pcanonical} provides an intuitive graphic interpretation for one-body canonical states which is illustrated in Fig.~\ref{fig::1BS_pcanonical}. In the left-hand part of the diagram, the rest state is represented with a single green arrow which stands for its spin projection. The state is then subjected to a rotation and a boost which, notably, alter its momentum, represented by a red arrow. After successive transformations, as expected, the state ends with a momentum direction $(\theta,\phi)$ and with a definite spin projection along the $z$ axis. Apart from its visualisation use, this expression, combined with relation \eqref{eq::1BS_pcanonical}, also enables easy derivations of transformation rules under operations such as rotations, boosts and parity. The interested reader is referred to \cite{suh71} for more details.

\subsubsection{The one-body helicity states}

\begin{figure}
\begin{center}
\begin{tikzpicture}
\draw[->] (0,0) - - (0,2);
\draw[->] (0,0) - - (2,0);
\draw[->] (0,0) - - (-0.6,-0.6);
\draw[-{Implies},double,very thick,teal] (0.2,0.3) - - (0.2,1.5);
\node[right, teal](s) at (0.2,0.9) {$\ket{s\,\lambda}$};
\draw[->,very thick] (2.8,0.35) - - (4.2,0.35);
\node[above](L) at (3.5,0.4) {$L_z(p)$};

\draw[->] (5,0) - - (5,2);
\draw[->] (5,0) - - (7,0);
\draw[->] (5,0) - - (4.4,-0.6);
\draw[-{Implies},double,very thick,teal] (5.2,0.3) - - (5.2,1.5);
\draw[-{Stealth},very thick,red] (4.8,0.3) - - (4.8,1.5);
\node[left, red](s) at (4.8,1.5) {$p\vec{e}_z$};
\draw[->,very thick] (7.8,0.35) - - (9.2,0.35);
\node[above](R) at (8.5,0.4) {$R(\phi,\theta,0)$};

\draw[->] (10,0) - - (10,2);
\draw[->] (10,0) - - (12,0);
\draw[->] (10,0) - - (9.4,-0.6);
\draw[-{Stealth},very thick,red] (10.1,0.4) - - (11,1.4);
\draw[-{Implies},double,very thick,teal] (10.1,0.1) - - (11,1.1);
\draw[dashed] (11,1.1) - - (11,-0.3);
\draw[dashed] (11,-0.3) - - (10,0);
\node[below](angles) at (11,-0.3) {$(\theta,\phi)$};
\end{tikzpicture}
\caption{Graphic illustration for the definition \eqref{eq::1BS_phelicity} of one-body $p$-helicity states.\label{fig::1BS_phelicity}}
\end{center}
\end{figure}
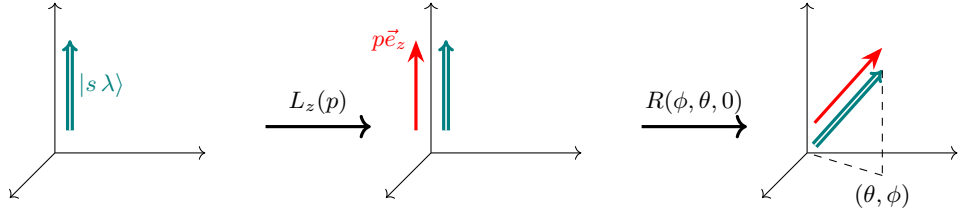

Although one-body canonical states are abundantly used in quantum mechanics, especially in the non-relativistic limit, this set can not encompass massless particles. For this reason, a second set of state is introduced. The states are still chosen as the common eigenstates of the four $P_\mu$ generators but, instead of $W_3$, these are taken as eigenstates of $\Lambda$, the helicity operator,
\begin{equation}
    \Lambda = \frac{\bm{J}\cdot \bm{P}}{\sqrt{\bm{P}^2}}.
\end{equation}
Above, $\bm{P}$ and $\bm{J}$ collect the spatial components of the momentum and angular momentum, respectively. These states are named \textit{one-body helicity states} and are denoted $\ket{m;p\theta\phi;s\lambda}$. Relations \eqref{eq::1BS_pcanMS} to \eqref{eq::1BS_pcanP_2} remain true in terms of one-body helicity states, but \eqref{eq::1BS_pcanW3} is replaced by
\begin{equation}
    \Lambda \ket{m;p\theta\phi;s\lambda} = \lambda\ket{m;p\theta\phi;s\lambda}.
\end{equation}
States now have a definite spin projection along their momentum direction. The completeness and orthonormality relations for one-body helicity states are similar to the ones for one-body canonical states (same conventions are used),
\begin{equation}
\braket{m;\bar p\bar\theta\bar\phi;s\bar\lambda|\,m; p\theta\phi;s\lambda} = 2w\,
\delta(p-\bar p)\delta(\phi-\bar \phi) \delta(\cos\theta-\cos\bar\theta) \delta_{\lambda\bar \lambda},
\end{equation}
\begin{equation}
\sum_{\lambda=-s}^s \int \ket{m;p\theta\phi;s\lambda} \frac{\mathrm{d}^3p}{2w} \bra{m;p\theta\phi;s\lambda} = \mathds{1}.
\end{equation}
As well as canonical states, helicity ones can also be expressed in terms of a rest state on which boosts and rotations are applied.
\begin{equation}
\begin{aligned}
    \ket{m;p\theta\phi;s\lambda} &= U(L_h(m,p,\theta,\phi))\ket{m;s\lambda} \\
    &= U(R(\phi,\theta,0) L_z(m,p))\ket{m;s\lambda}.
\end{aligned} \label{eq::1BS_phelicity}
\end{equation}
In comparison with relation \eqref{eq::1BS_pcanonical}, the canonical boost $L$ has here been replaced by an helicity boost $L_h$ which differs by a rotation. The graphic interpretation set up for relation \eqref{eq::1BS_pcanonical} in Fig.~\ref{fig::1BS_pcanonical} can be adapted to fit with formula \eqref{eq::1BS_phelicity}. The result is shown in Fig.~\ref{fig::1BS_phelicity}. Referring to the same graphical conventions, one can see that the state ends with a definite projection of the spin along the momentum direction instead of along the $z$ axis. This feature is in agreement with the definition of the helicity operator. Before to discuss the properties of one-body helicity states, let us draw reader's attention to the convention used for relation \eqref{eq::1BS_phelicity}. In the current definition, the third angle of the rotation has been settled to zero, a choice which will be denoted as the \textit{zero convention} in the following. Some other references consider $-\phi$ for this angle \cite{jaco59,wein95,mart70},
\begin{equation}
\ket{m;p\theta\phi;s\lambda}_{-\phi} = U(R(\phi,\theta,-\phi) L_z(m,p))\ket{m;s\lambda}.
\end{equation}
A priori, such a modified relation should result in a different definition for one-body helicity states. However, these two different definitions are shown to simply differ by a phase factor,
\begin{equation}
    \ket{m;p\theta\phi;s\lambda} = e^{-i\lambda\phi}\ket{m;p\theta\phi;s\lambda}_{-\phi}.\label{eq::1BS_phelicity_conv}
\end{equation}
Because physics is contained in the modulus squared of the state, this phase factor proves to be irrelevant in assessing a physical interpretation for one-body helicity states. Nevertheless, the user has to be conscious and consistent with the convention he employs, especially if he wants to superpose different helicity states. The current work will always use the zero convention.

The decomposition \eqref{eq::1BS_phelicity} also allows one to prove many properties about helicity states. Only a few of them will be recapped here. For more properties and demonstrations, the interested reader is referred to \cite{mart70,jaco59,gieb85,suh71}. The first property to be investigated is the transformation rule of helicity states under rotations. Simple algebra allows to show that
\begin{equation}
    U(R(\alpha,\beta,\gamma)) \ket{m;p\theta\phi;s\lambda} = \pm\, e^{-i\xi\lambda} \ket{m;p\theta'\phi';s\lambda}
    \label{eq::1BS_rotProp}
\end{equation}
where $\theta'$, $\phi'$ and $\xi$ satisfy $R(\phi',\theta',\xi) = R(\alpha,\beta,\gamma)R(\phi,\theta,0)$. The phase $\exp(-i\xi\lambda)$ comes out while ensuring that the state, initially in $0$ convention, ends in that same convention. Concerning the plus or minus sign, it is added to take into account the fact that, if at some point a $2\pi$ rotations is performed, it is to be identified to minus the identity for fermions. Property \eqref{eq::1BS_rotProp} is sometimes referred to as helicity rotational invariance, because the quantum number $\lambda$ remains unchanged despite the rotation. 

Secondly, the action of the parity transformation, $\Pi$, on one-body helicity states can be determined by making use of the commutation relations between parity, boosts and rotations. The following formula is obtained,
\begin{equation}
    \Pi\ket{m;p\theta\phi;s\lambda} = \eta(-1)^{-s} \ket{m;p(\pi-\theta)(\pi+\phi);s-\lambda}. \label{eq::1BS_parity}
\end{equation}
 Notice that the classical intuition that parity inverts particle's momentum and helicity is respected. Above, $\eta$ refers to the intrinsic parity of the particle. This quantity defines how parity acts on the particle rest state,
\begin{equation}
    \Pi \ket{m;s\lambda} = \eta \ket{m;s\lambda}. \label{eq::1BS_parityProp}
\end{equation}
It depends on the nature of the particle. In relation \eqref{eq::1BS_parity}, both left- and right-hand side states are written in the zero convention. However, in many references, the right-hand side is written in a different convention, such as the $\pi$ convention
\begin{align}
    &\begin{aligned}
    \ket{m;p(\pi-\theta)(\pi+\phi);s\lambda}_{\pi} =&\ U(R(\pi+\phi,\pi-\theta,\pi)) U(L_z(m,p))\ket{m;s\lambda}\\
    =&\ U(R(\phi,\theta,0)) U(R(0,\pi,0)) U(L_z(m,p))\ket{m;s\lambda}
    \end{aligned} \label{eq::1BS_oppPi}
\end{align}
or the opposite convention
\begin{equation}
    \begin{aligned}
    \ket{m;p(\pi-\theta)(\pi+\phi);s\lambda}_{-} =\ U(R(\phi,\theta,0))U(L_{-z}(m,p))\ket{m;s-\lambda}&\\
    \text{with }L_{-z}(m,p) = R(0,\pi,0)L_{z}(m,p)R(0,-\pi,0)&.
    \end{aligned} \label{eq::1BS_oppOpp}
\end{equation}
All these conventions being equal up to phase factors, they convey the same physical meaning,
\begin{equation}
\begin{aligned}
    \ket{m;p(\pi-\theta)(\pi+\phi);s\lambda}&\ = (-1)^\lambda \ket{m;p(\pi-\theta)(\pi+\phi);s\lambda}_{\pi}\\
    &\ = (-1)^{s} \ket{m;p(\pi-\theta)(\pi+\phi);s\lambda}_{-}.
\end{aligned}
\end{equation}
These various conventions are introduced to simplify calculations that mixes helicity states with opposed directions. Such situations will for example be encountered in Section \ref{ssec:2BS} where helicity states for two-body systems in their center-of-mass frame (CoMF) are introduced.

Third, we focus on the application of a general Lorentz transformation $L$ on an helicity state $\ket{m;p\theta\phi;s\lambda}$. The corresponding transformation rule reads as follows, 
\begin{equation}
    U(L) \ket{m;p\theta\phi;s\lambda} = \sum_{\lambda'=-s}^s D^s_{\lambda'\lambda}(\alpha_W,\beta_W,\gamma_W)\ket{m;p'\theta'\phi';s\lambda'}.\label{eq::1BS_canProp}
\end{equation}
Hereinabove, primed variables denote the components of the boosted momentum which are obtained by applying $L$ on the initial four-momentum. In addition, $(\alpha_W,\beta_W,\gamma_W)$ denotes the Euler angles of a rotation $R_W$, named Wigner rotation, defined by the following combination of boosts,
\begin{equation}
R_W = \left(L_h(m,p',\theta',\phi')\right)^{-1} L L_h(m,p,\theta,\phi). \label{eq::1BS_wignRot}
\end{equation}
Relation \eqref{eq::1BS_canProp} illustrates that, in general, the helicity of a massive state is not invariant under Lorentz transformations. The invariance of helicity under rotation \eqref{eq::1BS_rotProp} can be seen as a special case of property \eqref{eq::1BS_canProp}.

Finally, it is possible to switch from the one-body helicity basis to the one-body canonical one. Comparing relation \eqref{eq::1BS_phelicity} for helicity state to relation \eqref{eq::1BS_pcanonical} for canonical state and using relation \eqref{eq::1BS_restrot} provides the following transformation rule,
\begin{equation}
    \ket{m;p\theta\phi;s\lambda} = \sum_{m_s=-s}^s D^s_{m_s\lambda}(\phi,\theta,0)\ket{m;p\theta\phi;sm_s}.\label{eq::1BS_canToHel}
\end{equation}
This property allows the user to switch the complete set, based on its intended use.

Before closing this Section, let us insist on the fact that helicity states generalize very well to massless particles. In a common simplification of reality, massless particles are considered behaving as massive particles with only two spin projection, $+s$ and $-s$. This rule allows to infer the massless behavior from the massive one. \ref{app:massless} goes beyond this simplification and provides more details about the distinction between massive and massless particles.


\subsection{Two-body Helicity States}
\label{ssec:2BS}

One-body helicity states \eqref{eq::1BS_phelicity} can be used to build a complete set of states for two-body systems. Because, in the following, these many-body states will be used to describe composite particles, we will focus on obtaining helicity states in the CoMF of the entire many-body system (the ECoMF). Properties of composite particles are most easily obtained in this frame: mass is given by the total energy of the state while spin is given by its total angular momentum. Before to dive into this description, let us shorten a bit notations. In the previous section, one-body states at rest where denoted $\ket{m,sm_s}$, a notation in which the mass, the spin and the spin projection of the particle were specified. In the following, the mass of the particle will be frequently omitted. By default, it will be assumed that the state $\ket{s_i\lambda_i}$ possesses the mass of the $i^{\text{th}}$ particle. Similarly, in the following Sections, boosts along the $z$ axis will be written without specifying any mass parameter. By default, this parameters will be supposed equal to the mass of the state on which the boost acts.

\subsubsection{Helicity bases for two-body systems}

\begin{figure}[b]
\begin{center}
\begin{tikzpicture}
\draw[->] (0,0) - - (0,2);
\draw[->] (0,0) - - (1.5,0);
\draw[->] (0,0) - - (-0.6,-0.6);
\draw[-{Implies},double,very thick,teal] (0.2,0.3) - - (0.2,1.5);
\draw[-{Implies},double,very thick,asparagus] (0.4,0.3) - - (0.4,1.5);
\node[right, teal](s) at (0.4,1.1) {$\ket{s_1\,\lambda_1}$};
\node[right, asparagus](s) at (0.4,0.6) {$\ket{s_2\,\lambda_2}$};
\draw[->,very thick] (2,0.35) - - (4.3,0.35);
\node[above](L) at (3.1,0.3) {$L_{z}(p)$};
\node[below](L) at (3.1,0.3) {$R(0,\pi,0)L_{z}(p)$};

\draw[->] (5.5,0) - - (5.5,2);
\draw[->] (5.5,0) - - (7,0);
\draw[->] (5.5,0) - - (4.9,-0.6);
\draw[-{Implies},double,very thick,teal] (5.7,0.3) - - (5.7,1.5);
\draw[-{Implies},double,very thick,asparagus] (5.7,-0.3) - - (5.7,-1.5);
\draw[-{Stealth},very thick,red] (5.3,0.3) - - (5.3,1.5);
\draw[-{Stealth},very thick,burntorange] (5.3,-0.3) - - (5.3,-1.5);
\node[left, red](s) at (5.3,1.5) {$p\vec{e}_z$};
\node[left, burntorange](s) at (5.3,-1.5) {$-p\vec{e}_z$};
\draw[->,very thick] (7.4,0.35) - - (9.2,0.35);
\node[above](R) at (8.2,0.4) {$R(\phi,\theta,-\phi)$};

\draw[-{Stealth},very thick,red] (10.6,0.4) - - (11.3,1.6);
\draw[-{Implies},double,very thick,teal] (10.6,0.1) - - (11.3,1.3);
\draw[-{Stealth},very thick,burntorange] (10.3,-0.3) - - (9.75,-1.15);
\draw[-{Implies},double,very thick,asparagus] (10.5,-0.2) - - (9.95,-1.1);
\draw[->] (10.5,0) - - (10.5,2);
\draw[->] (10.5,0) - - (12,0);
\draw[->] (10.5,0) - - (9.9,-0.6);
\draw[dashed] (9.75,-1.15) - - (9.75,0.3);
\draw[dashed] (9.75,0.3) - - (10.5,0);
\draw[dashed] (11.3,1.3) - - (11.3,-0.3);
\draw[dashed] (11.3,-0.3) - - (10.5,0);
\node[below](angles) at (11.5,-0.3) {$(\theta,\phi)$};
\end{tikzpicture}
\caption{Schematic illustration for the definition of two-body $p$-helicity state \eqref{eq::2BS_pdef_1}.\label{fig::2BS_scheme}}
\end{center}
\end{figure}
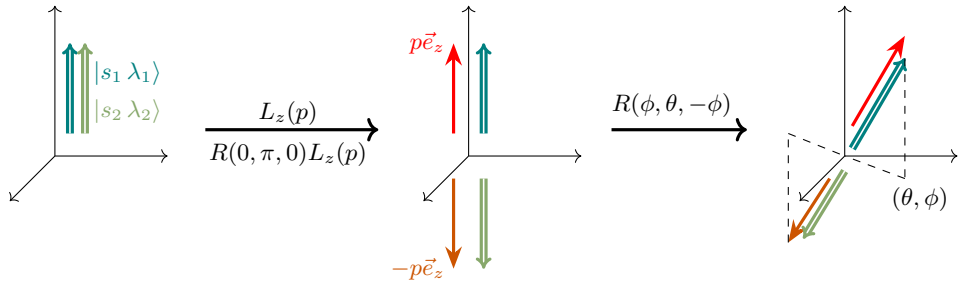

Again, our analysis starts at the level of massive particle states. Differences in presence of massless particles will be discussed subsequently. Two particles are brought in their CoMF by boosting them back-to-back,
\begin{subequations} \label{eq::2BS_pdef}
\begin{align}
    \ket{p\theta\phi ;s_1\lambda_1s_2\lambda_2} =(-1)^{\lambda_2-s_2} U(R(\phi,\theta,0))\left[U(L_z(p))\ket{s_1\lambda_1} \otimes
    U(R(0,\pi,0)L_z(p)) \ket{s_2\lambda_2}\right]&\label{eq::2BS_pdef_1}\\
    = (-1)^{\lambda_2-s_2}\, U(R(\phi,\theta,0)L_z(p))\ket{s_1\lambda_1} \otimes 
    U(R(\pi+\phi,\pi-\theta,\pi)L_z(p))\ket{s_2\lambda_2}&.\label{eq::2BS_pdef_2}
\end{align}
\end{subequations}
Above, $p$, $\theta$ and $\phi$ respectively denotes the modulus, the polar and the azimutal angle of the momentum of the first particle in the ECoMF. Helicities $\lambda_1$ and $\lambda_2$ are defined in that same frame. The phase in front of the definition is mainly conventional \cite{jaco59} and results from the choice to assign to the one-body state with opposed momentum the opposite convention \eqref{eq::1BS_oppOpp}. Fig.~\ref{fig::2BS_scheme} schematically decomposes the successive transformations in definition \eqref{eq::2BS_pdef_1}. By construction, states \eqref{eq::2BS_pdef_1} are eigenstates of particle $1$ and $2$ momentum operators. In the following, they will be referred to as \textit{two-body $p$-helicity states}. 
Their orthonormalisation relation is 
\begin{align}
    \braket{\bar p \bar \theta \bar \phi; s_1\bar\lambda_1 s_2\bar\lambda_2|p\theta\phi; s_1\lambda_1 s_2\lambda_2} =  \frac{4W}{p}\delta(\bar W-W)\delta(\cos\!\bar\theta-\cos\!\theta)\delta(\bar\phi-\phi)\delta_{\bar \lambda_1\lambda_1}\delta_{\bar \lambda_2\lambda_2}\label{eq::2BS_pOrtho1}&\\
    =  \frac{4w_1(p)w_2(p)}{p^2}\delta(\bar p-p)\delta(\cos\!\bar\theta-\cos\!\theta)\delta(\bar\phi-\phi)\delta_{\bar \lambda_1\lambda_1}\delta_{\bar \lambda_2\lambda_2} \label{eq::2BS_pOrtho2}&
\end{align}
where $w_i(p)$ serves as a notation shortcut for $\sqrt{m_i^2+p^2}$ and where $W=w_1(p)+w_2(p)$ is the total energy of the two-body state. If expression \eqref{eq::2BS_pdef_2} is more convenient for practical purposes, \eqref{eq::2BS_pdef_1} unveils that these states present a specific structure: they are obtained by rotating two-body states at rest with a reference orientation, here along the $z$ axis,
\begin{equation}
    \ket{p;s_1\lambda_1s_2\lambda_2} =(-1)^{\lambda_2-s_2} \left(U(L_z(p))\ket{s_1\lambda_1} \otimes
    U(R(0,\pi,0)L_z(p)) \ket{s_2\lambda_2}\right).
\end{equation}
A similar structure will reappear in the definition of helicity states for three-body systems in Section \ref{sec:3BS}. 

As already mentioned, states with a definite total angular momentum are expected. Such a property can be provided to two-body $p$-helicity states by weighting them with a Wigner $D$ matrix and by integrating the result on angular degrees-of-freedom,
\begin{equation}
\ket{p; JM; s_1\lambda_1 s_2\lambda_2} = \sqrt{\frac{2J+1}{4\pi}}\int \diff\!\cos\!\theta \diff \phi \, D^{J*}_{M\,\lambda_1-\lambda_2}(\phi,\theta,0)\ket{p\theta\phi;s_1\lambda_1 s_2\lambda_2}. \label{eq::2BS_Jdef}
\end{equation}
In the following, these states will be referred to as \textit{two-body $J$-helicity states}. The prefactor is introduced to ensure a standard normalisation,
\begin{align}
    \braket{\bar p; \bar J \bar M; s_1\bar\lambda_1 s_2\bar\lambda_2|p; JM; s_1\lambda_1 s_2\lambda_2} =  \frac{4W}{p}\delta(\bar W-W)\delta_{\bar JJ}\delta_{\bar MM}\delta_{\bar \lambda_1\lambda_1}\delta_{\bar \lambda_2\lambda_2}\label{eq::2BS_JOrtho1}&\\
    =  \frac{4w_1(p)w_2(p)}{p^2}\delta(\bar p-p)\delta_{\bar JJ}\delta_{\bar MM}\delta_{\bar \lambda_1\lambda_1}\delta_{\bar \lambda_2\lambda_2}&. \label{eq::2BS_JOrtho2}
\end{align}
Of course, definition \eqref{eq::2BS_Jdef} is only relevant for $J\geq|\lambda_1-\lambda_2|$. It can be shown that, under rotations, two-body $J$-helicity states follows the spin $J$ irreducible representation of $SO(3)$ \cite{wick62,suh71}, meaning that this state indeed has a total angular momentum $J$. For further use, let us mention that definition \eqref{eq::2BS_Jdef} can be inverted, expressing two-body $p$-helicity states as a linear combination of two-body $J$-helicity states \cite{gieb85},
\begin{equation}
\ket{p\theta\phi;s_1\lambda_1 s_2\lambda_2} = \sum_{J=|\lambda_1-\lambda_2|}^{\infty}\,\sum_{M=-J}^J \sqrt{\frac{2J+1}{4\pi}}\, D^{J}_{M\,\lambda_1-\lambda_2}(\phi,\theta,0)\ket{p;JM;s_1\lambda_1s_2\lambda_2}. \label{eq::2BS_pToJ}
\end{equation}

Other properties of these states are abundantly described in the literature \cite{jaco59,mart70,gieb85,math08,suh71}. Their behaviour under parity and permutation of particles, denoted $\mathbb{P}_{12}$, will be reminded here for further use,
\begin{align}
    & \Pi \ket{p; JM; s_1\lambda_1 s_2\lambda_2} = \eta_1 \eta_2 (-1)^{J-s_1-s_2} \ket{p; JM; s_1-\lambda_1 s_2-\lambda_2},\label{eq::2BS_parity}\\
    & \mathbb{P}_{12}\ket{p; JM; s\lambda_1 s\lambda_2} = (-1)^{J+2s} \ket{p; JM; s\lambda_2 s\lambda_1}.\label{eq::2BS_symmetry}
\end{align}
Above, $\eta_i$ is the intrinsic parity of the $i^{\text{th}}$ particle. Symmetry having to be implemented only for identical particles, $s_1$ and $s_2$ have been taken equal each-other in relation \eqref{eq::2BS_symmetry}. These two properties illustrate that two-body $J$-helicity states are neither parity eigenstates nor (anti)symmetric  by themselves. Parity eigenstates are obtained by superposing states with opposed helicity signs. Non-normalised symmetric (anti-symmetric) states are obtained by applying the two-body symmetriser $\mathbb{S}_2$ (anti-symmetriser $\mathbb{A}_2$),
\begin{align}
&\mathbb{S}_2=\mathds{1} + \mathbb{P}_{12},
&&\mathbb{A}_2=\mathds{1} - \mathbb{P}_{12}.
\end{align}
As agreed, helicity states have been written in the zero convention. Switching the convention to the "$-\phi$" one results in a different phase convention for $p$-helicity states. Nevertheless, provided that angles of the Wigner $D$ matrix in definition \eqref{eq::2BS_Jdef} are adapted consequently, it can be shown that two-body $J$-helicity states are not affected.

In presence of massless particles, the results presented above remains correct up to a few adjustments. First in definition \eqref{eq::2BS_pdef_1}-\eqref{eq::2BS_pdef_2}, the conventional phase has to be adapted to remove any spin occurrence, such a quantum number being formally undefined for massless particles. Following the usual misuse that massless particles have spin degrees-of-freedom with forbidden intermediary projections, it seems appropriate to replace $s_2$ by $|\lambda_2|$.  Secondly, relations \eqref{eq::2BS_parity} and \eqref{eq::2BS_symmetry} only accounts for massive particles. It can be shown that, for massless ones, the factor $(-1)^{-s_1-s_2}$ has to be replaced by $(-1)^{\lambda_1+\lambda_2}$ while the $(-1)^{2s}$ factor from \eqref{eq::2BS_symmetry} has to be replaced by $(-1)^{2|\lambda_1|}$. Notice that, as long as helicities are integers, one can naively use massive relations with $\lambda_i=\pm s_i$ to deal with massless particles. The case of helicity states for two massless particles is also briefly discussed at the end of \ref{app:massless}.

Two-body $p$- and $J$-helicity states are not the two only complete sets able to describe two-body systems at rest. For massive particles, one can replace in definition \eqref{eq::2BS_pdef_2} each occurences of $p$-helicity states by $p$-canonical ones. It results in the following definition for the so-called \textit{two-body $p$-canonical states} \cite{suh71},
\begin{equation}
    \begin{aligned}
\ket{p\phi\theta;s_1m_{s_1}s_2m_{s_2}}=\ &
U(R(\phi,\theta,0)L_z(p)R^{-1}(\phi,\theta,0))\ket{s_1m_{s_1}} \\
& \otimes U(R(\pi+\phi,\pi-\theta,0)L_{z}(p)R^{-1}(\pi+\phi,\pi-\theta,0))\ket{s_2m_{s_2}}.
\end{aligned} \label{eq::2BS_pCano}
\end{equation}
These states are then used to define \textit{two-body $J$-canonical states} through intermediary spin and spatial angular momentum couplings,
\begin{equation}
\begin{aligned}
\ket{p;JM;\ell s;s_1s_2}=\sum_{m_\ell,m_s,m_{s_1},m_{s_2}} (\ell m_\ell sm_s|JM)& (s_1m_{s_1}s_2m_{s_2}|sm_s) \\
\int \diff\!\cos\!\theta \diff\phi&\,Y^\ell_{m_\ell}(\theta,\phi) \ket{p\theta\phi;s_1m_{s_1}s_2m_{s_2}}.\label{eq::2BS_lsCano}
\end{aligned}
\end{equation}
Above, $(j_1m_1j_2m_2|j_3m_3)$ refers to a Clebsh-Gordan coefficient and $Y_{m_\ell}^\ell(\theta,\phi)$ to a spherical harmonic. Because of the appearance of an orbital angular momentum quantum number $\ell$, this definition is often used for in non-relativistic treatments. The interested reader will find a description of properties about two-body canonical states in reference \cite{suh71}. It is possible to relate two-body $J$-canonical states to two-body $J$-helicity states,
\begin{equation}
\begin{aligned}
    \ket{p; JM; s_1\lambda_1 s_2\lambda_2}=\sum_{s=|s_1-s_2|}^{s_1+s_2}\sum_{\ell=|J-s|}^{J+s} \sqrt{\frac{2\ell+1}{2J+1}} &(s_1\, \lambda_1\, s_2\, -\lambda_2\,|s\, \lambda_1-\lambda_2)\\
    (\ell&\, 0\, s\, \lambda_1-\lambda_2\,|J\, \lambda_1-\lambda_2)\ket{p;JM;\ell s;s_1s_2}.
    \label{eq::2BS_changeOfBasis}
\end{aligned}
\end{equation}
This relation allows to easily switch from one to the other set of states. For instance, it has been used in \cite{math08} to describe two-gluon glueballs in constituent approaches.

\subsubsection{Decomposition of a physical two-body state in \texorpdfstring{$J$}{Lg}-helicity states}

\label{ssec::2BS_decomPhysState}

Two-body $J$-helicity states can be used to model two-body composite particles. In the ECoMF, any two-body state with spin $J$ and helicity quantum numbers can be decomposed as an integral on internal momentum degree-of-freedom of two-body $p$-helicity states. For two-body systems, the internal motion is ruled by the relative momentum $\bm{p}=(\bm{p_1}-\bm{p_2})/2$ whose modulus and angles have already been denoted $p$, $\theta$ and $\phi$. Let us start by decomposing a generic two-body state in the ECoMF, denoted $\ket{\Phi;s_1\lambda_1s_2\lambda_2}$, as a combination of two-body $p$-helicity states. Using the completeness relation of the latter, one gets
\begin{equation}
    \ket{\Phi;s_1\lambda_1s_2\lambda_2} = \int \frac{p^2\diff p \diff\!\cos\!\theta\diff\phi}{4w_1(p)w_2(p)} \, \Phi(p,\theta,\phi) \ket{p\theta\phi;s_1\lambda_1s_2\lambda_2} \label{eq::2BS_protoWaveFunc}
\end{equation}
where
\begin{equation}
    \Phi(p,\theta,\phi) = \braket{p\theta\phi;s_1\lambda_1s_2\lambda_2|\Phi;s_1\lambda_1s_2\lambda_2}
\end{equation}
is the two-body helicity-momentum wave function of the state. The Jacobian factor in \eqref{eq::2BS_protoWaveFunc} ensure consistence with the normalisation of two-body $p$-helicity states \eqref{eq::2BS_pOrtho2}. Without additional requirements, this state has not the expected definite total angular momentum $J$. However, the definition of two-body $J$-helicity states proved that this property can be supplied by imposing the angular dependence of $\Phi(p,\theta,\phi)$,
\begin{equation}
    \Phi^{J}_{M}(p,\theta,\phi) = \sqrt{\frac{2J+1}{4\pi}}\Psi(p) D^{J*}_{M\,\lambda_1-\lambda_2}(\phi,\theta,0).
\end{equation}
The function $\Psi(p)$ can be understood as a radial helicity-momentum wave function. Replacing  $\Phi$ by $\Phi^{J}_{M}$ in expression \eqref{eq::2BS_protoWaveFunc} results in the decomposition of a generic two-body helicity state with total angular momentum $J$, denoted $\ket{\Psi;JM;s_1\lambda_1s_2\lambda_2}$, in two-body $J$-helicity states,
\begin{equation}
    \ket{\Psi;JM;s_1\lambda_1s_2\lambda_2} = \int \frac{p^2\diff p}{4w_1(p)w_2(p)}\, \Psi(p) \ket{p;JM;s_1\lambda_1s_2\lambda_2}. \label{eq::2BS_WaveFunc}
\end{equation}
States $\ket{\Psi;JM;s_1\lambda_1s_2\lambda_2}$ can be used to model spin $J$ two-body composite particles. If the composite state is made of identical particles or is expected to possess a parity quantum-number, the hereinabove discussion has to be slightly adapted. The structure \eqref{eq::2BS_WaveFunc} still applies but two-body $J$-helicity states in the right-hand side have to be replaced by parity and/or symmetry eigenstates which are obtained using properties \eqref{eq::2BS_parity} and \eqref{eq::2BS_symmetry}. Concerning normalisation, states $\ket{\Psi;JM;s_1\lambda_1s_2\lambda_2}$ are expected to have a unit normalisation,
\begin{equation}
    \braket{\Psi;JM;s_1\lambda_1s_2\lambda_2| \Psi;JM;s_1\lambda_1s_2\lambda_2}=1.
\end{equation}
Using expression \eqref{eq::2BS_WaveFunc} and orthonormalisation of two-body $J$-helicity states \eqref{eq::2BS_JOrtho2}, one can transfer this normalisation condition to the $\Psi(p)$ wave function,
\begin{equation}
    \int \frac{p^2\diff p}{4w_1(p)w_2(p)} \,|\Psi(p)|^2 = 1. \label{eq::2BS_norm}
\end{equation}
For convenience, it is worth introducing a modified radial helicity-momentum wave function, denoted $\Xi(p)$ that includes the Jacobian factors from \eqref{eq::2BS_norm},
\begin{equation}
    \Xi(p) = \frac{p\,\Psi(p)}{2\sqrt{w_1(p)w_2(p)}}
\end{equation}
In terms of $\Xi(p)$, the normalisation condition \eqref{eq::2BS_norm} significantly simplifies
\begin{equation}
    \int \diff p |\Xi(p)|^2 = 1, \label{eq::2BS_norm_xi}
\end{equation}
while \eqref{eq::2BS_WaveFunc} has to be slightly adapted,
\begin{equation}
    \ket{\Psi;JM;s_1\lambda_1s_2\lambda_2} = \int \frac{p\diff p}{2\sqrt{w_1(p)w_2(p)}}\, \Xi(p) \ket{p;JM;s_1\lambda_1s_2\lambda_2}. \label{eq::2BS_WaveFunc_xi}
\end{equation}
Notice that, $\Psi$ having to remain finite for all $p$, $\Xi$ must necessary cancel at $p=0$ for massive particles. Further tests show that this condition stays required in the massless case.

\subsubsection{Calculations with physical two-body states}

\label{ssec::2BS_calc}

The following developments aims to evaluate ME of observables, such as Hamiltonians, on composite particle states. It focuses on unsymmetrical two-body $J$-helicity states \eqref{eq::2BS_WaveFunc_xi} but can easily be adapted to symmetrised odd/even helicity states. States \eqref{eq::2BS_WaveFunc_xi} being written in momentum representation, observables that depends on momentum variables prove to be easier to evaluate. Let us consider an observable $\mathcal{O}(\hat p)$ that only depends on the modulus of the relative momentum (we are only interested in spherically symmetric variables). Let us evaluate $\mathcal{O}(\hat p)$ on $\ket{\Psi;JM;\lambda_1\lambda_2}$,
\begin{equation}
\begin{aligned}
    \bra{\Psi;JM;s_1\bar\lambda_1s_2\bar\lambda_2}\mathcal{O}(\hat p)\ket{\Psi;JM;s_1\lambda_1s_2\lambda_2}&\ =\\
    \int \frac{\bar p\diff \bar p}{2\sqrt{w_1(\bar p)w_2(\bar p)}}\,\frac{p\diff p}{2\sqrt{w_1(p)w_2(p)}} \,\Xi(\bar p)^*\,&\Xi(p)\bra{\bar p;JM;s_1\bar\lambda_1s_2\bar\lambda_2}\mathcal{O}(\hat p)\ket{ p;JM;s_1\lambda_1s_2\lambda_2}.    
\end{aligned}
\end{equation}
Two-body $J$-helicity states being by construction eigenstates of $p$, the action of $\mathcal{O}(p)$ reduces to a simple scalar multiplication. Using orthonormality relation \eqref{eq::2BS_JOrtho2}, one gets
\begin{equation}
    \begin{aligned}
    \bra{\Psi;JM;s_1\bar\lambda_1s_2\bar\lambda_2} O(\hat p) \ket{\Psi;JM;s_1\lambda_1s_2\lambda_2}= \delta_{\lambda_1\bar\lambda_1}\delta_{\lambda_2\bar\lambda_2} \int \diff p\,|\Xi(p)|^2 \mathcal{O}(p).
    \end{aligned} \label{eq::2BS_kinEnergy}
\end{equation}
A similar expression accounts for symmetrised odd/even helicity state as long as these are correctly orthonormalised.

The evaluation of operators that depend on position variables proves to be more difficult to handle. Let us consider $\mathcal{O}(\hat r)$ an operator that only depend on the modulus of the relative position (again, the emphasis is put on spherically symmetric observables), one may require to evaluate it on the composite particle state,
\begin{equation}
    \begin{aligned}
    \bra{\Psi\bar J\bar M;s_1\bar\lambda_1 s_2\bar\lambda_2} \mathcal{O}(\hat r) \ket{\Psi;JM;s_1\lambda_1 s_2\lambda_2}&\ =\\
    \int \frac{\bar p\diff \bar p}{2\sqrt{w_1(\bar p)w_2(\bar p)}}\frac{p\diff p}{2\sqrt{w_1(p)w_2(p)}}\,\Xi(\bar p)^*\,&\Xi( p)\bra{\bar p;\bar J\bar M;s_1\bar\lambda_1s_2\bar\lambda_2}\mathcal{O}(\hat r)\ket{p;JM;s_1\lambda_1s_2\lambda_2}. \label{eq::2BS_posMatElEv_0}
\end{aligned}
\end{equation}
Above, different total angular momenta $J$ and $\bar J$ were considered. This modification has been implemented for further use while studying the case of three-body systems. Two-body $J$-helicity states being not position eigenstate, the evaluation of $\mathcal{O}(\hat r)$ on $\ket{\bar p;JM;s_1\lambda_1s_2\lambda_2}$ requires more developments than the evaluation of $\mathcal{O}(\hat p)$. Let us first switch of basis and develop two-body $J$-helicity states into the canonical basis using relation \eqref{eq::2BS_changeOfBasis},
\begin{equation}
    \begin{aligned}
    &\bra{\bar p;\bar J\bar M;s_1\bar\lambda_1s_2\bar\lambda_2} \mathcal{O}(\hat r) \ket{ p;JM;s_1\lambda_1s_2\lambda_2}\ 
    = \\
    &\hspace{0.5cm}\sum_{\bar s=|s_1-s_2|}^{s_1+s_2}\sum_{\bar \ell=|J-\bar s|}^{J+\bar s} \mathcal{C}^{\bar J;s_1s_2}_{\bar \ell\bar s;\bar\lambda_1\bar\lambda_2} \sum_{s=|s_1-s_2|}^{s_1+s_2}\sum_{\ell=|J- s|}^{J+ s} \mathcal{C}^{ J;s_1s_2}_{\ell s;\lambda_1\lambda_2} \bra{\bar p;\bar J\bar M;\bar \ell\bar s;s_1s_2}\mathcal{O}(\hat r)
    \ket{p;JM;\ell s;s_1s_2}
    \end{aligned}\label{eq::2BS_posMatElEv_1}
\end{equation}
with the following notation shortcut 
\begin{equation}
\mathcal{C}^{J;s_1s_2}_{\ell s;\lambda_1\lambda_2} = \sqrt{\frac{2\ell+1}{2J+1}} (s_1\, \lambda_1\, s_2\, -\lambda_2\,|s\, \lambda_1-\lambda_2)
(\ell\, 0\, s\, \lambda_1-\lambda_2\,|J\, \lambda_1-\lambda_2). \label{eq::2BS_C_coeff}
\end{equation}
The problem now reduces to the evaluation of $\mathcal{O}(\hat r)$ on the more traditional canonical states. Using the definition \eqref{eq::2BS_lsCano} of two-body $J$-canonical states in terms of $p$-ones, one obtains
\begin{equation}
\begin{aligned}
    &\bra{\bar p;\bar J\bar M;\bar \ell\bar s;s_1s_2}\mathcal{O}(\hat r)
    \ket{p;JM;\ell s;s_1s_2}\\
    &\hspace{0.5cm}= \sum_{\bar m_\ell'\bar m_s'\bar m_{s_1}'\bar m_{s_2}'} (\bar \ell\bar m_\ell'\bar s\bar m_s'|\bar J\bar M) (s_1\bar m'_{s_1}s_2\bar m'_{s_2}|\bar s\bar m'_s) \int \diff\!\cos\!\bar\theta' \diff\bar\phi'\,Y^{\bar \ell*}_{\bar m_\ell'}(\bar \theta',\bar \phi') \\ &\hspace{1.5cm}\sum_{m_\ell'm_s'm'_{s_1}m'_{s_2}} (\ell m'_\ell s m'_s|JM)(s_1 m'_{s_1}s_2 m'_{s_2}| s m'_s) \int \diff\!\cos\!\theta' \diff\phi'\,Y^{ \ell}_{m_\ell'}(\theta',\phi') \\ &\hspace{5.7cm}\bra{\bar p\bar \theta'\bar \phi';s_1\bar m'_{s_1}s_2\bar m'_{s_2}}\mathcal{O}(\hat r)\ket{p\theta' \phi';s_1m'_{s_1}s_2m'_{s_2}}
\end{aligned}
\end{equation}
Because spin degrees-of-freedom are uncorrelated from spatial ones in canonical states and because the observable $\mathcal{O}(\hat r)$ does not affect them, the right-hand side residual ME are proportional to $\delta_{\bar m'_{s_1}m'_{s_1}}\delta_{\bar m'_{s_2}m'_{s_2}}$. Concerning spatial degrees-of-freedom, the action of the observable can be evaluated using a result from reference \cite{eyre86}\footnote{The result from \cite{eyre86} must slightly be adapted in two different ways. Firstly, our normalisation conventions being different, a kinematic factor have been added. Secondly, a property of Legendre polynomials is used to get spherical harmonics back.},
\begin{equation}
    \begin{aligned}
    \bra{\bar p\bar\theta'\bar\phi';s_1\bar m'_{s_1}s_2\bar m'_{s_2}} \mathcal{O}(\hat r) &\ket{p\theta'\phi';s_1m'_{s_1}s_2m'_{s_2}}\\
    = \delta_{\bar m'_{s_1}m'_{s_1}}\delta_{\bar m'_{s_2}m'_{s_2}} 4&\sqrt{w_1(\bar p)w_2(\bar p)w_1(p)w_2(p)} \sum_{\ell''m''_\ell}Y_{m''_\ell}^{\ell''}(\bar\theta',\bar\phi')Y_{m''_\ell}^{\ell''*}(\theta',\phi')\mathcal{O}_{\ell''}(\bar p,p)
    \end{aligned}
\end{equation}
where 
\begin{equation}
\mathcal{O}_{\ell}(\bar p,p)=\frac{2}{\pi} \int_0^\infty j_\ell(\bar pr)\mathcal{O}(r)j_\ell(pr)r^2\diff r,
\end{equation} $j_\ell(x)$ being spherical Bessel functions. Making use of spherical harmonics and Clebsh-Gordan coefficients properties, one gets
\begin{equation}
\begin{aligned}
    &\bra{\bar p;\bar J\bar M;\bar \ell\bar s;s_1s_2}\mathcal{O}(\hat r)
    \ket{p;JM;\ell s;s_1s_2} = 4\sqrt{w_1(\bar p)w_2(\bar p)w_1(p)w_2(p)}\, \mathcal{O}_{\ell}(\bar p, p) \delta_{\bar J J}\delta_{\bar M M}\delta_{\bar \ell \ell}\delta_{\bar s s}.
\end{aligned}
\end{equation}
This expression is inserted in \eqref{eq::2BS_posMatElEv_1} and leads to the following formula for the evaluation of $\mathcal{O}(\hat r)$ on two-body $J$-helicity states,
\begin{equation}
    \begin{aligned}
    &\bra{\bar p;\bar J\bar M;s_1\bar\lambda_1s_2\bar\lambda_2} \mathcal{O}(\hat r) \ket{ p;JM;s_1\lambda_1s_2\lambda_2} \\
    &\hspace{1cm} = \delta_{\bar J J}\delta_{\bar M M} \sum_{s=|s_1-s_2|}^{s_1+s_2}\sum_{\ell=|J- s|}^{J+ s} \mathcal{C}^{J;s_1s_2}_{\ell s;\bar \lambda_1\bar \lambda_2}\mathcal{C}^{J;s_1s_2}_{\ell s;\lambda_1\lambda_2}\,4\sqrt{w_1(\bar p)w_2(\bar p)w_1(p)w_2(p)}\, \mathcal{O}_{\ell}(\bar p,p).
    \end{aligned}\label{eq::2BS_posMatElEv_2}
\end{equation}
Finally, back to the state for the composite particle, expression \eqref{eq::2BS_posMatElEv_0} becomes
\begin{equation}
    \begin{aligned}
    &\bra{\Psi;\bar J\bar M;s_1\bar\lambda_1 s_2\bar\lambda_2} \mathcal{O}(\hat r) \ket{\Psi;JM;s_1\lambda_1 s_2\lambda_2} \\
    &\hspace{1cm} = \delta_{\bar J J}\delta_{\bar M M} \sum_{s=|s_1-s_2|}^{s_1+s_2}\sum_{\ell=|J- s|}^{J+ s} \mathcal{C}^{J;s_1s_2}_{\ell s;\bar\lambda_1\bar\lambda_2} \mathcal{C}^{J;s_1s_2}_{\ell s;\lambda_1\lambda_2} \int p\diff p\,\bar p\diff \bar p \ \Xi(\bar p)^*\,\Xi( p)\, \mathcal{O}_{\ell}(\bar p,p). \label{eq::2BS_posMatElEv_fin}
    \end{aligned}
\end{equation}
As expected due to angular momentum conservation, ME for central potentials are proportional to $\delta_{\bar J J}\delta_{\bar M M}$. Moreover, as soon as non-zero, these ME are also independent of the total angular momentum projection $M$. The evaluation of position ME on symmetrised odd/even helicity states only differs from the one on two-body $J$-helicity by the coefficients of the expansion in canonical states. Relation \eqref{eq::2BS_posMatElEv_fin} remains valid provided that the expression \eqref{eq::2BS_C_coeff} for $\mathcal{C}$ coefficients is correctly adapted. 

Analytical expressions of $\mathcal{O}_{l}(\bar p,p)$ for different potentials are given in \cite{eyre86}. For instance, in presence of a Yukawa potential $\mathcal{O}(r)=-\alpha e^{-\eta r}/r$, one gets
\begin{equation}
    \mathcal{O}_{\ell}(\bar p,p) = -\frac{\alpha}{\pi \bar p p}Q_\ell\left(\frac{\bar p^2+p^2+\eta^2}{2\bar p p}\right) \label{eq::2BS_partWavExpYuka}
\end{equation}
with $Q_l(x)$ a second kind Legendre function \cite{abra64}. This expression can be naively plugged into the integral from \eqref{eq::2BS_posMatElEv_fin} without any 	extra precaution, even for $\eta=0$. As a result, the treatment of Coulombic interactions only require the evaluation of the following integrals,
\begin{equation}
    \int  p\diff p\,\bar p\diff \bar p \ \Xi(\bar p)^*\,\Xi(p) \, \mathcal{O}_{\ell}(\bar p,p) = - \frac{\alpha}{\pi} \int \diff p\,\diff \bar p \ \Xi(\bar p)^*\,\Xi(p)\,Q_\ell\left(\frac{\bar p^2+p^2}{2\bar p p}\right). \label{eq::2BS_coulomb}
\end{equation}
Although they present a logarithmic divergence for $p=\bar p$, Gaussian quadrature allows for a reasonably efficient evaluation of these integrals. This is even easier by introducing a new set of coordinates,
\begin{align}
    &v = \bar p + p, &&\bar v=\bar p- p, && \diff v\diff \bar v = 2\diff p \diff \bar p. \label{eq::2BS_uv}
\end{align}
As a result, equation \eqref{eq::2BS_coulomb} becomes
\begin{equation}
\begin{aligned}
&\int  p\diff p\,\bar p\diff \bar p \ \Xi(\bar p)^*\,\Xi(p) \, \mathcal{O}_{\ell}(\bar p,p)\\
&\hspace{1.5cm} = - \frac{\alpha}{2\pi} \int_0^\infty \diff v\int_{-v}^v \diff \bar v \ \Xi\left(\frac{v+\bar v}{2}\right)^*\,\Xi\left(\frac{v-\bar v}{2}\right)\,Q_\ell\left(\frac{v^2+\bar v^2}{v^2-\bar v^2}\right). \label{eq::2BS_coulomb_uv}
\end{aligned}
\end{equation}
The situation is more delicate concerning linear potentials. The Fourier transform of these potentials are not traditional functions but are distributions \cite{hers93}. To overcome this difficulty, the $\mathcal{O}_\ell$ function associated to this potential is defined by screening the linear potential with a decreasing exponential,
\begin{equation}
\mathcal{O}(r)=\lambda r = \lim_{\eta \rightarrow 0}\lambda r e^{-\eta r}= -\lim_{\eta \rightarrow 0}\frac{\partial^2}{\partial\eta^2}\left(-\lambda \frac{e^{-\eta r}}{r}\right).
\end{equation}
The second equality relates the linear potential to a derivative of the Yukawa potential whose $\mathcal{O}_\ell$ function is provided hereinabove. Using this relation, before taking the limit for $\eta$, one gets
\begin{equation}
\begin{aligned}
    \mathcal{O}^\eta_{\ell}(\bar p,p) &= \frac{\lambda}{\pi \bar p p} 
    \frac{\partial^2}{\partial\eta^2}\left(Q_\ell\left(z_\eta\right)\right) \text{ with } z_\eta=\frac{\bar p^2+p^2+\eta^2}{2\bar p p}\\
    &= \frac{\lambda}{\pi \bar p p} \frac{\partial}{\partial\eta}\left(\frac{\eta Q_\ell'\left(z_\eta\right)}{p\bar p}\right) =  \frac{\lambda}{\pi} \left(\frac{ Q_\ell'\left(z_\eta\right)}{(p\bar p)^2}+\eta^2\frac{ Q_\ell''\left(z_\eta\right)}{(p\bar p)^3}\right)
\end{aligned}
\end{equation}
where $Q_\ell'$ and $Q_\ell''$ denotes the first and second derivatives of $Q_\ell$ with respect to its argument. The limit on $\eta$ is taken after integration and has to be considered carefully to ensure a finite result. Reference \cite{hers93} develops formulas for this ME with arbitrary $\ell$,
\begin{equation}
    \begin{aligned}
    \lim_{\eta\rightarrow0}\int & p\diff p\,\bar p\diff \bar p \ \Xi(\bar p)^*\,\Xi(p) \, \mathcal{O}_{\ell}^\eta(\bar p,p) \\
    & = \frac{\lambda}{\pi}\int_0^\infty \diff \bar p\ \Xi(\bar p)^* \,P\int_0^\infty \left(\frac{4\bar p^2\, \Xi(\bar p)}{(\bar p^2 - p^2)^2} + Q'_\ell\left(\frac{p^2+\bar p^2}{2p\bar p}\right)\frac{ \Xi(p)}{p \bar p}\right)\diff p \label{eq::2BS_HersFormula}
    \end{aligned}
\end{equation}
where $P$ denotes the Cauchy principal value of the integral. The left-hand side of relation \eqref{eq::2BS_HersFormula} presents an apparent difference with its equivalent in \cite{hers93}. This is because what is called $V^\ell_\eta(p,\bar p)$ in \cite{hers93} is actually equal to $p\bar p\,\mathcal{O}_{\ell}^\eta(\bar p,p)$. Integrals from \eqref{eq::2BS_HersFormula} are quite difficult to handle, mainly due to the Cauchy principal value. To overcome this problem, coordinates \eqref{eq::2BS_uv} are again introduced. Performing the change, \eqref{eq::2BS_HersFormula} becomes
\begin{equation}
    \begin{aligned}
    \lim_{\eta\rightarrow0}&\int  p\diff p\,\bar p\diff \bar p \ \Xi(\bar p)^*\,\Xi(p) \, \mathcal{O}_{\ell}^\eta(\bar p,p) \\
     = \frac{\lambda}{\pi}& \int_0^\infty \diff v\, P \int_{-v}^{v} \diff \bar v\ \Xi\left(\frac{v+\bar v}{2}\right)^* \left(\frac{(v+\bar v)^2}{2v^2\bar v^2}\,\Xi\left(\frac{v+\bar v}{2}\right) + Q'_\ell\left(\frac{v^2+\bar v^2}{v^2-\bar v^2}\right)\frac{2\,\Xi\left(\frac{v-\bar v}{2}\right)}{v^2-\bar v^2} \right).\label{eq::2BS_HersFormula_uv}
    \end{aligned}
\end{equation}
With these coordinates, the Cauchy principal value in the integration on $\bar v$ can naturally be taken into account by performing a Gauss-Legendre quadrature with an even number of points \cite{norb91,hers93}. Formulas \eqref{eq::2BS_coulomb} and \eqref{eq::2BS_HersFormula_uv} allows for an efficient evaluation of ME with \eqref{eq::2BS_posMatElEv_fin}. \ref{app:exem_momspace} illustrates the calculation of Hamiltonian matrix-elements using formulas \eqref{eq::2BS_kinEnergy}, \eqref{eq::2BS_coulomb} and \eqref{eq::2BS_HersFormula_uv}.


\section{Two-gluon Glueballs}
\label{sec:2GB}

The previous Section defined two-body helicity states and introduced a way to compute the corresponding ME. In the current Section, this formalism is applied to the description of glueballs in the framework of constituent approaches. Glueballs are modeled by colorless bound states of several constituent gluons. The latter are considered as bosons with helicity degrees of freedom $\lambda = \pm 1$ and negative intrinsic parity. The mass of the constituent gluon remains a controversial subject. Although formally massless, the QCD gluon has proven to acquire a dynamical mass in the non-perturbative regime \cite{corn82,cucc07,agui08}. This incited some studies to consider a massive kinetic energy for constituent gluons (with a mass around $0.5$ GeV) \cite{math08b,hou01}. On the other hand, reference \cite{math08} tends to indicate that the challenge in modeling glueballs consists more in the correct use of the helicity formalism than in the choice of the kinematics. In addition, it has already been observed in other constituent approaches that a modification in the kinematics of the system can be absorbed in modifications of the parameters from the potential \cite{sema92}. For these reasons, the current work will consider an ultrarelativistic kinetic energy for the constituent gluon. This Section limits the discussion to two-gluon bound states that produces the so-called two-gluon glueballs. The colorless constraint on the wave function implies for two-gluon glueballs a positive charge conjugation and a symmetric color part of the state \cite{boul08}. Gluons having a bosonic nature, the spin-space part of the two-body state must also be symmetrical.

To start with, let us focus on the construction of symmetrical two-body $J$-helicity states with $\lambda_i = \pm 1$ and having a definite parity. Making use of properties \eqref{eq::2BS_parity} and \eqref{eq::2BS_symmetry}, reference \cite{math08} constructs four sets of such symmetric parity eigenstates,
\begin{subequations}
\begin{align}
    &\ket{p;S_+;J^P=(2k)^+} = \frac{1}{\sqrt{2}}(\ket{p;JM;+1+1} + \ket{p;JM;-1-1}), \label{eq::2GB_S+}\\
    &\ket{p;S_-;J^P=(2k)^-} = \frac{1}{\sqrt{2}}(\ket{p;JM;+1+1} - \ket{p;JM;-1-1}),\label{eq::2GB_S-}\\
    &\ket{p;D_+;J^P=(2k+2)^+} = \frac{1}{\sqrt{2}}(\ket{p;JM;+1-1} + \ket{p;JM;-1+1}),\label{eq::2GB_D+}\\
    &\ket{p;D_-;J^P=(2k+3)^+} = \frac{1}{\sqrt{2}}(\ket{p;JM;+1-1} - \ket{p;JM;-1+1}).\label{eq::2GB_D-}
\end{align}
\end{subequations}
with $k \in \mathbb{N}$. For readability, $s_1$ and $s_2$ labels have been omitted from the notation. One can show that these states are correctly orthonormalised. In each of the four sets, parity and signature of $J$ had to be constrained to ensure a correct symmetry of the state. In addition, the constraint $J \geq |\lambda_1-\lambda_2|$ in definition \eqref{eq::2BS_Jdef} forbade the occurrence of total angular momenta smaller than two in $\ket{D_\pm;J^P}$ sets. The linear combinations of helicity states \eqref{eq::2GB_S+} to \eqref{eq::2GB_D-} can be expanded in terms of two-body $J$-canonical states making use of \eqref{eq::2BS_changeOfBasis}. This calculation has been performed in reference \cite{math08}. The resulting expansions are displayed in Table~\ref{tab::2GB_expStates}.
\begin{table}
\begin{tabular}{c}
\hline\hline \\[-1mm]
$\begin{aligned}
    &\begin{aligned}
    \ket{p;S_+;(2k)^+} = &\ \sqrt{\frac{2}{3}} \ket{p;\,^{1}2k_{2k}} - \sqrt{\frac{2k(2k+1)}{3(4k-1)(4k+3)}} \ket{p;\,^{5}2k_{2k}} \\
    & + \sqrt{\frac{k(2k-1)}{(4k+1)(4k-1)}} \ket{p;\,^{5}2k-2_{2k}} + \sqrt{\frac{(k+1)(2k+1)}{(4k+3)(4k+1)}} \ket{p;\,^{5}2k+2_{2k}},
    \end{aligned} \\
    &\ket{p;S_-;(2k)^-} = \sqrt{\frac{2k}{4k+1}} \ket{p;\,^{3}2k-1_{2k}} - \sqrt{\frac{2k+1}{4k+1}} \ket{p;\,^{3}2k+1_{2k}},\\
    &\begin{aligned}
    \ket{p;D_+;(2k+2)^+} = &\ \sqrt{\frac{(k+2)(2k+3)}{(4k+3)(4k+5)}} \ket{p;\,^{5}2k_{2k+2}} + \sqrt{\frac{6(k+2)(2k+1)}{(4k+3)(4k+7)}} \ket{p;\,^{5}2k+2_{2k+2}} \\
     & + \sqrt{\frac{(k+1)(2k+1)}{(4k+5)(4k+7)}} \ket{p;\,^{5}2k+4_{2k+2}},
    \end{aligned} \\
    &\ket{p;D_-;(2k+3)^+} = -\sqrt{\frac{2k+5}{4k+7}} \ket{p;\,^{5}2k+2_{2k+3}} - \sqrt{\frac{2(k+1)}{4k+7}} \ket{p;\,^{5}2k+4_{2k+3}}.
\end{aligned}$ 
\\[-1mm]\\[-1mm]
\hline\hline
\end{tabular}
\caption{Expansion of symmetrised parity eigenstate two-body $J$-helicity states in canonical $J$-helicity states from \cite{math08}. A condensed notation for $J$-canonical states has been used,
$\ket{p;\,^{2s+1}l_J} = \ket{p;JM;ls;s_1s_2}$.}
\label{tab::2GB_expStates}
\end{table}

As suggested in the previous Section, these two-gluon states $\ket{p;S_\pm/D_\pm;J^P}$ are integrated on their momentum degree-of-freedom to produce a generic glueball state,
\begin{align}
    \ket{\Psi;S_\pm/D_\pm;J^P} = \int \frac{\diff p}{2}\, \Xi(p) \ket{p;S_\pm/D_\pm;J^P}, \label{eq::2GB_WaveFunc_xi}
\end{align}
where $\Xi(p)$ satisfy the normalisation condition \eqref{eq::2BS_norm_xi}. Compared to \eqref{eq::2BS_WaveFunc_xi}, expression \eqref{eq::2GB_WaveFunc_xi} replaces $\sqrt{w_1(p)w_2(p)}$ by $p$ because both gluons are assumed to be massless. One can now go back over the demonstration of relations \eqref{eq::2BS_kinEnergy} and \eqref{eq::2BS_posMatElEv_fin} to generalise them for the symmetrical $\ket{\Psi;S_\pm/D_\pm;J^P}$ states. On the one hand, the generalisation of \eqref{eq::2BS_kinEnergy} is immediate. Symmetrical states being similarly orthonormalised, the demonstration does not require any modification and one gets,
\begin{equation}
    \begin{aligned}
    \bra{\Psi;S_\pm/D_\pm;J^P} O(\hat p) \ket{\Psi;S_\pm/D_\pm;J^P} = \int \diff p\,|\Xi(p)|^2 \mathcal{O}(p).
    \end{aligned} \label{eq::2GB_KME}
\end{equation}
Kinetic energy ME that mix states with different labels are shown to cancel. Because states are naturally written in momentum space, angular momentum does not play any role in the calculation. It results in a left-hand side of equation \eqref{eq::2GB_KME} independent of $S/D$ and $J$. This was not the case in \cite{math08} where calculations were performed in coordinate space. On the other hand, relation \eqref{eq::2BS_posMatElEv_fin} requires a slight modification. Because symmetrised states satisfies a different expansion in canonical states, coefficients $\mathcal{C}^{J;s_1s_2}_{\ell s;\lambda_1\lambda_2}$ from relation \eqref{eq::2BS_posMatElEv_fin} have to be replaced by the expansion coefficients from Table~\ref{tab::2GB_expStates}. Denoting  the latter $\mathcal{C}^{J}_{\ell s}$, one gets the following analog to \eqref{eq::2BS_posMatElEv_fin},
\begin{equation}
    \bra{\Psi;S_\pm/D_\pm;J^P} \mathcal{O}(\hat r) \ket{\Psi;S_\pm/D_\pm;J^P} = \sum_{\phantom{|}s\,=\,0\phantom{|}}^{2}\sum_{\ell=|J- s|}^{J+ s} \left(\mathcal{C}^{J}_{\ell s}\right)^2 \int p\diff p\,\bar p\diff \bar p \ \Xi(\bar p)^*\,\Xi( p)\, \mathcal{O}_{\ell}(\bar p,p).
\end{equation}

These two formulas can now be used to concretely compute a glueball spectrum by means of a variational approach.  A Gaussian shape supplemented by a $p$ factor that cancels at $p=0$ is suggested as a trial helicity-momentum wave function,
\begin{equation}
    \Xi_a(p) = A p e^{-ap^2}. \label{eq::2GB_trial_wave_func}
\end{equation}
In the following, the approximation provided by this trial state will be referred to as the \textit{single Gaussian approximation} (SGA). To enable comparison, the Hamiltonian considered is the same as the one in \cite{math08},
\begin{equation}
    \mathcal{H}_{\text{GB}} = 2\sqrt{p^2} + \frac{9\sigma}{4} r - 3 \frac{\alpha_s}{r}
\end{equation}
where $\alpha_s = 0.450$ is the strong coupling constant and $\sigma = 0.185\,$GeV$^2$ is the mesonic string tension. The factor $9/4$ comes from the Casimir scaling hypothesis \cite{sema04} and the factor $3$ corresponds to the color charge associated to a pair of constituent gluons in a color singlet. Instanton contributions, added in \cite{math08} to split the degeneracy of the lowest states, are not considered at first. Masses obtained using the SGA for the low-lying angular momenta are displayed in the second column of Table~\ref{tab::2GB_singGaussApp}. These results are compared to energies from reference \cite{math08} (instanton contribution are manually removed). At first sight, both values seem incompatible. Energies from the SGA, a variational method supposed to provide upper bounds, lies below the ones from reference \cite{math08}, where a very accurate resolution method has been used (namely, the Lagrange mesh method \cite{sema01}). This incompatibility is left noticing that reference \cite{math08} solves the same Hamiltonian using a fundamentally different approach. First of all, the total orbital angular momentum of the system is evaluated on the symmetric parity eigenstates from Table~\ref{tab::2GB_expStates}. The matrix obtained in this way turns out to be diagonal with respective diagonal elements $J(J+1)+2$ for $\ket{p;S_\pm,J^P}$ states and $J(J+1)-2$ for $\ket{p;D_\pm,J^P}$ states. Consequently, reference \cite{math08} suggests to replace the $\ket{p;S_\pm/D_\pm,J^P}$ states by canonical ones with effective angular momenta $\ell_{\text{eff}}$ such that
\begin{equation}
    \ell_{\text{eff}}(\ell_{\text{eff}}+1) = J(J+1)\pm 2. 
\end{equation}
The Hamiltonian matrix is then evaluated on these effective canonical states rather than on the true two-body helicity states. Mimicking this strategy with the SGA give rise to the third column of Table~\ref{tab::2GB_singGaussApp}. Such calculations require the evaluation of second kind Legendre functions for non-integer $\ell$ values. As expected, this modified SGA provides upper bounds of the energies from \cite{math08}. 

\begin{figure}
\centering
\includegraphics[scale=0.3]{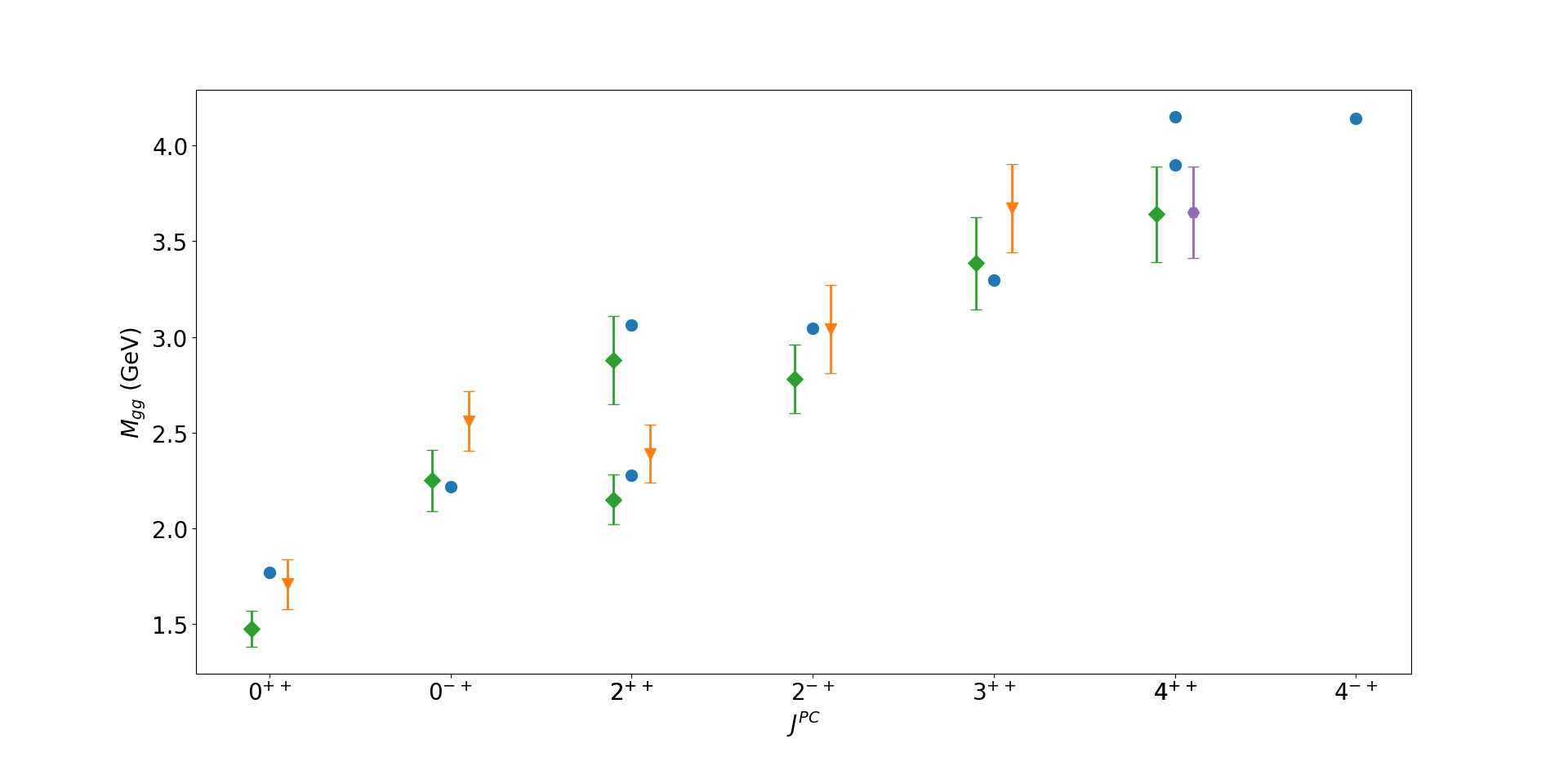}
\caption{Comparison of two-gluon glueball spectra. Upper bounds obtained with the SGA (blue circles) are compared to lattice QCD results from \cite{chen06} (orange triangles), \cite{meye05} (green diamonds) and \cite{liu02} (purple hexagon).}
\label{fig:2GB_comp}
\end{figure}

Second and fourth columns are compared with LQCD results in Table~\ref{tab::2GB_singGaussApp} and in Fig.~\ref{fig:2GB_comp}. First of all, one notices that most of the masses provided by the original SGA lies below the ones from \cite{math08}. Nevertheless, in most of the case, relative differences between both methods lies around a few percents. The main exception to this claim occurs for the $0^+$ state. In reference \cite{math08}, $0^+$ and $0^-$ states proves to be degenerated thereby requiring the introduction of instanton interactions to split both levels. The original SGA naturally raises this degeneracy. It does not mean that instanton does not contribute in the glueball spectrum but that their effects maybe overestimated by the use of the method from \cite{math08}. All results seem in global agreement with LQCD results \cite{morn99,chen06,meye05,liu02}, no matter which methodology is used.

\begin{table}
\begin{tabular}{lrrrrrrrr}
\hline
State\hspace{5mm} & \hspace{5mm} $E_{\text{or.SGA}}$ & \hspace{5mm} [$\delta$] & \hspace{5mm} $E_{\text{mod.SGA}}$ & \hspace{4mm} [$\delta$] & \hspace{5mm} Ref \cite{math08} & \hspace{20mm} LQCD \cite{chen06} & \hspace{20mm} LQCD \cite{meye05} \\
\hline\hline
$\ket{\Psi;S_+,0^+}$ &  $1.769$ & [$19\%$] &  $2.216$ & [$2\%$] & $2.174$ & $1.710\pm0.050\pm0.080$ & $1.475\pm0.030\pm0.065$ \\ 
$\ket{\Psi;S_-,0^-}$ &  $2.216$ & [$2\%$] & $2.216$ & [$2\%$] & $2.174$ & $2.560\pm0.035\pm0.120$ & $2.250\pm0.060\pm0.100$ \\ 
$\ket{\Psi;D_+,2^+}$ &  $2.279$ & [$12\%$] &  $2.651$ & [$2\%$] & $2.588$ & $2.390\pm0.030\pm0.120$ & $2.150\pm0.030\pm0.100$ \\ 
$\ket{\Psi;S_+,2^+}$ &  $3.060$ & [$1\%$] & $3.194$ & [$4\%$] & $3.077$ & N.A. & $2.880\pm0.100\pm0.130$ \\ 
$\ket{\Psi;S_-,2^-}$ &  $3.043$ & [$1\%$] &  $3.194$ & [$4\%$] & $3.077$ & $3.040\pm0.040\pm0.150$ & $2.780\pm0.050\pm0.130$ \\ 
$\ket{\Psi;D_-,3^+}$ &  $3.297$ & [$1\%$] & $3.393$ & [$4\%$] & $3.254$ & $3.670\pm0.050\pm0.180$ & $3.385\pm0.090\pm0.150$ \\
$\ket{\Psi;D_+,4^+}$ &  $3.897$ & [$3\%$] &   $3.981$ & [$6\%$] & $3.768$ & N.A. & $3.640\pm0.090\pm0.160$ \\ 
$\ket{\Psi;S_+,4^+}$ &  $4.150$ & [$5\%$] & $4.204$  & [$6\%$] & $3.961$ & N.A. & N.A. \\ 
$\ket{\Psi;S_-,4^-}$ &  $4.139$ & [$4\%$] & $4.204$ & [$6\%$] & $3.961$ & N.A. & N.A. \\
\hline
\end{tabular}
\caption{Comparison of two-gluon glueball spectra. Upper bounds obtained with the original SGA, $E_{\text{or.SGA}}$, are compared to the modified SGA, $E_{\text{mod.SGA}}$, and to the spectrum from \cite{math08} for which instanton contributions have been removed for the comparison. Some LQCD results from \cite{chen06,meye05} are added as points of comparison. A supplementary LQCD calculations \cite{liu02} that predicts a $\ket{\Psi;D_+,4^+}$ state of $3.650\pm0.060\pm0.180$ GeV can be mentioned. Energy results are provided in GeV. Relative differences with \cite{math08}, $\delta$, are indicated in square brackets.
}
\label{tab::2GB_singGaussApp}
\end{table}

It can be worth considering an extension of the SGA by incorporating a second Gaussian trial wave function \eqref{eq::2GB_trial_wave_func} within the variational approach \cite{suzu98}. The single Hamiltonian ME to compute in the SGA is replaced by the evaluation of $2$ by $2$ Hamiltonian and overlap matrices, which form the core of a generalised eigenvalue problem,
\begin{equation}
\begin{pmatrix}
\bra{\Psi_a}\mathcal{H}_{\text{GB}}\ket{\Psi_a} & \bra{\Psi_a}\mathcal{H}_{\text{GB}}\ket{\Psi_b} \\
\bra{\Psi_b}\mathcal{H}_{\text{GB}}\ket{\Psi_a} & \bra{\Psi_b}\mathcal{H}_{\text{GB}}\ket{\Psi_b} \\
\end{pmatrix}\begin{pmatrix}
C_1\\
C_2\\
\end{pmatrix} = E \begin{pmatrix}
\braket{\Psi_a|\Psi_a} & \braket{\Psi_a|\Psi_b} \\
\braket{\Psi_b|\Psi_a} & \braket{\Psi_b|\Psi_b}\\
\end{pmatrix} \begin{pmatrix}
C_1\\
C_2\\
\end{pmatrix}\label{eq::2GB_genEig}
\end{equation}
where $\ket{\Psi_a}$ serves as a shorthand for any symmetric helicity states $\ket{\Psi;S_\pm/D_\pm;J^P}$, with $a$ being the non-linear variational parameter considered. This double Gaussian approximation (DGA) serves two purposes: to assess the accuracy of the SGA, and to explore first radially excited states. The non-linear variational parameters of each Gaussian, denoted $a$ and $b$ above, are treated as independent, and optimisation is performed on both. Results are presented in Table~\ref{tab::2GB_doubleGaussApp}. For all considered states, the energy difference between the SGA and DGA does not exceed $0.1$ GeV (it even decreases to below $0.01$ GeV for high angular momenta). These findings suggest that the SGA is sufficiently accurate to study low-lying two-gluon glueball states. For radially excited states, as expected, the DGA results aligns in magnitude with those from reference \cite{math08}.

\begin{table}
\begin{tabular}{lrrrrr}
\hline
State\hspace{5mm} & \hspace{5mm} $E_{\text{or.SGA}}$ & \hspace{5mm} $E_{\text{DGA}}$ & \hspace{5mm} Ref \cite{math08} & \hspace{20mm} LQCD \cite{chen06,morn99} & \hspace{20mm} LQCD \cite{meye05} \\
\hline\hline
$\ket{\Psi_1;S_+,0^+}$ & $1.769$ & $1.668$ & $2.174$ & $1.710\pm0.050\pm0.080$ & $1.475\pm0.030\pm0.065$ \\ 
$\ket{\Psi_2;S_+,0^+}$ & N.A. & $2.808$ & $2.993$ & $2.670\pm0.180\pm0.130$ & $2.755\pm0.070\pm0.120$ \\ 
$\ket{\Psi_1;S_-,0^-}$ & $2.216$ & $2.202$ & $2.174$ & $2.560\pm0.035\pm0.120$ & $2.250\pm0.060\pm0.100$ \\
$\ket{\Psi_2;S_-,0^-}$ & N.A. & $3.069$ & $2.993$ & $3.640\pm0.060\pm0.180$ & $3.370\pm0.150\pm0.150$\\ 
$\ket{\Psi_1;D_+,2^+}$ & $2.279$ & $2.241$ & $2.588$ & $2.390\pm0.030\pm0.120$ & $2.150\pm0.030\pm0.100$ \\ 
$\ket{\Psi_2;D_+,2^+}$ & N.A. & $3.249$ & $3.325$ & N.A. & N.A.\\ 
$\ket{\Psi_1;S_+,2^+}$ & $3.060$ & $3.042$ & $3.077$ & N.A. & $2.880\pm0.100\pm0.130$  \\ 
$\ket{\Psi_2;S_+,2^+}$ & N.A. & $3.852$ & $3.732$ & N.A. & N.A. \\ 
$\ket{\Psi_1;S_-,2^-}$ & $3.043$ & $3.042$ & $3.077$ & $3.040\pm0.040\pm0.150$ & $2.780\pm0.050\pm0.130$ \\ 
$\ket{\Psi_2;S_-,2^-}$ & N.A. & $3.845$ & $3.732$ & $3.890\pm0.040\pm0.190$ & $3.480\pm0.140\pm0.160$\\ 
$\ket{\Psi_1;D_-,3^+}$ & $3.297$ & $3.295$ & $3.254$ & $3.670\pm0.050\pm0.180$ & $3.385\pm0.090\pm0.150$ \\
$\ket{\Psi_2;D_-,3^+}$ & N.A. & $4.067$ & $3.882$ & N.A. & N.A. \\
\hline
\end{tabular}
\caption{Comparison of two-gluon glueball spectra. Upper bounds obtained with the original SGA, $E_{\text{or.SGA}}$, are compared to the DGA, $E_{\text{DGA}}$, and to the spectrum from \cite{math08} from which instanton contributions have been removed for the comparison. Labels $\Psi_1$ and $\Psi_2$ respectively refer to the fundamental and to the first excited state. Some LQCD results from \cite{chen06,morn99,meye05} are added as points of comparison. Energy results are provided in GeV.}
\label{tab::2GB_doubleGaussApp}
\end{table}


\section{The Helicity Formalism for Three-body systems}
\label{sec:3BS}
As for two-body systems, efforts are made to produce complete sets of three-body helicity states possessing a total angular momentum $J$ in the ECoMF. These sets of three-body states are used to decompose the state of composite particles at rest. There are two main ways to define helicity states for three-body systems, both ways having their own strengths and weaknesses. These two definitions are reviewed in the following subsections.


\subsection{Berman's Definition}
\label{ssec:BD}

\begin{figure}[b]
\begin{center}
\begin{tikzpicture}
\tikzset{
    partial ellipse/.style args={#1:#2:#3}{
        insert path={+ (#1:#3) arc (#1:#2:#3)}}}

\draw[->] (0,0) - - (0,2);
\node[left](z) at (0,2) {$z$};
\draw[->] (0,0) - - (2,0);
\node[right](y) at (2,0) {$y$};
\draw[->] (0,0) - - (-0.6,-0.6);
\node[left](x) at (-0.6,-0.6) {$x$};
\draw[-{Implies},double,very thick,teal] (0.2,0.3) - - (0.2,1.5);
\draw[-{Implies},double,very thick,asparagus] (0.4,0.3) - - (0.4,1.5);
\draw[-{Implies},double,very thick,armygreen] (0.6,0.3) - - (0.6,1.5);
\node[right, teal](s) at (0.6,1.4) {$\ket{s_1\,\lambda_1}$};
\node[right, asparagus](s) at (0.6,0.9) {$\ket{s_2\,\lambda_2}$};
\node[right, armygreen](s) at (0.6,0.4) {$\ket{s_3\,\lambda_3}$};
\draw[->,very thick] (2.8,0.35) - - (5.2,0.35);
\node[above](L) at (4,0.4) {\scriptsize $R(\phi_i,\pi/2,0)L_z(p_i)$};

\draw[->] (7,0) - - (7,2);
\draw[->] (7,0) - - (8.5,0);
\draw[->] (7,0) - - (6.4,-0.6);
\draw[-{Stealth},very thick,blue(pigment)] (7,0) - - (7,1.4);
\node[left,blue(pigment)](pp) at (7,1.4) {\scriptsize $\vec{p}_1\times\vec{p}_2$};
\draw[dashed,bluebell] (7.45,-0.3) - - (8.05,0.3) - - (6.55,0.3) - - (5.95,-0.3) - - (7.45,-0.3);
\draw[-{Implies},double,very thick,armygreen] (6.9,0.1) - - (5.7,0.1);
\draw[-{Implies},double,very thick,teal] (7.15,-0.1) - - (7.75,-0.8);
\draw[-{Implies},double,very thick,asparagus] (7.3,0.1) - - (8.1,0.6);
\draw[-{Stealth},very thick,atomictangerine] (6.7,-0.05) - - (5.5,-0.05);
\node[below,atomictangerine](p3) at (5.5,-0) {\scriptsize $\vec{p}_3$};
\draw[-{Stealth},very thick,red] (7,-0.2) - - (7.6,-0.9);
\node[left,red](p1) at (7.6,-0.9) {\scriptsize $\vec{p}_1$};
\draw[-{Stealth},very thick,burntorange] (7.05,0.1) - - (7.85,0.6);
\node[above left,burntorange](p2) at (7.85,0.55) {\scriptsize $\vec{p}_2$};
\draw[->,very thick] (8.7,0.35) - - (10.5,0.35);
\node[above](R) at (9.5,0.4) {\scriptsize $R(\alpha,\beta,\gamma)$};

\draw[-{Stealth},very thick,blue(pigment)] (12,0) - - (11.3,1.2);
\node[left, blue(pigment)](angles) at (11.25,-0.2) {\scriptsize $(\alpha,\beta)$};
\draw[-{Straight Barb[length=0.3mm,width=0.6mm]},thick,blue(pigment)] (11.6,0.65) [partial ellipse=0:340:0.15cm and 0.09cm] ;
\node[above right, blue(pigment)](angles) at (11.6,0.65) {\scriptsize $\gamma$};
\draw[->] (12,0) - - (12,2);
\draw[->] (12,0) - - (13.5,0);
\draw[->] (12,0) - - (11.4,-0.6);
\draw[dashed,bluebell] (12.7,-0.25) - - (12.93,0.75) - - (11.43,0.23) - - (11.1,-0.72) - - (12.7,-0.25);
\draw[dashed,thick,blue(pigment)] (12,0) - - (11.2,-0.2) - - (11.3,1.3);

\end{tikzpicture}
\caption{Visual interpretation for the definition of Berman's $p$-helicity states \eqref{eq::BD_pdef1}.\label{fig::BD_f1}}
\end{center}
\end{figure}
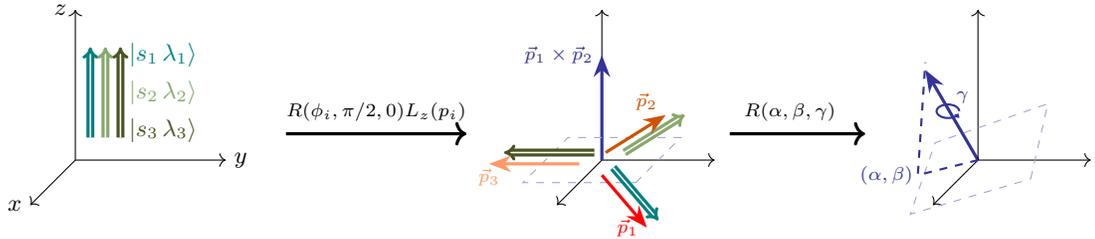

Berman's three-body helicity states definition is based on the observation that, in their CoMF, any set of three particles always lies and moves in a given plane. This feature allows the following geometrical construction for the tensorial product of three one-body helicity states in the ECoMF \cite{berm65,suh71},
\begin{align}
\ket{\alpha\beta\gamma;w_1w_2w_3;\lambda_1\lambda_2\lambda_3}=U(R(\alpha,\beta,\gamma))\big[&U(R(\phi_1,\pi/2,0)L_z(p_1))\ket{s_1\lambda_1} \nonumber \\
\otimes\ &U(R(\phi_2,\pi/2,0)L_z(p_2))\ket{s_2\lambda_2}  \label{eq::BD_pdef1}\\
\otimes\ &U(R(\phi_3,\pi/2,0)L_z(p_3))\ket{s_3\lambda_3}\big] \nonumber
\end{align}
where $p_i$ and $\phi_i$ are fixed combinations of particle's energies, $w_1$, $w_2$ and $w_3$, 
\begin{equation}
\begin{aligned}
&p_i=\sqrt{w_i^2-m_i^2}, &&\cos \varphi_{ij} = \frac{p_k^2-p_i^2-p_j^2}{2p_ip_j},
\end{aligned}\label{eq::BD_pdef2_1}
\end{equation}
\begin{equation}
\begin{aligned}
 &\phi_1=\varphi_{13}-\pi/2, &&\phi_2=\varphi_{13} + \varphi_{12} - \pi/2, &&\phi_3=3\pi/2.
\end{aligned} \label{eq::BD_pdef2_2}
\end{equation}
Above, $\varphi_{ij}$ angles are always to be taken in between $0$ and $\pi$. Individual masses and spins are not reminded in the notation for the sake of conciseness. As for two-body $p$-helicity states, definition \eqref{eq::BD_pdef1} can be decomposed in two pieces. Inside the square brackets, reference momenta are provided to each of the three particles using helicity convention. Their modulus are defined so that the $i$-th particle has an energy $w_i$, while their direction is chosen so that it respects four conditions.
\begin{enumerate}
\item Each momentum lies in the $xy$ plane.
\item The sum of the momenta is the null vector.
\item The momentum of the third particle is along the $y$ direction toward negatives.
\item The cross product of the momenta of particle $1$ and $2$ is along the $z$ axis toward positives.
\end{enumerate}
All these conditions are ensured by definitions \eqref{eq::BD_pdef2_1} and \eqref{eq::BD_pdef2_2}. This define what will be thereafter named \textit{Berman's reference state} and denoted as follows,
\begin{equation}
\begin{aligned}
\ket{w_1w_2w_3;\lambda_1\lambda_2\lambda_3} = U(R(\phi_1,\pi/2,0)L_z(p_1))\ket{s_1\lambda_1}\, &\otimes\, U(R(\phi_2,\pi/2,0)L_z(p_2))\ket{s_2\lambda_2}\\ 
&\otimes\, U(R(\phi_3,\pi/2,0)L_z(p_3))\ket{s_3\lambda_3}.
\end{aligned} \label{eq::BD_refdef}
\end{equation}
Once the reference state is built, in a second phase, this state is rotated with Euler angles $(\alpha,\beta,\gamma)$ to give its orientation to the plane in which lies the three particles. This decomposition provides its interpretation to the angles $(\alpha,\beta,\gamma)$, these are the angles of the normal to particles' plane. A visual interpretation of the aforementioned structure is proposed in Fig.~\ref{fig::BD_f1}. States defined in \eqref{eq::BD_pdef1} will be thereafter referred to as \textit{Berman's $p$-helicity states}.\footnote{This name refers to one of the authors of a pioneer work using this definition \cite{berm65}. Conventions in this work slightly differs from the ones used here, the third particle being taken along the $x$-axis towards negatives. For completeness, let us also mention the names of J.Werle and M.Jacob that abundantly contributed to this definition too \cite{werl63,berm65}.} These satisfies the following orthonormality relation \cite{berm65,suh71},
\begin{equation}
\begin{aligned}
\braket{\bar\alpha\bar\beta\bar\gamma ; \bar w_1 \bar w_2 \bar w_3; \bar\lambda_1 \bar\lambda_2 \bar\lambda_3|\alpha\beta\gamma ; w_1 w_2 w_3; \lambda_1 \lambda_2 \lambda_3} = 8 \delta(w_1-\bar w_1)\delta(w_2-\bar w_2)\delta(w_3-\bar w_3)&\\
\delta(\alpha-\bar\alpha)\delta(\cos\beta -\cos \,\bar\beta)\delta(\gamma-\bar\gamma)\delta_{\lambda_1 \bar\lambda_1 }\delta_{\lambda_2 \bar\lambda_2 }\delta_{\lambda_3 \bar\lambda_3}&,
\end{aligned}\label{eq::BD_pO}
\end{equation}
Normalisation announced in \cite{suh71} differ from this one by a multiplicative constant. It is due to a different choice of convention for normalisation of one-body helicity states (as a reminder, the convention used here is the one of \cite{mart70}).
\vspace{2mm}

The definition \eqref{eq::BD_pdef1} introduces three-body helicity states with defined directions for the momenta and therefore no good total angular momentum. It is as straightforward as for two-body $J$-helicity states to combine these states to overcome this deficiency \cite{berm65,suh71},
\begin{equation}
\begin{aligned}
\ket{JM\mu ; w_1 w_2 w_3 ; \lambda_1 \lambda_2 \lambda_3} = & \sqrt{\frac{2J+1}{8\pi^2}} \int \diff\alpha \diff\!\cos\!\beta \diff\gamma \, D^{J*}_{M\mu}(\alpha,\beta,\gamma)\ket{\alpha\beta\gamma ; w_1 w_2  w_3 ; \lambda_1 \lambda_2 \lambda_3}\\
= \sqrt{\frac{2J+1}{8\pi^2}} & \int \diff\alpha\diff\!\cos\!\beta\diff\gamma \, D^{J*}_{M\mu}(\alpha,\beta,\gamma)\,U\left(R\left(\alpha,\beta,\gamma\right)\right)\,\ket{ w_1 w_2 w_3 ; \lambda_1 \lambda_2 \lambda_3}.
\end{aligned}\label{eq::BD_Jdef}
\end{equation}
States defined hereinabove will be referred to as \textit{Berman's $J$-helicity states} and satisfy the following orthonormality relation \cite{suh71},
\begin{equation}
\begin{aligned}
\braket{\bar J \bar M \bar\mu ; \bar w_1 \bar w_2 \bar w_3 ; \bar \lambda_1 \bar \lambda_2 
\bar \lambda_3|JM\mu ; w_1 w_2 w_3; \lambda_1 \lambda_2 \lambda_3} = 8\delta(w_1-\bar w_1)\delta(w_2-\bar w_2)\delta(w_3-\bar w_3)&\\
 \delta_{J\bar J} \delta_{M\bar M}\delta_{\mu\bar\mu} \delta_{\lambda_1\bar\lambda_1}\delta_{\lambda_2\bar\lambda_2}\delta_{\lambda_3\bar\lambda_3}&.
\end{aligned}\label{eq::BD_JO}
\end{equation}
This relation can also be compared to the results in reference \cite{werl63}, where different orientation conventions are used, and in reference \cite{berm65}, where an expression only including space degrees-of-freedoms is provided. In addition to the total angular momentum $J$ and its projection $M$, a third quantum number $\mu$ is produced. This quantum number corresponds to the projection of the total angular momentum along the normal to the plane \cite{suh71}. Even if the direction of this normal is not fixed due to the integration on the $(\alpha,\beta,\gamma)$ angles. The projection of $J$ along the normal to the plane is well-defined. One may wonder if the choice to set up the reference state in the $xy$ plane really matters, the angles of this plane being integrated in Berman's $J$-helicity states anyway. It is demonstrated in \ref{app:prop_berm} that a modification of this reference plane truely impact the definition of Berman's $J$-helicity states,
\begin{equation}
\begin{aligned}
\ket{JM\mu ; w_1 w_2 w_3; \lambda_1 \lambda_2 \lambda_3} &= \sum_{\mu'=-J}^J D^{J}_{\mu\mu'}(\bar R^{-1}) \ket{JM\mu'; w_1 w_2 w_3; \lambda_1 \lambda_2 \lambda_3}_{\bar R} 
\end{aligned} \label{eq::BD_changeRefState}
\end{equation}
where the state denoted $\ket{JM\mu ; w_1 w_2 w_3 ; \lambda_1 \lambda_2 \lambda_3}_{\bar R}$ is built from a different initial configuration, with the rotation $\bar R$ relating the reference planes used to define the states on either side of equation \eqref{eq::BD_changeRefState}. The change of reference plane modified the definition of the $\mu$ quantum number.

The action of parity and permutation operators on Berman's states can be investigated. As for one- and two-body helicity states, Berman's $J$-helicity states are not parity eigenstates by themselves \cite{suh71},
\begin{align}
&\Pi \ket{JM\mu ; w_1 w_2 w_3 ; \lambda_1 \lambda_2 \lambda_3}= \eta_1\eta_2\eta_3(-1)^{-s_1-s_2-s_3-\mu} \ket{JM\mu ; w_1 w_2 w_3 ; -\lambda_1 -\lambda_2 -\lambda_3}. \label{eq::BD_parity}
\end{align}
Above, $\eta_i$ is still the intrinsic parity of the $i^{\text{th}}$ particle. To get parity eigenstates, linear combinations of Berman's $J$-helicity  states that mixes helicity signs have to be considered. Concerning permutations of particles, let $\mathbb{P}_{ij}$ be the operator that represents exchange operations between particles $i$ and $j$. Berman's $J$-helicity states are not eigenstates of permutations but
\begin{subequations}\begin{align}
&\mathbb{P}_{12} \ket{JM\mu ; w_1 w_2 w_3; \lambda_1 \lambda_2 \lambda_3}= (-1)^{J+\mu+\lambda_1+\lambda_2-\lambda_3} \ket{JM-\mu ; w_2 w_1 w_3; \lambda_2 \lambda_1 \lambda_3},\label{eq::BD_P12}\\
&\mathbb{P}_{13} \ket{JM\mu ; w_1 w_2 w_3; \lambda_1 \lambda_2 \lambda_3}= (-1)^{J-\mu-\lambda_1-\lambda_2-\lambda_3} e^{-i\varphi_{13}\mu} \ket{JM-\mu ; w_3 w_2 w_1; \lambda_3 \lambda_2 \lambda_1},\label{eq::BD_P13}\\
&\mathbb{P}_{23} \ket{JM\mu ; w_1 w_2 w_3; \lambda_1 \lambda_2 \lambda_3}= (-1)^{J+\mu+\lambda_1-\lambda_2-\lambda_3} e^{i\varphi_{23}\mu} \ket{JM-\mu ; w_1 w_3 w_2; \lambda_1 \lambda_3 \lambda_2},
\end{align}\label{eq::BD_P23}\end{subequations}
where expressions of $\varphi_{ij}$ are given in \eqref{eq::BD_pdef2_1}. The relation for $\mathbb{P}_{12}$ can be found in \cite{berm65,suh71} but results for $\mathbb{P}_{13}$ and $\mathbb{P}_{23}$ are new. The proof of \eqref{eq::BD_P13} is given in \ref{app:prop_berm}. This proof can easily be adapt to demonstrate \eqref{eq::BD_P23}. It has been checked that hereinabove relations are consistent with $S_3$ multiplication table, \begin{equation}
    \mathbb{P}_{23} = \mathbb{P}_{12}\mathbb{P}_{13}\mathbb{P}_{12}.
\end{equation}
Non-normalised symmetric (anti-symmetric) states can be obtained by applying the three-body symmetriser $\mathbb{S}_3$ (anti-symmetriser $\mathbb{A}_3$) on each Berman's $J$-helicity state \eqref{eq::BD_Jdef},
\begin{align}
\mathbb{S}_3&=\mathds{1} + \mathbb{P}_{12} + \mathbb{P}_{13} + \mathbb{P}_{12}\mathbb{P}_{13}\mathbb{P}_{12} + \mathbb{P}_{13}\mathbb{P}_{12} + \mathbb{P}_{12}\mathbb{P}_{13},\\
\mathbb{A}_3&=\mathds{1} - \mathbb{P}_{12} - \mathbb{P}_{13} - \mathbb{P}_{12}\mathbb{P}_{13}\mathbb{P}_{12} + \mathbb{P}_{13}\mathbb{P}_{12} + \mathbb{P}_{12}\mathbb{P}_{13}.
\end{align}
Symmeric and antisymmetric parity eigenstates for three-gluon systems are built in Section \ref{sec:3GB}.

\subsubsection{Decomposition of a physical three-body state in Berman's \texorpdfstring{$J$}{Lg}-helicity states}

\label{ssec::BD_decomPhysState}

In the same way as for two-body systems, any three-body bound state in the ECoMF with spin $J$ and with helicity quantum numbers can be decomposed as an integral on the internal momentum degrees-of-freedom of Berman's $p$-helicity states. But contrary to the two-body case, this integral can be written in many possible variables. Let us first introduce Jacobi coordinates. These coordinates, denoted $\bm{x}$ and $\bm{y}$, complemented with the center-of-mass position, $\bm{R}$, are abundantly used in the literature to deal with bound states of three identical bodies. In terms of individual positions, they reads
\begin{align}
    &\bm{x}=\bm{x_1} - \bm{x_2}, &&\bm{y}=\frac{\bm{x_1} + \bm{x_2}}{2} - \bm{r_3}, && \bm{R} = \frac{\bm{x_1} + \bm{x_2} + \bm{r_3}}{3}.
\end{align}
Berman's states being momentum eigenstates, Jacobi coordinates will take part in the following through their conjugated momenta,
\begin{align}
& \bm{p_x} = \frac{\bm{p_1} - \bm{p_2}}{2}, && \bm{p_y} = \frac{\bm{p_1}+\bm{p_2}-2\bm{p_3}}{3} && \bm{P}=\bm{p_1} +\bm{p_2} +\bm{p_3},\label{LME_jacobip}
\end{align}
where $\bm{P}$ is the total momentum of the system. This choice of coordinates has the advantage that it does not bring any Jacobian compared to individual momenta,
\begin{equation}
    \diff^3p_1\diff^3p_2\diff^3p_3 = \diff^3\!P\diff^3p_x\diff^3p_y.
\end{equation}
Another possible system of coordinates that may be used consists of three angles that describe the plane in which the momenta lies as well as the three momentum length. One will recognize in this set of coordinates the $\alpha$, $\beta$, $\gamma$, $p_1$, $p_2$ and $p_3$ variables used in Berman's definition of $p$-helicity states \eqref{eq::BD_pdef1}. Equivalently, one can replace the $p_i$ variables by the $w_i$ ones, these being related each other through relation \eqref{eq::BD_pdef2_1}. In the following, these coordinates will be called \textit{pseudo-momentum perimetric coordinates} (PMP-coordinates), in analogy with position perimetric coordinates, used, for instance, in \cite{hess02}. The full change of coordinates from Jacobi to PMP is not mandatory for the sake of this discussion. Let us simply mention that, for three massless particles, the modulus of Jacobi coordinates can be related to $w_i$ variables as follows
\begin{align}
    &p_x^2 = (2w_1^2+2w_2^2-w_3^2)/4, &&p_y^2=w_3^2,\label{eq::BD_jacobiToW}
\end{align}
and that, between these variables, a non-trivial Jacobian has to be taken into account,
\begin{equation}
    \diff^3p_x \diff^3p_y = w_1 w_2 w_3\, \diff w_1 \diff w_2 \diff w_3 \,\diff\alpha \diff\!\cos\!\beta \diff\gamma. \label{eq::BD_jacobiToW_J}
\end{equation}
This relation can be demonstrated by considering the intermediary system of coordinates that supplements $\alpha$, $\beta$, $\gamma$ with the respective moduli of $\bm{p_x}$ and $\bm{p_y}$ as well as the angle between these two vectors, denoted $\cos\theta_{xy}$ \cite{hess02}.

Out of the two possible sets of coordinates, the momentum perimetric ones clearly fits better with Berman's definition of helicity states. Making use of the completness relation of Berman's $p$-helicity states, a generic three-body helicity state in the ECoMF, $\ket{\Phi;\lambda_1\lambda_2\lambda_3}$, is naturally decomposed in coordinates $w_1$, $w_2$, $w_3$, $\alpha$, $\beta$ and $\gamma$,
\begin{equation}
    \ket{\Phi;\lambda_1 \lambda_2 \lambda_3} = \int \frac{\diff w_1 \diff w_2 \diff w_3 \; \diff\alpha \diff\!\cos\!\beta \diff\gamma}{8} \;\Phi(\alpha,\beta,\gamma,w_1,w_2,w_3)\ket{\alpha\beta\gamma; w_1w_2w_3; \lambda_1\lambda_2\lambda_3}\label{eq::BD_protoWaveFunc}
\end{equation}
where
\begin{equation}
   \Phi(\alpha,\beta,\gamma,w_1,w_2,w_3) = \braket{\alpha\beta\gamma; w_1w_2w_3; \lambda_1\lambda_2\lambda_3|\Phi;\lambda_1 \lambda_2 \lambda_3} 
\end{equation}
is the three-body helicity-momentum wave function of the state. Above, $w_1$ and $w_2$ are integrated along the set of positive real while $w_3$ is bounded by $|w_1-w_2|$ and $w_1+w_2$ so that $\bm P = \bm 0$ is possible. This relation is the three-body equivalent to the two-body decomposition \eqref{eq::2BS_protoWaveFunc}. Here again, this state has not the expected definite total angular momentum $J$. But reminding Berman's definition of $J$-helicity states, this property can be supplied to the state by imposing the angular dependence of $\Phi(\alpha,\beta,\gamma,w_1,w_2,w_3)$,
\begin{equation}
    \Phi^{J}_{M\mu}(\alpha,\beta,\gamma,w_1,w_2,w_3) = \sqrt{\frac{2J+1}{8\pi^2}}\Psi(w_1,w_2,w_3) D^{J*}_{M\mu}(\alpha,\beta,\gamma). \label{eq::BD_JWaveFunc}
\end{equation}
Replacing  $\Phi$ by $\Phi^{J}_{M\mu}$ in expression \eqref{eq::BD_protoWaveFunc} results in the decomposition of a generic three-body helicity state with total angular momentum $J$, denoted $\ket{\Psi;JM\mu;\lambda_1\lambda_2\lambda_3}$, in Berman's $J$-helicity states,
\begin{equation}
    \ket{\Psi;JM\mu;\lambda_1\lambda_2\lambda_3} = \int \frac{\diff w_1 \diff w_2 \diff w_3}{8}\;\Psi(w_1,w_2,w_3)\ket{JM\mu;w_1w_2w_3;\lambda_1\lambda_2\lambda_3}.\label{eq::BD_JPhysState}
\end{equation}
Imposing a unit normalisation for  $\ket{\Psi;JM\mu;\lambda_1\lambda_2\lambda_3}$ and making use the orthonormalisation of Berman's $J$-helicity \eqref{eq::BD_JO} allows to infer the normalisation condition on $\Psi(w_1,w_2,w_3)$,
\begin{equation}
\braket{\Psi;JM\mu;\lambda_1\lambda_2\lambda_3|\Psi;JM\mu;\lambda_1\lambda_2\lambda_3} = \int \frac{\diff w_1 \diff w_2 \diff w_3}{8}\;|\Psi(w_1,w_2,w_3)|^2 = 1.\label{eq::BD_physStateNorm}
\end{equation}
States $\ket{\Psi;JM\mu;\lambda_1\lambda_2\lambda_3}$ can be used to model spin $J$ three-body composite particles made of three constituents. Although the aforementioned developments considered three identical particles, they can be easily generalised to three different particles. The same commentary than for two-body systems (cf. Section \ref{ssec::2BS_decomPhysState}) about symmetry and parity applies. In presence of identical particles and/or if a parity quantum-number is expected, Berman's $J$-helicity states have to be replaced by parity and/or symmetry eigenstates which can be obtained using properties \eqref{eq::BD_parity} to \eqref{eq::BD_P23}. This construction will be performed in the special case of three-gluon systems in Section \ref{sec:3GB}.


\subsection{Wick's definition} 
\label{ssec::WD}

Next to Berman's proposal to define three-body helicity states, Wick suggests another scheme based on an intermediate two-body coupling. First, following relation \eqref{eq::2BS_Jdef}, particles $1$ and $2$ are coupled in a two-body $J$-helicity state at rest, 
\begin{align}
\ket{p_{12}; j_{12}\lambda_{12}; s_1\lambda_1' s_2\lambda_2'} = \sqrt{\frac{2j_{12}+1}{4\pi}}\int \diff\!\cos\!\theta_{12} \diff \phi_{12} \, D^{j_{12}*}_{\lambda_{12}\, \lambda_1'-\lambda_2'}(\phi_{12},\theta_{12},0)\ket{p_{12}\theta_{12}\phi_{12};s_1\lambda_1' s_2\lambda_2'}.\label{eq::WD_sub2BS}
\end{align}
Above, $p_{12}$, $\theta_{12}$ and $\phi_{12}$ respectively denotes the momentum modulus, polar and azimutal angle of particle $1$ in the CoMF of particles $1$ and $2$ ($12$-CoMF). Helicities $\lambda_1'$ and $\lambda_2'$ are also defined in this frame. By construction, states defined by equation \eqref{eq::WD_sub2BS} possess a definite total angular momentum for particles $1$ and $2$, denoted $j_{12}$. The third particle is then taken into account. The subsystem of particle $1$ and $2$ is considered as a composite particle of spin $j_{12}$, of spin projection $\lambda_{12}$ and of mass $m_{12}(p_{12})$ with
\begin{equation}
    m_{12}(p_{12}) = \sqrt{p_{12}^{2}+m_1^2} + \sqrt{p_{12}^{2}+m_2^2}. \label{eq::WD_invMass_12}
\end{equation}
This mass is necessarily non-zero. The composite particle is then coupled with the third particle in the same way it has been done for particles $1$ and $2$. Because masses of the different states are less explicit than before, these quantities are temporary brought back in the notation of the boosts along the $z$ axis. The composite particle and the third one are boosted in their own center of mass frame, which coincide with the ECoMF,
\begin{equation}
\begin{aligned}
\ket{p\theta\phi;j_{12}\lambda_{12}s_3\lambda_3;p_{12}s_1\lambda_1's_2\lambda_2'} = (-1)^{\lambda_3-s_3}\, U(R(\phi,\theta,0)L_z(m_{12}(p_{12}),p))\ket{p_{12};j_{12}\lambda_{12};s_1\lambda_1's_2\lambda_2'} & \\
\otimes\, U(R(\pi+\phi,\pi-\theta,\pi)L_z(m_3,p))\ket{s_3\lambda_3}&. \label{eq::WD_3bdef_1}
\end{aligned}
\end{equation}
Above, $p$, $\theta$ and $\phi$ respectively denotes the modulus, the polar and the azimutal angle of the momentum of the composite particle in the ECoMF. By construction, this momentum is opposite to that of the third particle, both thereby having the same modulus. For this reason, in the following, the notation $p_3$ will supplant $p$. States \eqref{eq::WD_3bdef_1} will be referred to as \textit{Wick's $p$-helicity states}. Even if quantum numbers that describe the internal motion of the subsystem have been included in the notations, definition \eqref{eq::WD_3bdef_1} has the exact same structure than \eqref{eq::2BS_pdef_1}. Therefore, a total angular momentum can be provided to the whole system by integrating on momentum angles as well,
\begin{equation}
\begin{aligned}
&\ket{p_3; JM; j_{12}\lambda_{12} s_3\lambda_3; p_{12}s_1\lambda_1's_2\lambda_2'}\\
&\hspace{2cm} = \sqrt{\frac{2J+1}{4\pi}}\int \diff\!\cos\!\theta \diff \phi \, D^{J*}_{M\, \lambda_{12}-\lambda_3}(\phi,\theta,0)\ket{p_3\theta\phi;j_{12}\lambda_{12}s_3\lambda_3;p_{12}s_1\lambda_1's_2\lambda_2'}.\label{eq::WD_3bdef_2}
\end{aligned}    
\end{equation}
These three-body helicity states will be referred as \textit{Wick's $J$-helicity states}. Their orthonormality relation is the following,
\begin{equation}    
\begin{aligned}
&\braket{\bar{p}_3;\bar{J}\bar{M};\bar{j}_{12}\bar{\lambda}_{12} s_3\bar{\lambda}_3;\bar{p}_{12} s_1\bar{\lambda}_1' s_2\bar{\lambda}_2'|p_3;JM;j_{12}\lambda_{12}s_3\lambda_3;p_{12}s_1\lambda_1' s_2\lambda_2'} \\
&\hspace{3cm}=\frac{8W(p_{12},p_3)}{p_3p_{12}}\,\delta(\bar{W}-W)\delta(\bar{m}_{12}-m_{12})\delta_{\bar{J}J}\delta_{\bar{M}M}\delta_{\bar{j}_{12}j_{12}}\delta_{\bar{ \lambda}_{12}\lambda_{12}}
\delta_{\bar{ \lambda}_{1}'\lambda_{1}'}\delta_{\bar{\lambda}_{2}'\lambda_{2}'}
\delta_{\bar{ \lambda}_{3}\lambda_{3}}.\\
\end{aligned}\label{eq::WD_norm_massive}
\end{equation}
Above, $W$ is the total energy of the three-body system in its CoMF,
\begin{equation}
    W(p_{12},p_3) = \sqrt{m_{12}(p_{12})^2+p^2_3} + \sqrt{m_3^2+p_3^2}.
\end{equation}
As mentioned above, $p_{12}$ is defined in the 12-CoMF, whereas $p_3$ is a momentum in the overall CoMF. Relation \eqref{eq::WD_norm_massive} is to be compared to the one provided in reference \cite{wick62}. Both are not directly equivalent because the definition of Wick's $J$-helicity states from this reference differs by an additional $(p_3p_{12}/Wm_{12})^{1/2}/4$ kinematic factor from ours. For further use, this relation is specified for the case of three massless particles. In that case, the relations between $p_{12}$, $p_3$, $m_{12}$ and $W$ simplifies
\begin{equation}
\begin{cases}
W= p_3+\sqrt{4p_{12}^{\,2}+p_3^2},\\
m_{12}=2p_{12}
\end{cases} \Longleftrightarrow \hspace{2mm} \begin{cases}
p_3= (W^2-m_{12}^2)/(2W),\\
p_{12}=m_{12}/2
\end{cases} \label{eq::WD_Wm_pp}
\end{equation}
As a result, in the massless case, relation \eqref{eq::WD_norm_massive} is shown to become
\begin{equation}    
\begin{aligned}
&\braket{\bar{p}_3;\bar{J}\bar{M};\bar{j}_{12}\bar{\lambda}_{12} s_3\bar{\lambda}_3;\bar{p}_{12} s_1\bar{\lambda}_1' s_2\bar{\lambda}_2'|p_3;JM;j_{12}\lambda_{12}s_3\lambda_3;p_{12}s_1\lambda_1' s_2\lambda_2'} \\
&\hspace{3cm}=\frac{2^5W^2}{(W^2-m_{12}^2)m_{12}}\,\delta(\bar{W}-W)\delta(\bar{m}_{12}-m_{12})\delta_{\bar{J}J}\delta_{\bar{M}M}\delta_{\bar{j}_{12}j_{12}}\delta_{\bar{ \lambda}_{12}\lambda_{12}}
\delta_{\bar{ \lambda}_{1}'\lambda_{1}'}\delta_{\bar{\lambda}_{2}'\lambda_{2}'}
\delta_{\bar{ \lambda}_{3}\lambda_{3}}.\\
\end{aligned}\label{eq::WD_norm}
\end{equation}
For the current purpose, this relation will prove more comfortable to use in terms of $p_3$ and $p_{12}$, these variables being directly momenta of particles in different frames,\footnote{The Jacobian determinant related to this change is $\diff W \diff m_{12} = 2W(p_{12},p_3)/\sqrt{p_3^2+4p_{12}^2}\  \diff p_3 \diff p_{12}$}
\begin{equation}    
\begin{aligned}
&\braket{\bar{p}_3;\bar{J}\bar{M};\bar{j}_{12}\bar{\lambda}_{12} s_3\bar{\lambda}_3;\bar{p}_{12} s_1\bar{\lambda}_1' s_2\bar{\lambda}_2'|p_3;JM;j_{12}\lambda_{12}s_3\lambda_3;p_{12}s_1\lambda_1's_2\lambda_2'} \\
&\hspace{3cm}=\frac{4\sqrt{p_3^2+4p_{12}^{\,2}}}{p_3p_{12}}\,\delta(\bar{p_3}-p_3)\delta(\bar p_{12}-p_{12})\delta_{\bar{J}J}\delta_{\bar{M}M}\delta_{\bar{j}_{12}j_{12}}\delta_{\bar{ \lambda}_{12}\lambda_{12}}
\delta_{\bar{ \lambda}_{1}'\lambda_{1}'}
\delta_{\bar{\lambda}_{2}'\lambda_{2}'}
\delta_{\bar{ \lambda}_{3}\lambda_{3}}.\\
\end{aligned}\label{eq::WD_norm_1}
\end{equation} 

Whereas Berman's $J$-helicity states \eqref{eq::BD_Jdef} allows for an easy implementation of symmetry through relations \eqref{eq::BD_parity} to \eqref{eq::BD_P23}, Wick's states, thanks to the intermediary two-body coupling, are more convenient to compute ME for operators related to the internal motion. To exploit the advantages of both definitions, a relationship between the two different sets of states can be constructed,
\begin{equation}
\begin{aligned}
\ket{JM\mu ; w_1 w_2 w_3 ; \lambda_1 \lambda_2 \lambda_3} &= \sum_{\lambda_1',\lambda_2'}\;D^{s_1}_{\lambda_1^\prime\,\lambda_1}(R_W^1) D^{s_2}_{\lambda_2^\prime\,\lambda_2}(R_W^2)\, e^{i(2s_2+2s_3+\lambda_2'-\lambda_1'-\mu)\pi/2} \\
& \sum_{j_{12}=|\lambda_1'-\lambda_2'|}^{\infty} \,\sum_{\lambda_{12}=-j_{12}}^{j_{12}}  e^{i\lambda_{12}\pi/2}\,\sqrt{\frac{2j_{12}+1}{2}}\,
d^{j_{12}}_{\lambda_{12}\,\lambda_1'-\lambda_2'}(\pi/2-\phi_{12})d^{J}_{\mu\,\lambda_{12}-\lambda_3}(\pi/2)\\
&\hspace{6.27cm}  \ket{p_3;JM;j_{12}\lambda_{12}s_3\lambda_3;p_{12}s_1\lambda_1s_2\lambda_2}. 
\end{aligned}\label{eq::BtoW_fin}
\end{equation}
A demonstration of this property is provided in \ref{app:BtoW}. Above, $\phi_{12}$ is the azimutal angle of the first particle in the $12$-CoMF where the momentum of the third particle is along the $y$ axis towards negatives. This quantity, as well as every other dynamical quantity in this relation, is to be understood as depending on $w_1$, $w_2$ and $w_3$. Rotations $R_W^1$ and $R_W^2$ are Wigner rotations given by the following combinations of boosts and rotations,
\begin{equation}
\begin{aligned}
&R_W^1=(R(\phi_{12},\pi/2,0) L_z(m_1,p_{12}))^{-1}\,L_3\, (R(\phi_1,\pi/2,0) L_z(m_1,p_1)),\\
&R_W^2=(R(\pi+\phi_{12},\pi/2,0) L_z(m_2,p_{12}))^{-1}\,L_3\,(R(\phi_2,\pi/2,0) L_z(m_2,p_2)).
\end{aligned}\label{eq::BtoW_wignerRot_12}
\end{equation}
where $L_3=R(3\pi/2,\pi/2,0)L_z(m_{12},p_3)R^{-1}(3\pi/2,\pi/2,0)$. The fact that helicities of particle $1$ and $2$ are summed in the formula keeps track that helicities $\lambda_i$ and $\lambda'_i$ are not defined in the same frame. The appearance of an infinite sum over $j_{12}$ is also an expected feature. To obtain a given total angular momentum  $J$, one can always choose an arbitrarily large relative angular momentum between particle $1$ and $2$ and then compensate it with the relative angular momentum between this subsystem and the third particle. Let us mention that formula \eqref{eq::BtoW_fin} looks rather similar to a result obtained by Wick in \cite{wick62}. In this reference, Wick rewrites its states in a more symmetrical way, closer to Berman's definition but where the $yz$ plane is chosen as reference. This rewriting introduces two Wigner rotations and a Wigner $d$-matrix which depends on a dynamical angle. The formula introduced in the current work presents the same components supplemented by a second Wigner $D$-matrix that rotates the reference plane, in accordance with property \eqref{eq::BD_changeRefState}. Formula \eqref{eq::BtoW_fin} can be seen as a descendant to Wick's rewriting.

In Section \ref{sec:3GB}, formula \eqref{eq::BtoW_fin} will be used to describe systems of three massless gluons. For massless particles, both previous Wigner rotations simplifies following expression \eqref{eq::1BS_D_ISO2}. The change of basis formula reduces to
\begin{equation}
\begin{aligned}
\ket{JM\mu ; w_1 w_2 w_3 ; \lambda_1 \lambda_2 \lambda_3} &= e^{i(2s_2+2s_3+\lambda_2-\lambda_1-\mu)\pi/2} e^{i(\theta_1\lambda_1+\theta_2\lambda_2)} \hspace{3cm}\\
\phantom{\bigg[}  &\sum_{j_{12}=|\lambda_1-\lambda_2|}^{\infty} \,\sum_{\lambda_{12}=-j_{12}}^{j_{12}} e^{i\pi\lambda_{12}/2} \sqrt{\frac{2j_{12}+1}{2}}\,d^{j_{12}}_{\lambda_{12}\,\lambda_1-\lambda_2}(\pi/2-\phi_{12}) d^{J}_{\mu\,\lambda_{12}-\lambda_3}(\pi/2)\\
\phantom{\bigg[} &\hspace{6.22cm}\ket{p_3;JM;j_{12}\lambda_{12}s_3\lambda_3;p_{12}s_1\lambda_1s_2\lambda_2}. 
\end{aligned}\label{eq::BtoW_fin_massless}
\end{equation}
As expected, since helicity is Lorentz invariant for massless particles, the corresponding quantum numbers are no longer summed. Values for both $\theta_i$ angles are founded by applying the methodology suggested in reference \cite{lind03}. These two angles are shown to cancel and the phase factor reduces, 
\begin{equation}
e^{i(2s_2+2s_3+\lambda_2-\lambda_1-\mu)\pi/2} e^{i(\theta_1\lambda_1+\theta_2\lambda_2)} = (-1)^{s_3+s_2}\,e^{i(\pi/2)\left(\lambda_2-\lambda_1-\mu\right)}.
\end{equation}
This phase is manifestly independent of the $w_1$, $w_2$ or $w_3$ energies. In the case of massless particles, the expression of $\phi_{12}$, $p_{12}$ and $p_3$ in terms of $w_1$, $w_2$ and $w_3$ also simplifies,
\begin{align}
&\cos(\pi/2-\phi_{12}) = (w_1-w_2)/w_3, && 2p_{12} = \sqrt{(w_1 + w_2)^2-w_3^2}, &&p_3=w_3.
\label{eq::BtoW_um_w1w2}
\end{align}
Inverting these relations, one gets
\begin{align}
&w_1 = (\sqrt{4p_{12}^2 + w_3^2} + u w_3)/2, && w_2 = (\sqrt{4p_{12}^2 + w_3^2} - u w_3)/2, && w_3=p_3, \label{eq::BtoW_w1w2_um}
\end{align}
where $u=\cos(\pi/2-\phi_{12})= \sin\phi_{12}$. A consistency check of formula \eqref{eq::BtoW_fin_massless} is provided in \ref{app:BtoW}.


\section{Three-gluon Glueballs}
\label{sec:3GB}

Using the helicity formalism for three-body systems introduced in the previous section, a three-gluon glueball spectrum can be obtained following a similar methodology than for the two-gluon case. The developments are carried out in three main steps. First, symmetry and parity are implemented in Berman's basis so that this complete set is ready to be used in the description of three-gluon bound states. Secondly, this basis is used to develop trial states that should approximate reasonably well the true three-gluon glueball states. Third, Hamiltonian ME are evaluate on these trial state to obtain the aforementioned spectrum.

\subsection{Berman's basis for three-gluon systems}

\label{ssec::3GB_sym} 

Let us start with the acquisition of symmetric parity eigenstates from Berman's $J$-helicity states. Dealing with three spin $1$ massless particles, the helicity quantum number $\lambda_i$ can only take two values, $-1$ and $+1$. Therefore, there are only height possible triplets of helicities $(\lambda_1,\lambda_2,\lambda_3)$. In Berman's definition, any consistent set of $J$, $M$, $\mu$, $w_1$, $w_2$, $w_3$ quantum numbers can be built from each of these triplets. In other words, for any given energy and angular momentum quantum numbers, eight three-gluon $J$-helicity states can be obtained :
\begin{equation}
\begin{aligned}
&\ket{JM\mu ; w_1 w_2 w_3 ; +++}, &&\ket{JM\mu ; w_1 w_2 w_3; --+}, &&\ket{JM\mu ; w_1 w_2 w_3 ; -++}, &&\ket{JM\mu ; w_1 w_2 w_3; -+-},\\
&\ket{JM\mu ; w_1 w_2 w_3 ; +-+}, &&\ket{JM\mu ; w_1 w_2 w_3; +--},&&\ket{JM\mu ; w_1 w_2 w_3 ; ++-}, &&\ket{JM\mu ; w_1 w_2 w_3; ---}.
\end{aligned}\label{eq::3GB_8set1}
\end{equation}
For shortness, only the sign of helicities have been kept in the notation. Let us start by implementing parity quantum numbers in these eight states. Considering that gluons have negative intrinsic parity and using relation \eqref{eq::BD_parity}, the eight states \eqref{eq::3GB_8set1} can be recombined into eight parity eigenstates,
\begin{align}
&\begin{aligned}
&\ket{JM\mu ; w_1 w_2 w_3 ; +++} + \ket{JM\mu ; w_1 w_2 w_3; ---},\\
&\ket{JM\mu ; w_1 w_2 w_3 ; -++} + \ket{JM\mu ; w_1 w_2 w_3; +--},\\
&\ket{JM\mu ; w_1 w_2 w_3 ; +-+} + \ket{JM\mu ; w_1 w_2 w_3; -+-},\\
&\ket{JM\mu ; w_1 w_2 w_3; ++-} + \ket{JM\mu ; w_1 w_2 w_3; --+},
\end{aligned}
&\begin{aligned}
&\ket{JM\mu ; w_1 w_2 w_3 ; +++} - \ket{JM\mu ; w_1 w_2 w_3; ---},\\
&\ket{JM\mu ; w_1 w_2 w_3 ; -++} - \ket{JM\mu ; w_1 w_2 w_3; +--},\\
&\ket{JM\mu ; w_1 w_2 w_3 ; +-+} - \ket{JM\mu ; w_1 w_2 w_3; -+-},\\
&\ket{JM\mu ; w_1 w_2 w_3; ++-} - \ket{JM\mu ; w_1 w_2 w_3; --+}.
\end{aligned}\label{eq::3GB_8set2}
\end{align}
Their parity eigenvalue is $(-1)^{-\mu}$ for the four left states and $(-1)^{1-\mu}$ for the four right ones. In addition to parity, symmetry is also to be implemented. Depending on the expected charge conjugation of the system, three-gluon states have to be symmetrised (negative charge conjugation) or anti-symmetrised (positive charge conjugation)~\cite{boul08}. Applying both the symmetriser and the anti-symmetriser on the eight states \eqref{eq::3GB_8set2} provides states with a definite symmetry under exchange of particles. The result of these applications is presented in Table~\ref{tab::3GB_symStates}. Fixing respectively $\sigma=+1$ or $\sigma=-1$ provides symmetric or anti-symmetric states. Because permutation operators change the sign of $\mu$, (anti-)symmetric states mix different states with opposite $\mu$ quantum numbers. To avoid any redundancy, $\mu$ must be understood as positive in Table~\ref{tab::3GB_symStates}.

\begin{table}
\centering
-\begin{tabular}{c}
\hline \hline \\[-1mm] 
$\begin{aligned}
&\left(\ket{JM\mu ; w_1 w_2 w_3 ; +++} + \ket{JM\mu ; w_1 w_2 w_3 ; ---}\right) \\
+ \sigma (-1)^{J+\mu+1}&\left(\ket{JM-\mu ; w_2 w_1 w_3 ; +++} + \ket{JM-\mu ; w_2 w_1 w_3  ; ---}\right) \\
+ \sigma (-1)^{J+\mu+1}e^{-i\varphi_{13}\mu}&\left(\ket{JM-\mu ; w_3 w_2 w_1 ; +++} + \ket{JM-\mu ; w_3 w_2 w_1 ; ---}\right) \\
+ \sigma (-1)^{J+\mu+1}e^{i\varphi_{23}\mu}&\left(\ket{JM-\mu ; w_1 w_3 w_2 ; +++} + \ket{JM-\mu ; w_1 w_3 w_2 ; ---}\right)  \\
+ e^{-i\varphi_{13}\mu}&\left(\ket{JM\mu ; w_2 w_3 w_1 ; +++} + \ket{JM\mu ; w_2 w_3 w_1 ; ---}\right) \\
+ e^{i\varphi_{23}\mu}&\left(\ket{JM\mu ; w_3 w_1 w_2 ; +++} + \ket{JM\mu ; w_3 w_1 w_2 ; ---}\right) \\
\end{aligned}$ \vspace{2mm} \\
\hline \hline \\[-1mm] 
$\begin{aligned}
&\left(\ket{JM\mu ; w_1 w_2 w_3 ; -++} + \ket{JM\mu ; w_1 w_2 w_3; +--}\right) \\
+ \sigma (-1)^{J+\mu+1}&\left(\ket{JM-\mu ; w_2 w_1 w_3; +-+} + \ket{JM-\mu ; w_2 w_1 w_3; -+-}\right) \\
+ \sigma (-1)^{J+\mu+1}e^{-i\varphi_{13}\mu}&\left(\ket{JM-\mu ; w_3 w_2 w_1 ; ++-} + \ket{JM-\mu ; w_3 w_2 w_1; --+}\right) \\
+ \sigma (-1)^{J+\mu+1}e^{i\varphi_{23}\mu}&\left(\ket{JM-\mu ; w_1 w_3 w_2; -++} + \ket{JM-\mu ; w_1 w_3 w_2 ; +--}\right)  \\
+ e^{-i\varphi_{13}\mu}&\left(\ket{JM\mu ; w_2 w_3 w_1 ; ++-} + \ket{JM\mu ; w_2 w_3 w_1; --+}\right) \\
+ e^{i\varphi_{23}\mu}&\left(\ket{JM\mu ; w_3 w_1 w_2 ; +-+} + \ket{JM\mu ; w_3 w_1 w_2 ; -+-}\right) \\
\end{aligned} $ \vspace{2mm} \\
\hline \hline \\[-1mm] 
$\begin{aligned}
&\left(\ket{JM\mu ; w_1 w_2 w_3; +++} - \ket{JM\mu ; w_1 w_2 w_3; ---}\right) \\
+ \sigma (-1)^{J+\mu+1}&\left(\ket{JM-\mu ; w_2 w_1 w_3 ; +++} - \ket{JM-\mu ; w_2 w_1 w_3 ; ---}\right) \\
+ \sigma (-1)^{J+\mu+1}e^{-i\varphi_{13}\mu}&\left(\ket{JM-\mu ; w_3 w_2 w_1 ; +++} - \ket{JM-\mu ; w_3 w_2 w_1 ; ---}\right) \\
+ \sigma (-1)^{J+\mu+1}e^{i\varphi_{23}\mu}&\left(\ket{JM-\mu ; w_1 w_3 w_2 ; +++} - \ket{JM-\mu ; w_1 w_3 w_2 ; ---}\right)  \\
+ e^{-i\varphi_{13}\mu}&\left(\ket{JM\mu ; w_2 w_3 w_1 ; +++} - \ket{JM\mu ; w_2 w_3 w_1 ; ---}\right) \\
+ e^{i\varphi_{23}\mu}&\left(\ket{JM\mu ; w_3 w_1 w_2 ; +++} - \ket{JM\mu ; w_3 w_1 w_2 ; ---}\right) \\
\end{aligned}$ \vspace{2mm} \\
\hline \hline \\[-1mm] 
$\begin{aligned}
&\left(\ket{JM\mu ; w_1 w_2 w_3 ; -++} - \ket{JM\mu ; w_1 w_2 w_3 ; +--}\right) \\
+ \sigma (-1)^{J+\mu+1}&\left(\ket{JM-\mu ; w_2 w_1 w_3 ; +-+} - \ket{JM-\mu ; w_2 w_1 w_3 ; -+-}\right) \\
+ \sigma (-1)^{J+\mu+1}e^{-i\varphi_{13}\mu}&\left(\ket{JM-\mu ; w_3 w_2 w_1 ; ++-} - \ket{JM-\mu ; w_3 w_2 w_1 ; --+}\right) \\
+ \sigma (-1)^{J+\mu+1}e^{i\varphi_{23}\mu}&\left(\ket{JM-\mu ; w_1 w_3 w_2 ; -++} - \ket{JM-\mu ; w_1 w_3 w_2 ; +--}\right)  \\
+ e^{-i\varphi_{13}\mu}&\left(\ket{JM\mu ; w_2 w_3 w_1 ; ++-} - \ket{JM\mu ; w_2 w_3 w_1 ; --+}\right) \\
+ e^{i\varphi_{23}\mu}&\left(\ket{JM\mu ; w_3 w_1 w_2 ; +-+} - \ket{JM\mu ; w_3 w_1 w_2 ; -+-}\right) \\
\end{aligned}
$ \vspace{2mm} \\
\hline \hline  
\end{tabular}
\caption{Combinaisons of Berman's $J$ helicity states that present a given parity and symmetry. Depending on whether $\sigma$ is chosen to be $+1$ or $-1$ the state is symmetric or anti-symmetric. The parity eigenvalue of the two first sets of states is $(-1)^{-\mu}$ while it is $(-1)^{1-\mu}$ for the two last one.}
\label{tab::3GB_symStates}
\end{table}

\subsection{Three-gluon glueball states}

\label{ssec::3GB_trial}
The eight states from Table~\ref{tab::3GB_symStates} will now be used to construct states that model $J^{PC}$ three-gluon glueballs. The procedure for this construction was outlined at the end of Section \ref{ssec:BD}. Specifically, this involves replacing $\ket{JM\mu;w_1w_2w_3;\lambda_1\lambda_2\lambda_3}$ in the right-hand side of equation \eqref{eq::BD_JPhysState} with one of the states from Table~\ref{tab::3GB_symStates}. Due to their similarities, states form the first and third lines from Table~\ref{tab::3GB_symStates} can be treated together, as can those from the second and fourth sets. Notably, due to symmetrization, each glueball state combines multiple Berman's states where the energy variables $w_1$, $w_2$ and $w_3$ appear in different orders. This difference can be transferred to the $\Psi(w_1,w_2,w_3)$ wave function by appropriately exchanging integration variables. The resulting combinations are presented in Table~\ref{tab::3GB_physStates}. At this stage, the normalisation of the symmetrised states is not guaranteed.

Table~\ref{tab::3GB_physStates} shows that any $J^{PC}$ quantum numbers can, in principle, be realized by a three-gluon system. However, it may be reasonable to assume that low-lying glueball states correspond to symmetric helicity-momentum wave functions,
\begin{equation}
    \forall\,i,j,k \in \{1,2,3\}, \ \Psi(w_i,w_j,w_k)=\Psi(w_1,w_2,w_3). \label{eq::3GB_hyp_sym}
\end{equation}
Imposing this symmetry reduces the states in Table~\ref{tab::3GB_physStates} to those in Table~\ref{tab::3GB_physStates_sym}. Unlike the general case, states with a symmetric helicity-momentum wave function exhibit a selection rule for $\mu=0$. Specifically, when $\sigma(-1)^J = 1$, terms involving $+\mu$ systematically cancel those with $-\mu$. This implies that states with $\mu=0$ and negative (positive) charge conjugation only exist for odd (even) $J$ values. Since the following discussions primarily consider states with negative charge conjugation, $\mu=0$ will always imply an odd $J$. This feature highlights two interesting properties of the spectrum. First, because the even value $J=0$ can only be achieved by setting $\mu=0$, no state with symmetric wave function, null total angular momentum and negative charge conjugation can be constructed. Color-singlet two-gluon states have only positive charge conjugation, therefore a $0^{--}$ bound state of pure glue would either contradict hypothesis \eqref{eq::3GB_hyp_sym} or require at least four constituent gluons, a requirement consistent with group theory arguments \cite{boul08}. Secondly, negative charge conjugation states with $J=2$ must at least have $\mu=1$. Qualitatively, one may suppose that higher $\mu$ projection of the total angular momentum would result in higher energy states. If this is correct, the $\mu=0$ selection rule would push $J=2$ states to higher masses.

\begin{figure}
    \centering \includegraphics{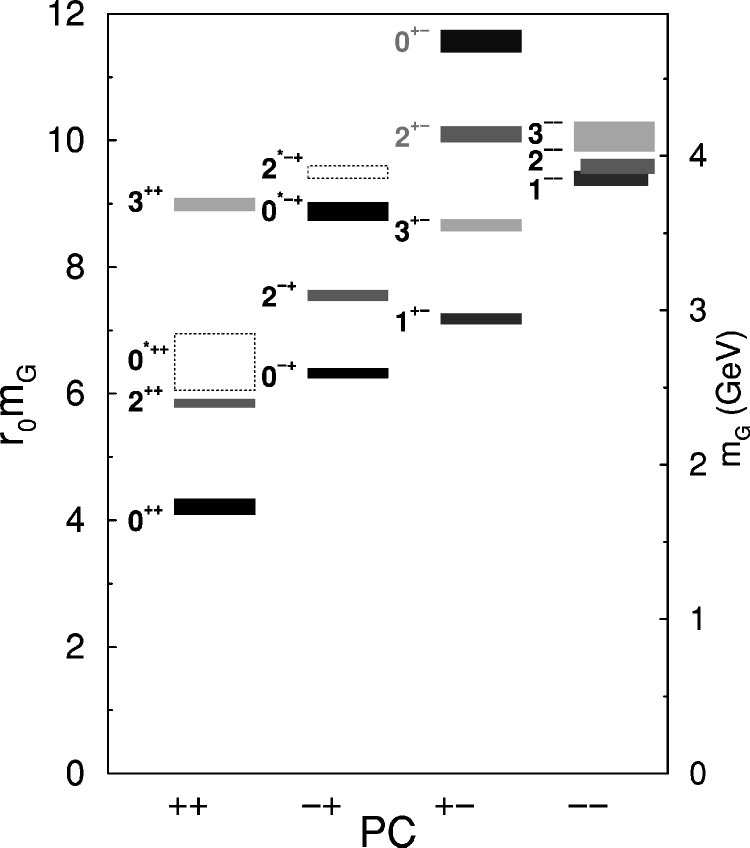}
    \caption{Glueball spectrum obtained using lattice QCD. The picture is taken from reference \cite{morn99}.}
    \label{fig::3GB_lQCD}
\end{figure}

This analysis is supported by glueball spectrum calculations from other approaches. For instance, Figure \ref{fig::3GB_lQCD}, taken from reference reference \cite{morn99}, shows glueball masses calculated using LQCD. In this spectrum, the first $0^{P-}$ state appears above $4.5$ GeV, significantly higher than the lowest $J^{P-}$ state, which lies below $3$ GeV. This aligns with the preceding analysis. The $J=0$ state may be interpreted as a first four-gluon state or a three-gluon state with a non-symmetric helicity-momentum wave function $\Psi(w_1,w_2,w_3)$. Beyond $J=0$, the hierarchy of even states follows the expected pattern: the $J=1$ and $J=3$ states, which allow $\mu=0$, appear in order, while the $J=2$ state is shifted to higher energies. However, the situation for odd states remains less clear. Further discussion is deferred until quantitative results are obtained. 

States from Table~\ref{tab::3GB_physStates} and \ref{tab::3GB_physStates_sym} are also consistent with results from references \cite{peas50} and \cite{fumi53}. In References \cite{peas50}, it is argued that no general selection rules govern the decay of a particle into three massless spin-$1$ particles. This is consistent with the absence of selection rules in Table \ref{tab::3GB_physStates}. Reference \cite{fumi53} refines this discussion for symmetric decays into three photons, concluding that such a decay is not possible for $J=0$ particles. This agrees with the analysis of states from Table \ref{tab::3GB_physStates_sym}. Moreover, the helicity triplets combinations constructed in \cite{fumi53} closely resemble those in the current work. For this reason, the labels $A_2'$ and $A_2''$ from reference \cite{fumi53} have been adopted in Table~\ref{tab::3GB_physStates} and \ref{tab::3GB_physStates_sym} to distinguish these states.

\begin{table}
\begin{tabular}{c}
\hline\hline \\[-1mm]
$
\begin{aligned}
\ket{\Psi;A'_2;JM;C=-\sigma;P=\pm(-1)^\mu} = \int \frac{\diff w_1 \diff w_2 \diff w_3}{8} \left(\Psi(w_1,w_2,w_3)+ e^{-i\varphi_{23}\mu}\Psi(w_3,w_1,w_2)+ e^{i\varphi_{13}\mu}\Psi(w_2,w_3,w_1)\right)&\\
\hspace{3.1cm}\left(\ket{JM\mu ; w_1 w_2 w_3 ; +++} \pm \ket{JM\mu ; w_1 w_2 w_3 ; ---}\right) &\\
\hspace{0.4cm}- \sigma (-1)^{J+\mu} \int \frac{\diff w_1 \diff w_2 \diff w_3}{8} \left(\Psi(w_2,w_1,w_3)+e^{-i\varphi_{13}\mu}\Psi(w_3,w_2,w_1)+e^{i\varphi_{23}\mu}\Psi(w_1,w_3,w_2)\right)&\\
\hspace{3.8cm}\left(\ket{JM-\mu ; w_1 w_2 w_3 ; +++} \pm \ket{JM-\mu ; w_1 w_2 w_3 ; ---}\right)&
\end{aligned}$ \vspace{2mm}\\
\hline\hline \\[-1mm]
$\begin{aligned}
\ket{\Psi;A''_2;JM;C=-\sigma;P=\pm(-1)^\mu} = \int \frac{\diff w_1 \diff w_2 \diff w_3}{8}
\Big(\Psi(w_1,w_2,w_3) \left(\ket{JM\mu ; w_1 w_2 w_3; -++} \pm \ket{JM\mu ; w_1 w_2 w_3; +--}\right)& \\
+ e^{-i\varphi_{23}\mu} \Psi(w_3,w_1,w_2) \left(\ket{JM\mu ; w_1 w_2 w_3; ++-} \pm \ket{JM\mu ; w_1 w_2 w_3; --+}\right) &\\
+ e^{i\varphi_{13}\mu} \Psi(w_2,w_3,w_1) \left(\ket{JM\mu ; w_1 w_2 w_3; +-+} \pm \ket{JM\mu ; w_1 w_2 w_3; -+-}\right)&\!\Big)\\
- \sigma (-1)^{J+\mu}\int \frac{\diff w_1 \diff w_2 \diff w_3}{8}
\Big(\Psi(w_2,w_1,w_3)\left(\ket{JM-\mu ; w_1 w_2 w_3; +-+} \pm \ket{JM-\mu ; w_1 w_2 w_3; -+-}\right) &\\
+ e^{-i\varphi_{13}\mu}\Psi(w_3,w_2,w_1)\left(\ket{JM-\mu ; w_1 w_2 w_3; ++-} \pm \ket{JM-\mu ; w_1 w_2 w_3; --+}\right) &\\
+ e^{i\varphi_{23}\mu}\Psi(w_1,w_3,w_2)\left(\ket{JM-\mu ; w_1 w_2 w_3; -++} \pm \ket{JM-\mu ; w_1 w_2 w_3; +--}\right)&\!\Big)
\end{aligned}
$\vspace{2mm} \\
\hline\hline
\end{tabular}
\caption{Total angular momentum, parity and charge conjugation eigenstates for three-gluon systems. A generic helicity-momentum wave function $\Psi(w_1,w_2,w_3)$ is considered. Labels $A_2'$ and $A_2''$ differentiate the two symmetrical combinations of helicity triplets possible. These labels are inspired by the notations in reference \cite{fumi53} and originate from crystallography \cite{hame89}. \label{tab::3GB_physStates}}
\end{table}

\begin{table}
\begin{tabular}{c}
\hline\hline \\[-1mm]
$
\begin{aligned}
\ket{\Psi;A'_2;JM;C=-\sigma;P=\pm(-1)^\mu} = \int \frac{\diff w_1 \diff w_2 \diff w_3}{8} \, \Psi(w_1,w_2,w_3)\left(1+e^{-i\varphi_{23}\mu}+e^{i\varphi_{13}\mu}\right)\hspace{4cm}&\\
\left(\ket{JM\mu ; w_1 w_2 w_3 ; +++} \pm \ket{JM\mu ; w_1 w_2 w_3 ; ---}\right)& \\
- \sigma (-1)^{J+\mu} \int \frac{\diff w_1 \diff w_2 \diff w_3}{8} \, \Psi(w_1,w_2,w_3) \left(1+e^{-i\varphi_{13}\mu}+e^{i\varphi_{23}\mu}\right)\hspace{1.7cm}&\\
\left(\ket{JM-\mu ; w_1 w_2 w_3 ; +++} \pm \ket{JM-\mu ; w_1 w_2 w_3 ; ---}\right)&
\end{aligned}
$\vspace{2mm} \\
\hline\hline \\[-1mm]
$
\begin{aligned}
\ket{\Psi;A''_2;JM;C=-\sigma;P=\pm(-1)^\mu} = \int \frac{\diff w_1 \diff w_2 \diff w_3}{8} \, \Psi(w_1,w_2,w_3) \Big(\left(\ket{JM\mu ; w_1 w_2 w_3; -++} \pm \ket{JM\mu ; w_1 w_2 w_3; +--}\right)& \\
+ e^{-i\varphi_{23}\mu} \left(\ket{JM\mu ; w_1 w_2 w_3; ++-} \pm \ket{JM\mu ; w_1 w_2 w_3; --+}\right)& \\
+ e^{i\varphi_{13}\mu} \left(\ket{JM\mu ; w_1 w_2 w_3; +-+} \pm \ket{JM\mu ; w_1 w_2 w_3; -+-}\right)&\!\Big)\\
- \sigma (-1)^{J+\mu}\int \frac{\diff w_1 \diff w_2 \diff w_3}{8} \, \Psi(w_1,w_2,w_3) \Big(\left(\ket{JM-\mu ; w_1 w_2 w_3; +-+} \pm \ket{JM-\mu ; w_1 w_2 w_3; -+-}\right)& \\
+ e^{-i\varphi_{13}\mu}\left(\ket{JM-\mu ; w_1 w_2 w_3; ++-} \pm \ket{JM-\mu ; w_1 w_2 w_3; --+}\right) &\\
+ e^{i\varphi_{23}\mu}\left(\ket{JM-\mu ; w_1 w_2 w_3; -++} \pm \ket{JM-\mu ; w_1 w_2 w_3; +--}\right)&\!\Big)
\end{aligned}
$\vspace{2mm}\\
\hline\hline
\end{tabular}
\caption{Total angular momentum, parity and charge conjugation eigenstates for three-gluon systems. A symmetric helicity-momentum wave function $\Psi(w_1,w_2,w_3)$ is considered. Labels $A_2'$ and $A_2''$ differentiate the two possible symmetrical combinations of helicity triplets. These labels are inspired by the notations in reference \cite{fumi53} and originate from crystallography \cite{hame89}.\label{tab::3GB_physStates_sym}}
\end{table}

\subsection{Low-lying three-gluon Glueballs with negative charge conjugation}

The current description of the glueball spectrum has remained qualitative. To move toward quantitative analysis, the methodology based on the variational theorem, applied to two-gluon glueballs in Section \ref{sec:2GB}, will be extended to three-gluon glueballs. This approach involves, on one hand, constructing a set of trial states designed to approximate three-gluon glueballs accurately and, on the other, evaluating Hamiltonian ME on this set. For the former, appropriate trial helicity-momentum wave functions $\Psi$ must be chosen. For the latter, formulas to compute ME on three-body helicity states are required. This subsection addresses both these tasks for the low-lying three-gluon glueball states with negative charge conjugation and with a symmetric $\Psi$ function. Angular momenta up to $J=3$ and $\mu$ projections up to $\mu=1$ are considered. To keep equations concise, reduced notations are introduced for Berman's $J$-helicity states and for the three-body helicity states defined in \eqref{eq::BD_JPhysState},
\begin{align}
&\ket{\lambda_1\lambda_2\lambda_3}^{J}_\mu = \ket{JM\mu;w_1w_2w_3;\lambda_1\lambda_2\lambda_3},
&\ket{\Psi;\lambda_1\lambda_2\lambda_3}^J_\mu = \ket{\Psi;JM\mu;\lambda_1\lambda_2\lambda_3}.
\end{align}
In $\ket{\lambda_1\lambda_2\lambda_3}^{J}_\mu$, the order of energy labels is tacitly assumed to be always $w_1w_2w_3$. For clarity in plain text, three-body states defined in equation \eqref{eq::BD_JPhysState} will be referred to as \textit{unsymmetrical states}, while those from Table~\ref{tab::3GB_physStates_sym} will be referred to as \textit{symmetrical states}. Symmetrical states corresponding to the above-named low-lying three-gluon glueballs can be made explicit using these notations. Considering $\mu=0$, the following states are obtained,
\begin{align}
&\ket{\Psi;A'_2;\mu=0;J^{PC} = (2k+1)^{\pm-}} = \frac{1}{\sqrt{2}}\left(\ket{\Psi;+++}^{2k+1}_0 \pm \ket{\Psi;---}^{2k+1}_0 \right), \label{eq::3GB_sym_10_+++} \\
&\begin{aligned}
\ket{\Psi;A''_2;\mu=0;J^{PC} = (2k+1)^{\pm-}} = \frac{1}{\sqrt{6}}\Big(\ket{\Psi;-++}^{2k+1}_0 \pm \ket{\Psi;+--}^{2k+1}_0 + \ket{\Psi;++-}^{2k+1}_0 & \\
\pm \ket{\Psi;--+}^{2k+1}_0 + \ket{\Psi;+-+}^{2k+1}_0 \pm \ket{\Psi;-+-}^{2k+1}_0 &\Big)
\end{aligned}\label{eq::3GB_sym_10_+-+}
\end{align}
where $k\in\mathbb{N}$ (even angular momenta are forbidden for $\mu=0$). Notations of the states have slightly been adapted in comparison with those from Table~\ref{tab::3GB_physStates_sym}. Considering $\mu=\pm1$, the following states are obtained,
\begin{equation}
\begin{aligned}
\ket{\Psi;A'_2;|\mu|=1;J^{PC} = J^{\mp-}} = \Big(\ket{\Psi\left(1+e^{-i\varphi_{23}}+e^{i\varphi_{13}}\right);+++}^J_1\pm \ket{\Psi\left(1+e^{-i\varphi_{23}}+e^{i\varphi_{13}}\right);---}^J_1&\\
 + (-1)^{J} \ket{\Psi\left(1+e^{-i\varphi_{13}}+e^{i\varphi_{23}}\right);+++}^J_{-1}&\\ \pm (-1)^{J} \ket{\Psi\left(1+e^{-i\varphi_{13}}+e^{i\varphi_{23}}\right);---}^J_{-1} &\Big), 
\end{aligned}\label{eq::3GB_sym_11_+++}
\end{equation}
\begin{equation}
\begin{aligned}
\ket{\Psi;A''_2;|\mu|=1;J^{PC} = J^{\mp-}} = \frac{1}{\sqrt{12}}\Big(\ket{\Psi;-++}^J_1\pm \ket{\Psi;+--}^J_1 + \ket{\Psi e^{i\varphi_{13}}};+-+^J_1 \pm \ket{\Psi e^{i\varphi_{13}};-+-}&^J_1\\
\phantom{\Big)}+ \ket{\Psi e^{-i\varphi_{23}};++-}^J_1 \pm \ket{\Psi e^{-i\varphi_{23}};--+}&^J_1\\
\phantom{\Big)}+ (-1)^J \big(\ket{\Psi;+-+}^J_{-1} \pm \ket{\Psi;-+-}^J_{-1} + \ket{\Psi e^{i\varphi_{23}};-++}^J_{-1} \pm \ket{\Psi e^{i\varphi_{23}};+--}&^J_{-1}\\
+ \ket{\Psi e^{-i\varphi_{13}};++-}^J_{-1} \pm  \ket{\Psi e^{-i\varphi_{13}};--+}&^J_{-1}\big)\!\Big). 
\label{eq::3GB_sym_11_+-+}
\end{aligned}
\end{equation}
In general, symmetrical states are linear combinations of unsymmetrical ones whose  helicity-momentum wave function is potentially modified by an additional kinematic factor introduced during the symmetrisation process. In \eqref{eq::3GB_sym_10_+++}, \eqref{eq::3GB_sym_10_+-+} and \eqref{eq::3GB_sym_11_+-+}, square root factors have been added to ensure that  symmetric states are normalised as long as unsymmetrical ones are. The question of the normalisation of equation \eqref{eq::3GB_sym_10_+++} is deferred to Section \ref{ssec::3GB_spec}.

\subsubsection{Selection of trial helicity-momentum wave functions}

Let us start by selecting a suitable structure to use for trial helicity-momentum wave functions $\Psi(w_1,w_2,w_3)$. The most convenient choice is probably a Gaussian shaped wave function for which calculations are reasonably simple and which often offers great convergence properties. Symmetrical states \eqref{eq::3GB_sym_10_+++} to \eqref{eq::3GB_sym_11_+-+} being linear combinations of unsymmetrical ones, this discussion takes place at the level of the latter. Gaussians may be constructed using different sets of coordinates, thereby leading to non-equivalent structures. The most straightforward choice is probably to consider a Gaussian shaped helicity-momentum wave function in PMP-coordinates,
\begin{equation}
    \Psi_{\text{PMP}}(w_1,w_2,w_3) = Ae^{-a((w_1-b)^2+(w_2-b)^2+(w_3-b)^2)}. \label{eq::3GB_ME_trialFunc_MPG}
\end{equation}
Above, $A$ is a normalisation constant used to ensure that \eqref{eq::BD_physStateNorm} is respected. If an expansion with more than one trial state is to be used, for instance adapting equation \eqref{eq::2GB_genEig}, the normalisation condition is replaced by the evaluation of an overlap matrix \cite{suzu98}. In that case, the constant $A$ can be omitted. The structure \eqref{eq::3GB_ME_trialFunc_MPG} incorporates two non-linear variational parameters, $a$ and $b$. The former encodes the spreading in energy and has dimensions of inverse energy squared. The latter encodes the position of the energy peak and has energy dimension. Different constant for each terms in the exponential cannot be used to keep the trial wave function symmetrical. The choice of wave function is, of course, arbitrary, and many modifications can be proposed. After several tests, adding a square root factor $\sqrt{8w_1w_2w_3}$ in front of the Gaussian was found to improve convergence. As a result, the following trial wave function is suggested,
\begin{equation}
    \Psi_{a,b}(w_1,w_2,w_3) = A\,\sqrt{8w_1w_2w_3}\,e^{-a((w_1-b)^2+(w_2-b)^2+(w_3-b)^2)}, \label{eq::3GB_ME_trialFunc}
\end{equation}
and the corresponding unsymmetrical state reads
\begin{equation}
\ket{\Psi_{\text{PMP}};\lambda_1\lambda_2\lambda_3}^J_\mu = A \int \frac{\diff w_1 \diff w_2 \diff w_3}{8}\,\sqrt{8w_1w_2w_3}\, e^{-a((w_1-b)^2+(w_2-b)^2+(w_3-b)^2)}
\ket{\lambda_1\lambda_2\lambda_3}^J_\mu.\label{eq::3GB_ME_trialState_MPG}
\end{equation} 
Further calculations will require to switch for the system of coordinates depicted in relation \eqref{eq::BtoW_w1w2_um}, namely $u =\sin(\phi_{12})$, $p_{12}$ and $p_3$. In these new coordinates, $\Psi_{a,b}(w_1,w_2,w_3)$ becomes
\begin{equation}
    \Psi_{a,b}(u,p_{12},p_3) = A\,\sqrt{p_3\left(8p_{12}^2+2 p_{3}^2 \left(1-u^2\right)\right)}\,e^{-\frac{a}{2}\left(6 b^2-4 b \left(\sqrt{4p_{12}^2+p_3^2}+p_3\right)+4 p_{12}^2+3 p_3^2\right)}\,e^{-\frac{a}{2}u^2p_3^2}. \label{eq::3GB_ME_trialFunc_ump}
\end{equation}

One can already anticipate why this system of coordinates will be helpful: the variable $u$ is by definition the cosine of the angle that occurs in the change of basis formula \eqref{eq::BtoW_fin_massless},
while the variable $p_{12}$ is the relative momentum between particle $1$ and $2$ in their CoMF, a variable of great interest for the evaluation of two-body potential ME. Symmetric states from equations \eqref{eq::3GB_sym_11_+++} and \eqref{eq::3GB_sym_11_+-+} supplement the helicity-momentum wave function with kinematics factors that depend on the angles $\varphi_{ij}$ defined in \eqref{eq::BD_pdef2_1}. To complete the picture of the wave function, these additional kinematic factors should be made explicit in $u,p_{12},p_3$ coordinates,
\begin{equation}
\begin{gathered}
\begin{aligned}
&\cos\varphi_{12} = \frac{\left(1-u^2\right) p_3^2 - 4p_{12}^2}{\left(1-u^2\right) p_3^2+4p_{12}^2},
&&\cos\varphi_{13} = -\frac{u \sqrt{4 p_{12}^2+p_3^2}+p_3}{\sqrt{4 p_{12}^2+p_3^2}+u p_3},
\end{aligned}\\
\cos\varphi_{23} = \frac{u \sqrt{4 p_{12}^2+p_3^2}-p_3}{\sqrt{4 p_{12}^2+p_3^2}-u p_3}.
\end{gathered}
\end{equation}
On one hand, in states \eqref{eq::3GB_sym_11_+++}, the following combinations of $\varphi_{ij}$ appear,
\begin{equation}
1+e^{\pm i\varphi_{13}}+e^{\mp i\varphi_{23}}= \frac{(1-u^2)(p_3^2 - 2\sqrt{(4p_{12}^2+p_3^2)p_3^2}) + 4p_{12}^2}{(1-u^2)p_3^2+4p_{12}^2} \mp i\,\frac{4p_{12}u\sqrt{(1-u^2)p_3^2}}{(1-u^2)p_3^2+4p_{12}^2}. \label{eq::3GB_spinJ_1e13e23}
\end{equation}
On the other hand, in states \eqref{eq::3GB_sym_11_+-+}, $\varphi_{ij}$ angles are taken in complex exponential,
\begin{equation}
\begin{aligned}
&e^{i\varphi_{13}} = -\frac{u \sqrt{4 p_{12}^2+p_3^2}+p_3}{\sqrt{4 p_{12}^2+p_3^2}+u p_3} + i \frac{2p_{12}\sqrt{1-u^2}}{\sqrt{4p_{12}^2+p_3^2}+up_3}, \\
&e^{i\varphi_{23}} = \frac{u \sqrt{4 p_{12}^2+p_3^2}-p_3}{\sqrt{4 p_{12}^2+p_3^2}-u p_3} + i \frac{2p_{12}\sqrt{1-u^2}}{\sqrt{4p_{12}^2+p_3^2}-up_3}.
\end{aligned}\label{eq::3GB_spinJ_e}
\end{equation}

\subsubsection{Evaluation of Matrix Elements on unsymmetrical states}

To acquire an approximate energy spectrum by using the variational theorem requires the evaluation of Hamiltonian ME on trial states. Therefore, it requires deriving formulas for performing such calculations on symmetrical states. Since symmetrical states decompose as linear combinations of unsymmetrical ones (cf. equations \eqref{eq::3GB_sym_10_+++} to \eqref{eq::3GB_sym_11_+-+}), establishing formulas for the latter will give support to the evaluation of ME for the former. Extension to symmetrical states will be discussed further in the next subsection.

Because helicity states are momentum eigenstates, the evaluation of kinetic energy operators is the most straightforward to perform. To encompass many different kinematics at once, a generic $T(w_1,w_2,w_3)$ kinetic energy is considered for now. This function is only supposed to depend on the three $w$ energies and not on the angles $\alpha$, $\beta$, $\gamma$. First, definition \eqref{eq::BD_JPhysState} can be used for both the bra and the ket,
\begin{equation}
\begin{aligned}
\leftindex^{J}_{\bar\mu}{ \bra{\bar\Psi;\bar\lambda_1\bar\lambda_2\bar\lambda_3}}T\ket{\Psi;\lambda_1\lambda_2\lambda_3}^J_\mu = \int \frac{\diff \bar w_1 \diff \bar w_2 \diff \bar w_3}{8} \frac{\diff  w_1 \diff  w_2 \diff w_3}{8} \, \bar{\Psi}^*(\bar w_1,\bar w_2,\bar w_3)\Psi(w_1,w_2,w_3)& \\
\leftindex^{J}_{\bar\mu}{\bra{\bar\lambda_1\bar\lambda_2\bar\lambda_3}}T
\ket{\lambda_1\lambda_2\lambda_3}^J_\mu&.
\end{aligned}
\end{equation}
Berman's $J$-helicity states being eigenstates of the three $w$ energies, the evaluation of $T(w_1,w_2,w_3)$ on these comes down to a simple scalar multiplication,
\begin{equation}
\begin{aligned}
\leftindex^J_{\bar\mu}{\bra{\bar\Psi;\bar\lambda_1\bar\lambda_2\bar\lambda_3}}T\ket{\Psi;\lambda_1\lambda_2\lambda_3}^J_\mu = \int \frac{\diff \bar w_1 \diff \bar w_2 \diff \bar w_3}{8} \frac{\diff  w_1 \diff  w_2 \diff w_3}{8} \, \bar{\Psi}^*(\bar w_1,\bar w_2,\bar w_3)\Psi(w_1,w_2,w_3)&\\
\,T(w_1,w_2,w_3)\,\leftindex^{J}_{\bar\mu}{\braket{\bar\lambda_1\bar\lambda_2\bar\lambda_3|\lambda_1\lambda_2\lambda_3}}^J_\mu&.
\end{aligned}
\end{equation}
Finally, using the orthonormalisation relation of Berman's $J$-helicity states, three Dirac deltas are produced and removes the integrations on bar variables,
\begin{equation}
\begin{aligned}
\leftindex^J_{\bar\mu}{\bra{\bar\Psi;\bar\lambda_1\bar\lambda_2\bar\lambda_3}}T\ket{\Psi;\lambda_1\lambda_2\lambda_3}^J_\mu = \delta_{\bar\mu\mu}\, \delta_{\bar\lambda_1\lambda_1}\delta_{\bar\lambda_2\lambda_2}\delta_{\bar\lambda_3\lambda_3} \int \frac{\diff  w_1 \diff  w_2 \diff w_3}{8} \, \bar\Psi^*(w_1,w_2,w_3)\Psi(w_1,w_2,w_3)&\\
T(w_1,w_2,w_3)&.
\end{aligned}
\label{eq::3GB_kinFinal}
\end{equation}
As a reminder, both $w_1$ and $w_2$ vary from $0$ to $+\infty$ while $w_3$ varies between $|w_1-w_2|$ and $w_1+w_2$. This formula allows for reasonably easy evaluations of kinetic energy ME on unsymmetrical states.

Let us turn to the evaluation of two-body potential ME. For symmetric states, ME of the three two-body interactions are shown to be equal each-other. As a result, computing a single two-body interaction ME and multiplying it by three is sufficient to evaluate the entire potential energy of the system. Because the current goal is to perform evaluations on such symmetric states, calculations will only be developed for the interaction between particles $1$ and $2$. One may expect to re-use the method set up to compute potential ME for two-body systems. To do so, the trial states have to be developed in Wick's helicity basis in order introduce two-body sub-couplings. From definition \eqref{eq::BD_JPhysState}, one can use the expansion \eqref{eq::BtoW_fin_massless} to develop the state in Wick's helicity basis,
\begin{equation}
\begin{aligned}
\ket{\Psi;\lambda_1\lambda_2\lambda_3}^J_\mu = \ &e^{i(\pi/2)\left(\lambda_2-\lambda_1-\mu\right)} \sum_{j_{12}=|\lambda_1-\lambda_2|}^{\infty} \,\sum_{\lambda_{12}=-j_{12}}^{j_{12}} i^{\lambda_{12}} \sqrt{\frac{2j_{12}+1}{2}} d^{J}_{\mu\,\lambda_{12}-\lambda_3}(\pi/2)\\
&\int \frac{\diff w_1 \diff w_2 \diff w_3}{8}\, \Psi(w_1,w_2,w_3) \,d^{j_{12}}_{\lambda_{12}\,\lambda_1-\lambda_2}(\pi/2-\phi_{12}) \ket{p_3;JM;j_{12}\lambda_{12}\lambda_3;p_{12}\lambda_1\lambda_2}.
\end{aligned} 
\end{equation}
For the sake of readability, summation ranges are omitted in the next expressions. For further convenience, variables $u$, $p_{12}$ and $p_3$ are introduced in the integral (cf. relations \eqref{eq::BtoW_w1w2_um} and \eqref{eq::BtoW_um_w1w2_Jaco}),
\begin{equation}
\begin{aligned}
\ket{\Psi;\lambda_1\lambda_2\lambda_3}^J_\mu  = \ & e^{i(\pi/2) \left(\lambda_2-\lambda_1-\mu\right)} \sum_{j_{12}} \,\sum_{\lambda_{12}} i^{\lambda_{12}} \sqrt{\frac{2j_{12}+1}{2}} d^{J}_{\mu\,\lambda_{12}-\lambda_3}(\pi/2)\\
& \int \frac{p_3p_{12} \, \diff u \diff p_{12} \diff p_3}{4\sqrt{4p_{12}^2+p_3^2}}\, \Psi(u,p_{12},p_3) \,d^{j_{12}}_{\lambda_{12}\,\lambda_1-\lambda_2}(\arccos\,u) \ket{p_3;JM;j_{12}\lambda_{12}\lambda_3;p_{12}\lambda_1\lambda_2}.
\end{aligned} \label{eq::3GB_sym_dev}
\end{equation}
As a reminder, integration ranges are $-1$ to $1$ for $u$ while it is $0$ to $\infty$ for both $p_3$ and $p_{12}$. Because this important result is relatively cumbersome, the following notation shortcuts are introduced,
\begin{align}
&\mathcal{C}_{J\mu;\lambda_1\lambda_2\lambda_3}^{j_{12}\lambda_{12}} = e^{i(\pi/2) \left(\lambda_2-\lambda_1-\mu+\lambda_{12}\right)} \sqrt{\frac{2j_{12}+1}{2}} d^{J}_{\mu\,\lambda_{12}-\lambda_3}(\pi/2),\\
& \Psi_{\lambda_{12}\,\lambda_1-\lambda_2}^{j_{12}}(u,p_{12},p_3) =  \Psi(u,p_{12},p_3)\,d^{j_{12}}_{\lambda_{12}\,\lambda_1-\lambda_2}(\arccos\,u).
\end{align}
The above $\mathcal{C}$ coefficients should not be confused with those for two-body systems, defined in relation \eqref{eq::2BS_C_coeff}. The difference should be clear depending on the context and looking at the index structure. With these notations, equation \eqref{eq::3GB_sym_dev} shortens,
\begin{equation}
\begin{aligned}
&\ket{\Psi;\lambda_1\lambda_2\lambda_3}^J_\mu = \sum_{j_{12}} \,\sum_{\lambda_{12}} \mathcal{C}_{J\mu;\lambda_1\lambda_2\lambda_3}^{j_{12}\lambda_{12}} \int \frac{p_3p_{12} \, \diff u \diff p_{12} \diff p_3}{4\sqrt{4p_{12}^2+p_3^2}} \Psi_{\lambda_{12}\,\lambda_1-\lambda_2}^{j_{12}}(u,p_{12},p_3) \ket{p_3;JM;j_{12}\lambda_{12}\lambda_3;p_{12}\lambda_1\lambda_2}.
\end{aligned} \label{eq::3GB_sym_dev_short}
\end{equation}
This formula will be very helpful to evaluate two-body potential ME associated with particle $1$ and $2$. This interaction, denoted $\mathcal{O}$, is supposed to solely depend on the relative distance between these two particles, $r_{12}=|\bm{r_1}-\bm{r_2}|$. This distance is assumed to be defined in the $(12)$-CoMF. First, equation \eqref{eq::3GB_sym_dev_short} is used on both the bra and the ket,
\begin{equation}
\begin{aligned}
&\leftindex^J_{\bar\mu}{\bra{\bar\Psi;\bar\lambda_1\bar\lambda_2\bar\lambda_3}}\mathcal{O}(r_{12})\ket{\Psi;\lambda_1\lambda_2\lambda_3}^J_\mu = \sum_{\bar j_{12}} \,\sum_{\bar \lambda_{12}} \sum_{j_{12}} \,\sum_{\lambda_{12}} \left(\mathcal{C}_{J\bar\mu;\bar\lambda_1\bar\lambda_2\bar\lambda_3}^{\bar j_{12}\bar\lambda_{12}}\right) ^*\,\mathcal{C}_{J\mu;\lambda_1\lambda_2\lambda_3}^{j_{12}\lambda_{12}} \\
&\hspace{3.25cm} \int \frac{\bar p_3\bar p_{12} \, \diff \bar u \diff \bar p_{12} \diff \bar p_3}{4\sqrt{4\bar p_{12}^2+\bar p_3^2}} \left(\bar\Psi_{\bar\lambda_{12}\,\bar\lambda_1-\bar\lambda_2}^{\bar j_{12}}(\bar u,\bar p_{12},\bar p_3)\right)^* \int \frac{p_3p_{12} \, \diff u \diff p_{12} \diff p_3}{4\sqrt{4p_{12}^2+p_3^2}}\, \Psi_{\lambda_{12}\,\lambda_1-\lambda_2}^{j_{12}}(u,p_{12},p_3)\\
&\hspace{6.9cm} \bra{\bar p_3;JM;\bar j_{12}\bar\lambda_{12}\bar\lambda_3;\bar p_{12}\bar\lambda_1\bar\lambda_2}\mathcal{O}(r_{12})\ket{p_3;JM;j_{12}\lambda_{12}\lambda_3;p_{12}\lambda_1\lambda_2}. 
\end{aligned}\label{eq::3GB_TBP_int1}
\end{equation}
By doing so, the evaluation of $\mathcal{O}(r_{12})$ on Berman's $J$-helicity states comes down to its evaluation on Wick's $J$-helicity states. The latter, thanks to the intermediary coupling in Wick's definition, consists of evaluating on $\mathcal{O}(r_{12})$ on two-body $J$-helicity states. Dividing up the normalisation factor from \eqref{eq::WD_norm_1} between both state and ensuring a normalisation for two-body states consistent with Sections \ref{sec:1BS2BS} and \ref{sec:2GB}, one gets
\begin{equation}
\begin{aligned}
&\bra{\bar p_3;JM;\bar j_{12}\bar\lambda_{12}\bar\lambda_3;\bar p_{12}\bar\lambda_1\bar\lambda_2}\mathcal{O}(r_{12})\ket{p_3;JM;j_{12}\lambda_{12}\lambda_3;p_{12}\lambda_1\lambda_2}\\
&\hspace{1.5cm} =\frac{\sqrt[4]{\bar p_3^2+4\bar p_{12}^{\,2}}\sqrt[4]{ p_3^2+4 p_{12}^{\,2}}}{\sqrt{\bar p_3\bar p_{12}}\sqrt{p_{12}p_3}}\,\delta(\bar p_3-p_3)\delta_{\bar\lambda_3\lambda_3} \bra{\bar p_{12};\bar j_{12}\bar\lambda_{12};\bar\lambda_1\bar\lambda_2}\mathcal{O}(r_{12})\ket{p_{12};j_{12}\lambda_{12};\lambda_1\lambda_2}.\label{eq::3GB_TBP_int2}
\end{aligned}
\end{equation}
Because $r_{12}$ is defined in the $(12)$-CoMF, the residual ME on two-body states can be computed by using formulas developed in Section \ref{sec:1BS2BS}. Such ME for such central potentials have proven to cancel for $\bar j_{12}\neq j_{12}$ or $\bar \lambda_{12}\neq\lambda_{12}$ in equation \eqref{eq::2BS_posMatElEv_2}. It has also been enhanced in this section that these ME, as soon as non-zero, does not truly dependent on the total angular momentum projection $\lambda_{12}$. Equation \eqref{eq::3GB_TBP_int2} can be plugged into equation \eqref{eq::3GB_TBP_int1},
\begin{equation}
\begin{aligned}
&\leftindex^J_{\bar\mu}{\bra{\bar\Psi;\bar\lambda_1\bar\lambda_2\bar\lambda_3}}\mathcal{O}(r_{12})\ket{\Psi;\lambda_1\lambda_2\lambda_3}^J_\mu = \delta_{\bar\lambda_3\lambda_3} \sum_{j_{12}}\,\sum_{\lambda_{12}}\left(\mathcal{C}_{J\bar\mu;\bar\lambda_1\bar\lambda_2\lambda_3}^{ j_{12}\lambda_{12}}\right)^*\mathcal{C}_{J\mu;\lambda_1\lambda_2\lambda_3}^{j_{12}\lambda_{12}}\\
&\hspace{2.75cm} \int \frac{p_3\sqrt{\bar p_{12}p_{12}}  \ \diff \bar u \diff \bar p_{12}\,\diff u \diff p_{12}\, \diff p_3 }{4\sqrt[4]{ p_3^2+4\bar p_{12}^{\,2}}4\sqrt[4]{p_3^2+4 p_{12}^{\,2}}}\,\left(\bar\Psi_{\lambda_{12}\,\bar\lambda_1-\bar\lambda_2}^{j_{12}}(\bar u,\bar p_{12}, p_3)\right)^* \Psi_{\lambda_{12}\lambda_1-\lambda_2}^{j_{12}}(u,p_{12},p_3) 
 \\
&\hspace{8.3cm} \bra{\bar p_{12}; j_{12}\lambda_{12};\bar\lambda_1\bar\lambda_2}\mathcal{O}(r_{12})\ket{p_{12};j_{12}\lambda_{12};\lambda_1\lambda_2}.
\end{aligned}\label{eq::3GB_ME_Vunsym}
\end{equation}
The lower boundary in the summation on $j_{12}$ must now fulfill both the constraints $j_{12}\geq|\lambda_1-\lambda_2|$ and $j_{12}\geq|\bar\lambda_1-\bar\lambda_2|$. This integral can be reorganised as follows,
\begin{equation}
\begin{aligned}
&\leftindex^J_{\bar\mu}{\bra{\bar\Psi;\bar\lambda_1\bar\lambda_2\bar\lambda_3}}\mathcal{O}(r_{12})\ket{\Psi;\lambda_1\lambda_2\lambda_3}^J_\mu\ = \delta_{\bar\lambda_3\lambda_3} \sum_{j_{12}}\,\sum_{\lambda_{12}}\left(\mathcal{C}_{J\bar\mu;\bar\lambda_1\bar\lambda_2\lambda_3}^{ j_{12}\lambda_{12}}\right)^*\mathcal{C}_{J\mu;\lambda_1\lambda_2\lambda_3}^{j_{12}\lambda_{12}}\\
&\hspace{2.75cm} \int p_3 \diff p_3 \int \frac{\diff \bar p_{12}}{2}\frac{\diff p_{12}}{2}\, \tilde\Psi(\bar p_{12},p_3;j_{12},\lambda_{12},\bar\lambda_1-\bar\lambda_2;\bar\Psi^*) \tilde\Psi(p_{12}, p_3;j_{12},\lambda_{12},\lambda_1-\lambda_2;\Psi)\\
&\hspace{8.75cm}\bra{\bar p_{12}; j_{12}\lambda_{12};\bar\lambda_1\bar\lambda_2}\mathcal{O}(r_{12})\ket{p_{12};j_{12}\lambda_{12};\lambda_1\lambda_2}
\end{aligned}\label{eq::3GB_ME_Vunsym_fin}
\end{equation}
where 
\begin{equation}
\begin{aligned}
&\tilde\Psi(p_{12},p_3;j_{12},\lambda_{12},\Delta\lambda;\Psi) = \frac{\sqrt{p_{12}}}{2\sqrt[4]{4 p_{12}^2+p_3^2}}   \int \diff u\,\Psi^{j_{12}}_{\lambda_{12}\,\Delta\lambda}(u,p_{12},p_3)
\end{aligned}\label{eq::3GB_ME_modPsi}
\end{equation}
and where the product of three-body $\mathcal{C}$ coefficients can be made a bit more explicit,
\begin{equation}
\begin{aligned}
\left(\mathcal{C}_{J\bar\mu;\bar\lambda_1\bar\lambda_2\lambda_3}^{ j_{12}\lambda_{12}}\right)^*\mathcal{C}_{J\mu;\lambda_1\lambda_2\lambda_3}^{j_{12}\lambda_{12}} = e^{i \frac{\pi}{2}\left((\lambda_2-\bar\lambda_2) + (\bar\lambda_1-\lambda_1) + (\bar\mu-\mu)\right)} \frac{2j_{12}+1}{2}\, d_{\bar\mu\,\lambda_{12}-\lambda_3}^{J}(\pi/2)d_{\mu\,\lambda_{12}-\lambda_3}^{J}(\pi/2)&\\
= (-1)^{(\lambda_2-\bar\lambda_2)/2 + (\bar\lambda_1-\lambda_1)/2 + (\bar\mu-\mu)/2}\,\frac{2j_{12}+1}{2}\, d_{\bar\mu\,\lambda_{12}-\lambda_3}^{J}(\pi/2)d_{\mu\,\lambda_{12}-\lambda_3}^{J}(\pi/2)&.\\
\end{aligned} \label{eq::3GB_ME_coefficients}
\end{equation}
In equation \eqref{eq::3GB_ME_Vunsym_fin}, the function $\tilde\Psi$ acts like an unnormalised two-body helicity-momentum wave function that would depend on three parameters ($a$, $b$ and $p_3$). Therefore, the evaluation of the integrals on $p_{12}$ and $\bar p_{12}$ is fully analogous to what has been done to compute potential ME for two-body systems. 

Equations \eqref{eq::3GB_ME_Vunsym_fin} to \eqref{eq::3GB_ME_coefficients} are key formulas for the current work as they allow computation of two-body potential ME on three-body helicity states. However, these formulas involves the evaluation of an infinite series of four dimensional integrals. Even when truncating this series, calculating such a large number of integrals remains a significant challenge. Gaining deeper insight into the components of this formula may help mitigate this complexity. \ref{app::simpl} is dedicated to this analysis.

\subsubsection{Evaluation of matrix elements on symmetrical states}

The previous subsection worked at obtaining formulas to compute Hamiltonian ME on unsymmetrical states. In the current one, this technology is used to compute ME for symmetric states. Starting with kinetic ME, a formula for both symmetric states having $\mu=0$ can easily be obtained thanks to the absence of mixing between helicities in \eqref{eq::3GB_kinFinal}. The situation is even simpler because non-zero terms prove to be independent of the actual value of $\lambda_1$, $\lambda_2$ or $\lambda_3$. Only a single integral has to be evaluated,
\begin{equation}
\begin{aligned}
\bra{\bar\Psi;A'_2;0;(2k+1)^{\pm-}}T\ket{\Psi;A'_2;0;(2k+1)^{\pm-}} & = \bra{\bar\Psi;A''_2;0;(2k+1)^{\pm-}}T\ket{\Psi;A''_2;0;(2k+1)^{\pm-}}\\
& =  \int \frac{\diff w_1 \diff  w_2 \diff w_3}{8} \, \bar\Psi^*(w_1,w_2,w_3)\Psi(w_1,w_2,w_3)T(w_1,w_2,w_3).
\end{aligned} \label{eq::3GB_ME_TA0}
\end{equation}
Concerning states based on $\mu=\pm1$, the presence of Kronecker deltas in \eqref{eq::3GB_kinFinal} still simplifies the evaluation of kinetic energy. For the $A_2'$ states, cf. Eq.~\eqref{eq::3GB_sym_11_+++}, due to the additional kinematic factors arising from the symmetrisation, the result requires the evaluation of a slightly different integral,
\begin{equation}
\begin{aligned}
&\bra{\bar\Psi;A'_2;1;J^{\mp-}}T\ket{\Psi;A'_2;1;J^{\mp-}} \\
&\hspace{1.5cm} = \int \frac{\diff w_1 \diff  w_2 \diff w_3 }{2}\, \bar\Psi^*(w_1,w_2,w_3)\Psi(w_1,w_2,w_3)|1+e^{-i\varphi_{13}}+e^{i\varphi_{23}}|^2\, T(w_1,w_2,w_3). \label{eq::3GB_ME_TA'1}
\end{aligned}
\end{equation}
For the $A''_2$ states, cf. Eq.~\eqref{eq::3GB_sym_11_+-+}, the kinetic energy ME are again provided by the previous integral,
\begin{equation}
\begin{aligned}
&\bra{\bar\Psi;A''_2;1; J^{\mp-}}T\ket{\Psi;A''_2;1; J^{\mp-}} = \int \frac{\diff w_1 \diff  w_2 \diff w_3}{8} \, \bar\Psi^*(w_1,w_2,w_3)\Psi(w_1,w_2,w_3)T(w_1,w_2,w_3).
\end{aligned} \label{eq::3GB_ME_TA''1}
\end{equation}
For each symmetric state, kinetic energy ME can be evaluated in a single integral. Notice that equations \eqref{eq::3GB_ME_TA0} to \eqref{eq::3GB_ME_TA''1} allows to infer formulas for the overlap between symmetrical states by plugging $T(w_1,w_2,w_3)=1$.

Let us move on to the case of two-body potential ME. Due to the $\delta_{\bar\lambda_3\lambda_3}$ factor in \eqref{eq::3GB_ME_Vunsym_fin}, some ME on unsymmetrical states are directly shown to cancel. For state \eqref{eq::3GB_sym_10_+++}, the resulting expression only requires to evaluate two ME,
\begin{equation}
\begin{aligned}
\bra{\bar\Psi;A'_2;0; (2k+1)^{\pm-}}\mathcal{O}(r_{12})\ket{\Psi;A'_2;0; (2k+1)^{\pm-}} 
=\frac{1}{2}\big(\leftindex^{2k+1}_0{\bra{\bar\Psi;+++}}\mathcal{O}(r_{12})\ket{\Psi;+++}^{2k+1}_0&\\
+ \leftindex^{2k+1}_0{\bra{\bar\Psi;---}}\mathcal{O}(r_{12})\ket{\Psi;---}^{2k+1}_0&\big),
\end{aligned} \label{eq::3GB_ME_VA'0A'0}
\end{equation}
One can immediately observe that potential ME for these states are parity degenerated. Concerning state \eqref{eq::3GB_sym_10_+-+}, this sum is composed of twelve ME, including non-diagonal ones,
\begin{equation}
\begin{aligned}
\bra{\bar\Psi;A''_2;0;(2k+1)^{\pm-}}\mathcal{O}(r_{12})\ket{\Psi;A''_2;0;(2k+1)^{\pm-}} \hspace{7.5cm} &\\
=\frac{1}{6}\big(\leftindex^{2k+1}_0{\bra{\bar\Psi;-++}}\mathcal{O}(r_{12})\ket{\Psi;-++}^{2k+1}_0 + \leftindex^{2k+1}_0{\bra{\bar\Psi;+-+}}\mathcal{O}(r_{12})\ket{\Psi;+-+}^{2k+1}_0 &\\
+ \leftindex^{2k+1}_0{\bra{\bar\Psi;--+}}\mathcal{O}(r_{12})\ket{\Psi;--+}^{2k+1}_0 + \leftindex^{2k+1}_0{\bra{\bar\Psi;-+-}}\mathcal{O}(r_{12})\ket{\Psi;-+-}^{2k+1}_0& \\ 
+ \leftindex^{2k+1}_0{\bra{\bar\Psi;+--}}\mathcal{O}(r_{12})\ket{\Psi;+--}^{2k+1}_0 + \leftindex^{2k+1}_0{\bra{\bar\Psi;++-}}\mathcal{O}(r_{12})\ket{\Psi;++-}^{2k+1}_0&\big) \\
+ \frac{1}{3}\big(\leftindex^{2k+1}_0{\bra{\bar\Psi;+-+}}\mathcal{O}(r_{12})\ket{\Psi;-++}^{2k+1}_0 + \leftindex^{2k+1}_0{\bra{\bar\Psi;+--}}\mathcal{O}(r_{12})\ket{\Psi;-+-}^{2k+1}_0 & \\
\pm \leftindex^{2k+1}_0{\bra{\bar\Psi;--+}}\mathcal{O}(r_{12})\ket{\Psi;-++}^{2k+1}_0 \pm \leftindex^{2k+1}_0{\bra{\bar\Psi;--+}}\mathcal{O}(r_{12})\ket{\Psi;+-+}^{2k+1}_0 & \\
\pm \leftindex^{2k+1}_0{\bra{\bar\Psi;-+-}}\mathcal{O}(r_{12})\ket{\Psi;++-}^{2k+1}_0 \pm \leftindex^{2k+1}_0{\bra{\bar\Psi;++-}}\mathcal{O}(r_{12})\ket{\Psi;+--}^{2k+1}_0&\big). 
\end{aligned} \label{eq::3GB_ME_VA''0A''0}
\end{equation}
Finally, concerning states with $\mu = \pm 1$, calculations will only be illustrated with $\ket{\Psi;A_2';1;J^{-\pm}}$ because it is less cumbersome to decompose than $\ket{\Psi;A_2'';1;J^{-\pm}}$,  
\begin{equation}
\begin{aligned}
&\bra{\bar\Psi;A'_2;1;J^{\mp-}}\mathcal{O}(r_{12})\ket{\Psi;A'_2;1;J^{\mp-}} \\
&\hspace{2.68cm} = \big(\leftindex^J_1{\bra{\left(1+e^{i\varphi_{13}}+e^{-i\varphi_{23}}\right)\bar\Psi};+++}\mathcal{O}(r_{12})\ket{\left(1+e^{i\varphi_{13}}+e^{-i\varphi_{23}}\right)\Psi;+++}^J_1\\
&\hspace{2.95cm}+\leftindex^J_1{\bra{\left(1+e^{i\varphi_{13}}+e^{-i\varphi_{23}}\right)\bar\Psi};---}\mathcal{O}(r_{12})\ket{\left(1+e^{i\varphi_{13}}+e^{-i\varphi_{23}}\right)\Psi;---}^J_1\\
&\hspace{2.75cm} + \leftindex^J_{-1}{\bra{\left(1+e^{-i\varphi_{13}}+e^{i\varphi_{23}}\right)\bar\Psi;+++}}\mathcal{O}(r_{12})\ket{\left(1+e^{-i\varphi_{13}}+e^{i\varphi_{23}}\right)\Psi;+++}^J_{-1}\\
&\hspace{2.75cm} + \leftindex^J_{-1}{\bra{\left(1+e^{-i\varphi_{13}}+e^{i\varphi_{23}}\right)\bar\Psi;---}}\mathcal{O}(r_{12})\ket{\left(1+e^{-i\varphi_{13}}+e^{i\varphi_{23}}\right)\Psi;---}^J_{-1} \\
&\hspace{1.8cm} + 2(-1)^J \leftindex^J_1{\bra{\left(1+e^{i\varphi_{13}}+e^{-i\varphi_{23}}\right)\bar\Psi;+++}}\mathcal{O}(r_{12})\ket{\left(1+e^{-i\varphi_{13}}+e^{i\varphi_{23}}\right)\Psi;+++}^J_{-1} \\
&\hspace{1.8cm} + 2(-1)^J \leftindex^J_1{\bra{\left(1+e^{i\varphi_{13}}+e^{-i\varphi_{23}}\right)\bar\Psi};---}\mathcal{O}(r_{12})\ket{\left(1+e^{-i\varphi_{13}}+e^{i\varphi_{23}}\right)\Psi;---}^J_{-1}\big). 
\end{aligned} \label{eq::3GB_ME_VA'1A'1}
\end{equation}
In general, the symmetry properties depicted in \ref{app::simpl} facilitate in evaluating these ME. They show that some ME on unsymmetrical states are strictly equal, while some others involve the same integrals but combine them differently (see equation \eqref{eq::3GB_ME_Vunsym_fin}). This significantly reduces the overall computational cost. This is especially true if the helicity-momentum wave function $\Psi$ is even in $u$. Apart from diagonal ME, symmetric states are also shown to mix with each other. For instance, mixing ME between $\ket{\Psi;A_2';0;(2k)^{\pm-}}$ and $\ket{\Psi;A_2'';0;(2k)^{\pm-}}$ are given by
\begin{equation}
\begin{aligned}
\bra{\bar\Psi;A'_2;0; (2k+1)^{\pm-}}\mathcal{O}(r_{12})\ket{\Psi;A''_2;0; (2k+1)^{\pm-}} \hspace{8.cm} &\\
=\frac{1}{\sqrt{12}}\big(\leftindex^{2k+1}_0{\bra{\bar\Psi;+++}}\mathcal{O}(r_{12})\ket{\Psi;-++}^{2k+1}_0 + \leftindex^{2k+1}_0{\bra{\bar\Psi;+++}}\mathcal{O}(r_{12})\ket{\Psi;+-+}^{2k+1}_0 & \\
\pm \leftindex^{2k+1}_0{\bra{\bar\Psi;+++}}\mathcal{O}(r_{12})\ket{\Psi;--+}^{2k+1}_0 + \leftindex^{2k+1}_0{\bra{\bar\Psi;---}}\mathcal{O}(r_{12})\ket{\Psi;+--}^{2k+1}_0 & \\
\phantom{\frac{1}{2}}\pm \leftindex^{2k+1}_0{\bra{\bar\Psi;---}}\mathcal{O}(r_{12})\ket{\Psi;++-}^{2k+1}_0 + \leftindex^{2k+1}_0{\bra{\bar\Psi;---}}\mathcal{O}(r_{12})\ket{\Psi;-+-}^{2k+1}_0&\big).
\end{aligned} \label{eq::3GB_ME_VA'0A0''}
\end{equation}
However, it will be shown in the next section that this mixing is quite small for three-gluon glueballs and only affects masses very slightly.  

\subsubsection{Determination of the low-lying Gluball Spectrum}
\label{ssec::3GB_spec}

\begin{figure}
\centering
\begin{tikzpicture}
\usetikzlibrary{decorations.pathmorphing} 
\usetikzlibrary{calc}

\coordinate (G1) at ({2*cos(90)}, {2*sin(90)});
\coordinate (G2) at ({2*cos(210)}, {2*sin(210)});
\coordinate (G3) at ({2*cos(330)}, {2*sin(330)});

\fill (G1) circle (5pt);
\fill (G2) circle (5pt);
\fill (G3) circle (5pt);

\draw[->, >=stealth, very thick, black] (G1) -- ($(G1)!0.21!(G2)$);
\draw[-, very thick, black] (G1) -- ($(G1)!0.3!(G2)$) node[midway, above left] {$3$};
\draw[->, >=stealth reversed, very thick, black] (G2) -- ($(G2)!0.19!(G1)$);
\draw[-, very thick, black] (G2) -- ($(G2)!0.3!(G1)$) ;

\draw[->, >=stealth, very thick, black] (G2) -- ($(G2)!0.21!(G3)$);
\draw[-, very thick, black] (G2) -- ($(G2)!0.3!(G3)$);
\draw[->, >=stealth reversed, very thick, black] (G3) -- ($(G3)!0.19!(G2)$);
\draw[-, very thick, black] (G3) -- ($(G3)!0.3!(G2)$) ;

\draw[->, >=stealth, very thick, black] (G3) -- ($(G3)!0.21!(G1)$);
\draw[-, very thick, black] (G3) -- ($(G3)!0.3!(G1)$);
\draw[->, >=stealth reversed, very thick, black] (G1) -- ($(G1)!0.19!(G3)$);
\draw[-, very thick, black] (G1) -- ($(G1)!0.3!(G3)$) node[midway, above right] {$\bar{3}$};

\end{tikzpicture}
\caption{Schematic representation of the $\Delta$ junction modeling three-gluon glueballs in the current work. Gluons bounds two by two by pooling $3$ and $\bar 3$ flux tubes.}
\label{fig::3GB_spec_delta}
\end{figure}
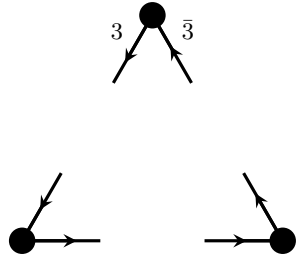

Trial states and formulas set up all along the previous subsections will now be used to compute a true mass spectrum for $1^{\pm-}$ three-gluon glueballs. It will require to evaluate ME of a Hamiltonian that should model three-gluon glueballs. This work considers the following structure for the latter,
\begin{equation}
    H = T(w_1,w_2,w_3) + V(r_{12}) + V(r_{13}) + V(r_{23}). \label{eq::3GB_spec_ham}
\end{equation}
with $r_{ij}=|\bm{r}_i-\bm{r}_j|$. Different kinematics $T$ can be envisaged for gluons. Although these particles are formally massless, they acquire a constituent mass in some potential models. In the following, calculations will be performed considering ultra-relativistic kinematics,
\begin{equation}
    T(w_1,w_2,w_3)=w_1+w_2+w_3.\label{eq::3GB_spec_ham_kin}
\end{equation}
Concerning potential, Cornell interactions will be assumed,
\begin{equation}
    V(r) = \sigma r - \frac{3\alpha_s}{2r}.\label{eq::3GB_spec_ham_pot}
\end{equation}
Above, $\sigma$ is the meson string tension, generally set around $0.185\,$GeV$^2$, and $\alpha_s$ is the strong coupling constant. In the following, the latter is fixed at $0.450$ (as in Section \ref{sec:2GB} and reference \cite{math08}). The mesonic string tension is used as confinement constant because each gluon is supposed to provide both a $3$ and a $\bar 3$ flux tube. The six flux tubes join two by two in a $\Delta$ shape, as illustrated in Fig.~\ref{fig::3GB_spec_delta}. Because each junction connects $3$ and $\bar 3$ flux tubes, the meson string tension has not to be rescaled. The factor $3/2$ is the color factor corresponding with three gluons in a color singlet. The use of this potential and its comparison with a $Y$ junction is explained in \ref{app:pot}.
To avoid reproducing many evaluations for different mesonic string tensions $\sigma$, dimensionless variables are introduced by rescaling the current ones with the square-root of the string tension,
\begin{align}
    &\mathbbmss{w}_{i} = w_i/\sqrt{\sigma}, && \mathbbmss{r}_{ij} = \sqrt{\sigma}\,r_{ij}.
\end{align}
Dimensionless equivalent for $p_3$ and $p_{12}$ are also defined
\begin{align}
    &\mathbbmss{p}_{3} = p_{3}/\sqrt{\sigma}, && \mathbbmss{p}_{12} = p_{12}/\sqrt{\sigma}.
\end{align}
Because $u$ is already dimensionless, it remains unchanged. Relation \eqref{eq::BtoW_w1w2_um} still relates variables $\mathbbmss{w}_1, \mathbbmss{w}_2,\mathbbmss{w}_3$ and $u,\mathbbmss{p}_{12},\mathbbmss{p}_{3}$.
Using these new coordinates, a dimensionless version $\mathbbmss{H}$ of the Hamiltonian \eqref{eq::3GB_spec_ham} is introduced,
\begin{equation}
    \mathbbmss{H} = H/\sqrt{\sigma}, = \mathbbmss{T}(\mathbbmss{w}_1,\mathbbmss{w}_2,\mathbbmss{w}_3) + \mathbbmss{V}(\mathbbmss{r}_{12}) + \mathbbmss{V}(\mathbbmss{r}_{13}) + \mathbbmss{V}(\mathbbmss{r}_{23}) \label{eq::3GB_ham_dimless}
\end{equation}
where
\begin{align}
&\mathbbmss{T}(\mathbbmss{w}_1,\mathbbmss{w}_2,\mathbbmss{w}_3) = \mathbbmss{w}_1 + \mathbbmss{w}_2 + \mathbbmss{w}_3, &&\mathbbmss{V}(\mathbbmss{r})=\mathbbmss{r}-\frac{3\alpha_s}{2\mathbbmss{r}}.
\end{align}
This Hamiltonian will be evaluated on different symmetric trial states. To start with, a single trial state is considered. As a reminder, the reasonable trial wave function that had been chosen is
\begin{equation}
    \Psi_{a,b}(w_1,w_2,w_3) = A\sqrt{8w_1w_2w_3}e^{-a((w_1-b)^2+(w_2-b)^2+(w_3-b)^2)}.
\end{equation}
For states \eqref{eq::3GB_sym_10_+++} and \eqref{eq::3GB_sym_10_+-+}, this trial wave function remains unmodified. For states \eqref{eq::3GB_sym_11_+++} and \eqref{eq::3GB_sym_11_+-+}, it has to be multiplied by different kinematic factors to ensure the right symmetry properties. In the following, calculations will be made explicit for a generic kinematic factor, denoted $\Gamma(w_1,w_2,w_3)$. These trial wave functions are to be turned into the new dimensionless coordinate system. The change simply rescales the $a$ parameter by a factor $\sigma$ and modify the value of the normalisation constant $A$,
\begin{align}
&\mathbbmss{a}=\sigma a, &&\mathbbmss{b}=b/\sqrt{\sigma}, &&\mathbbmss{A} = \sigma^{\frac{3}{2}} A
\end{align}
The rescaling factor for the normalisation constant has been chosen to eliminate the dimensions of the wave function. The use of a dimensionless wave function also suggests to rescale Berman's and Wick's bases to make them dimensionless,
\begin{align}
&\ket{JM\mu;\mathbbmss{w}_1\mathbbmss{w}_2\mathbbmss{w}_3;\lambda_1\lambda_2\lambda_3} = \sigma^{\frac{3}{4}} \ket{JM\mu;w_1w_2w_3;\lambda_1\lambda_2\lambda_3},\\
&\ket{\mathbbmss{p}_3;JM;j_{12}\lambda_{12}\lambda_3;\mathbbmss{p}_{12}\lambda_1\lambda_2} = \sigma^{\frac{3}{4}} \ket{p_3;JM;j_{12}\lambda_{12}\lambda_3;p_{12}\lambda_1\lambda_2}.
\end{align}
All the formulas set up to evaluate ME can now be switched to the new coordinates. Taking every change into account, all the rescaling factors cancels each others. As a result, each formula can be naively turned into dimensionless coordinates by replacing $A$, $a$, $w_i$, $p_3$ and/or $p_{12}$ by their dimensionless counterpart.

At first, let us investigate the normalisation of the symmetric trial states. For states \eqref{eq::3GB_sym_10_+++}, \eqref{eq::3GB_sym_10_+-+} and \eqref{eq::3GB_sym_11_+-+}, the normalisation constant $A$ is fixed by imposing the normalisation of the unsymmetrical trial states, implying that $\Psi_{a,b}$ must satisfy \eqref{eq::BD_physStateNorm}. Switching to dimensionless coordinates, one has to evaluate the following integral,
\begin{equation}
|\mathbbmss{A}|^2 = \left(\int \diff \mathbbmss{w}_1 \diff \mathbbmss{w}_2 \diff \mathbbmss{w}_3\, \mathbbmss{w}_1\mathbbmss{w}_2\mathbbmss{w}_3\, e^{-2\mathbbmss{a} ((\mathbbmss{w}_1-\mathbbmss{b})^2+(\mathbbmss{w}_2-\mathbbmss{b})^2+(\mathbbmss{w}_3-\mathbbmss{b})^2)}\right)^{-1},
\end{equation}
where both $\mathbbmss{w}_1$ and $\mathbbmss{w}_2$ are integrated between $0$ and $\infty$ while $\mathbbmss{w}_3$ is integrated between $|\mathbbmss{w}_1-\mathbbmss{w}_2|$ and $\mathbbmss{w}_1+\mathbbmss{w}_2$. Generic multidimensional integration techniques provide fast and accurate evaluations of $\mathbbmss{A}$. As inferred from \eqref{eq::3GB_ME_TA'1}, the situation is slightly more complicated for state \eqref{eq::3GB_sym_11_+++} due to the additional kinematic factor. For the $A'_2$ state, the normalization involves the integral
\begin{equation}
\begin{aligned}
    &|\mathbbmss{A}|^2 = \left(4\int \diff \mathbbmss{w}_1 \diff \mathbbmss{w}_2 \diff \mathbbmss{w}_3\,\mathbbmss{w}_1\mathbbmss{w}_2\mathbbmss{w}_3\, e^{-2\mathbbmss{a} ((\mathbbmss{w}_1-\mathbbmss{b})^2+(\mathbbmss{w}_2-\mathbbmss{b})^2+(\mathbbmss{w}_3-\mathbbmss{b})^2)}|1+e^{-i\varphi_{13}}+e^{i\varphi_{23}}|^2.\right)^{-1}
\end{aligned}
\end{equation}
It is possible to evaluate these two integrals numerically, with any desired accuracy.

The evaluation of kinetic energy ME is analogous to the normalisation. States \eqref{eq::3GB_sym_10_+++}, \eqref{eq::3GB_sym_10_+-+} and \eqref{eq::3GB_sym_11_+-+} requires to evaluate a single three-dimensional integral, as demonstrated in relations \eqref{eq::3GB_ME_TA0} and \eqref{eq::3GB_ME_TA''1}. For the current trial wave function, it equates to compute the following integral
\begin{equation}
\begin{aligned}
\braket{\mathbbmss{T}}=|\mathbbmss{A}|^2 \int \diff \mathbbmss{w}_1 \diff  \mathbbmss{w}_2 \diff \mathbbmss{w}_3 \, \mathbbmss{w}_1\mathbbmss{w}_2\mathbbmss{w}_3\,e^{-2\mathbbmss{a}((\mathbbmss{w}_1-\mathbbmss{b})^2+(\mathbbmss{w}_2-\mathbbmss{b})^2+(\mathbbmss{w}_3-\mathbbmss{b})^2)}(\mathbbmss{w}_1+\mathbbmss{w}_2+\mathbbmss{w}_3).
\end{aligned}
\end{equation}
On the other hand, kinetic energy ME for state \eqref{eq::3GB_sym_11_+++} are provided by relation \eqref{eq::3GB_ME_TA'1}, which complements the integrand with a kinematic factor,
\begin{equation}
\begin{aligned}
&\bra{\Psi_{a,b};A'_2;\pm  ;1^{\pm-}}\mathbbmss{T}\ket{\Psi_{a,b};A'_2;\pm; 1^{\pm-}} \hspace{8cm} \\
&\hspace{0.75cm} = 4|\mathbbmss{A}|^2 \int \diff \mathbbmss{w}_1 \diff \mathbbmss{w}_2 \diff \mathbbmss{w}_3 \, \mathbbmss{w}_1\mathbbmss{w}_2\mathbbmss{w}_3\,e^{-2\mathbbmss{a}((\mathbbmss{w}_1-\mathbbmss{b})^2+(\mathbbmss{w}_2-\mathbbmss{b})^2+(\mathbbmss{w}_3-\mathbbmss{b})^2)} |1+e^{-i\varphi_{13}}+e^{i\varphi_{23}}|^2\, (\mathbbmss{w}_1+\mathbbmss{w}_2+\mathbbmss{w}_3).
\end{aligned}
\end{equation}

It remains to evaluate potential ME. As already mentioned, thanks to the symmetry of the state, the evaluation for the interaction between particles $1$ and $2$ is sufficient to infer the full potential energy of the $\Delta$ junction,
\begin{equation}
    \braket{\,\mathbbmss{V}(\mathbbmss{r}_{12})+\mathbbmss{V}(\mathbbmss{r}_{13})+\mathbbmss{V}(\mathbbmss{r}_{23})} = 3\braket{\,\mathbbmss{V}(\mathbbmss{r}_{12})}. \label{eq::3GB_spec_v=3v}
\end{equation}
Relations \eqref{eq::3GB_ME_VA'0A'0}, \eqref{eq::3GB_ME_VA''0A''0} and \eqref{eq::3GB_ME_VA'1A'1} reduce evaluations on symmetric states to several evaluations on unsymmetrical states, eventually with additional kinematic factors. The latter evaluations can be performed using formula \eqref{eq::3GB_ME_Vunsym_fin}. Specifying for the current trial wave function and using variables $u$, $p_{12}$ and $p_3$ defined in equation \eqref{eq::BtoW_um_w1w2}, one gets
\begin{equation}
\begin{aligned}
&\leftindex_{\mu}^J {\bra{\Psi_{a,b}\bar\Gamma;\bar\lambda_1\bar\lambda_2\bar\lambda_3}}\mathbbmss{V}(\mathbbmss{r}_{12})\ket{\Psi_{a,b}\Gamma;\lambda_1\lambda_2\lambda_3}_\mu^J = \delta_{\bar\lambda_3\lambda_3} \sum_{j_{12}} \,\sum_{\lambda_{12}}  \left(\mathcal{C}_{J\mu;\bar\lambda_1\bar\lambda_2\lambda_3}^{ j_{12}\lambda_{12}}\right)^*\mathcal{C}_{J\mu;\lambda_1\lambda_2\lambda_3}^{j_{12}\lambda_{12}}\\
&\hspace{2.5cm} 
\int \mathbbmss{p}_3\diff \mathbbmss{p}_3 \int \frac{\diff\bar{\mathbbmss{p}}_{12}}{2} \frac{\diff \mathbbmss{p}_{12}}{2}\, \tilde\Psi^*_{\mathbbmss{a},\mathbbmss{b}}(\bar{\mathbbmss{p}}_{12},\mathbbmss{p}_3;j_{12},\lambda_{12},\bar \lambda_1-\bar \lambda_2;\bar\Gamma) \tilde\Psi_{\mathbbmss{a},\mathbbmss{b}}(\mathbbmss{p}_{12},\mathbbmss{p}_3;j_{12},\lambda_{12},\lambda_1-\lambda_2;\Gamma)\\
& \hspace{9.20cm} \bra{\bar{\mathbbmss{p}}_{12}; j_{12}\lambda_{12};\bar\lambda_1\bar\lambda_2}\mathbbmss{V}(\mathbbmss{r}_{12})\ket{\mathbbmss{p}_{12};j_{12}\lambda_{12};\lambda_1\lambda_2}\\
\end{aligned}\label{eq::3GB_spec_integral}
\end{equation}
where
\begin{equation}
\begin{aligned}
\tilde\Psi_{\mathbbmss{a},\mathbbmss{b}}(\mathbbmss{p}_{12},\mathbbmss{p}_3;j_{12},\lambda_{12},\Delta\lambda;\Gamma) =\ & \mathbbmss{A}\,\frac{\sqrt{\mathbbmss{p}_{12}\mathbbmss{p}_3}}{2\sqrt[4]{4 \mathbbmss{p}_{12}^2+\mathbbmss{p}_3^2}}\, e^{-\frac{\mathbbmss{a}}{2}\left(6 \mathbbmss{b}^2-4 \mathbbmss{b} \left(\sqrt{4\mathbbmss{p}_{12}^2+\mathbbmss{p}_3^2}+\mathbbmss{p}_3\right)+4 \mathbbmss{p}_{12}^2+3 \mathbbmss{p}_3^2\right)} \\
& \int_{-1}^1 \diff u\,\sqrt{8\mathbbmss{p}_{12}^2+2 \mathbbmss{p}_{3}^2 \left(1-u^2\right)}\, d^{j_{12}}_{\lambda_{12}\,\Delta\lambda}(\arccos\,u) \,e^{-\frac{a}{2}\mathbbmss{p}_3^2u^2} \Gamma(u,\mathbbmss{p}_{12},\mathbbmss{p}_3).
\end{aligned}\label{eq::3GB_spec_tildePsi}
\end{equation}

These integrals must be evaluated. Performing efficient numerical integrations requires to slightly modify the variables and to refine the definition of $\tilde\Psi$. First, because the integral on $\mathbbmss{p}_3$ will be performed using a generalised Gauss-Laguerre quadrature, the variable $\mathbbmss{p}_3$ is modified as follows
\begin{align}
    & x = 3\mathbbmss{a}\mathbbmss{p}_3^2, && \mathbbmss{p}_3 = \sqrt{\frac{x}{3\mathbbmss{a}}},  &&\mathbbmss{p}_3\diff \mathbbmss{p}_3 = \frac{\diff x}{6\mathbbmss{a}}.
\end{align}
After replacement and a bit of reorganisation, the integrals to evaluate become
\begin{equation}
\begin{aligned}
& |\mathbbmss{A}|^2 \sqrt{\frac{2}{(6\mathbbmss{a})^3}} \int \sqrt{x}\,e^{-x} \diff x \int \frac{\diff\bar{\mathbbmss{p}}_{12}}{2} \frac{\diff \mathbbmss{p}_{12}}{2}\, \Theta_{\mathbbmss{a}, \mathbbmss{b}}(\bar{\mathbbmss{p}}_{12},x;j_{12},\lambda_{12},\bar \lambda_1-\bar \lambda_2;\bar\Gamma^*) \\
& \hspace{2.4cm} \Theta_{\mathbbmss{a},\mathbbmss{b}}(\mathbbmss{p}_{12},x;j_{12},\lambda_{12},\lambda_1-\lambda_2;\Gamma) \bra{\bar{\mathbbmss{p}}_{12}; j_{12}\lambda_{12};\bar\lambda_1\bar\lambda_2}\mathbbmss{V}(\mathbbmss{r}_{12})\ket{\mathbbmss{p}_{12};j_{12}\lambda_{12};\lambda_1\lambda_2}
\end{aligned} \label{eq::3GB_spec_integral_num}
\end{equation}
with a function $\Theta$ that replaces $\tilde{\Psi}$,
\begin{equation}
\begin{aligned}
&\Theta_{\mathbbmss{a},\mathbbmss{b}}(\mathbbmss{p}_{12},x;j_{12},\lambda_{12},\Delta\lambda;\Gamma) = \mathcal{F}_{\mathbbmss{a},\mathbbmss{b}}(\mathbbmss{p}_{12},x) \int \diff u\ \mathcal{K}_{\mathbbmss{a}}(u,\mathbbmss{p}_{12},x;j_{12},\lambda_{12},\Delta\lambda;\Gamma)
\end{aligned}\label{eq::3GB_spec_theta}
\end{equation}
where 
\begin{equation}
\mathcal{F}_{\mathbbmss{a},\mathbbmss{b}}(\mathbbmss{p}_{12},x) = \frac{\sqrt{\mathbbmss{p}_{12}}}{2\sqrt[4]{4 \mathbbmss{p}_{12}^2+\frac{x}{3\mathbbmss{a}}}}\,e^{-\frac{\mathbbmss{a}}{2}\left(6 \mathbbmss{b}^2-4 \mathbbmss{b} \left(\sqrt{4\mathbbmss{p}_{12}^2+x/(3\mathbbmss{a})}+\sqrt{x/(3\mathbbmss{a})}\right)+4 \mathbbmss{p}_{12}^2\right)} \label{eq::3GB_spec_Factor}
\end{equation}
and where
\begin{equation}
\begin{aligned}
&\mathcal{K}_{\mathbbmss{a}}(u,\mathbbmss{p}_{12},x;j_{12},\lambda_{12},\Delta\lambda;\Gamma) = \sqrt{8\mathbbmss{p}_{12}^2+\frac{2x}{3\mathbbmss{a}} \left(1-u^2\right)}\, d^{j_{12}}_{\lambda_{12}\,\Delta\lambda}(\arccos\,u) \,e^{-\frac{x}{6}u^2} \Gamma\left(u,\mathbbmss{p}_{12},\sqrt{x/(3\mathbbmss{a})}\right).
\end{aligned}\label{eq::3GB_spec_Kernel}
\end{equation}

Symmetry properties of $\Theta$ under modifications of indices $\lambda_{12}$ and $\Delta\lambda$ are the same as the ones of $\tilde\Psi$. They are compiled in \ref{app::simpl}. Integrals \eqref{eq::3GB_spec_integral_num}-\eqref{eq::3GB_spec_theta} can be computed using a series of Gauss quadratures. The following procedure is applied.
\begin{itemize}
    \item Choose a definite number of points to use in each of the integrals on $x$, $\mathbbmss{p}_{12}$, $\bar{\mathbbmss{p}}_{12}$ and $u$. Notice that the integrals on $\mathbbmss{p}_{12}$ and $\bar{\mathbbmss{p}}_{12}$ will be turned into integrals on $\mathbbmss{v}=\mathbbmss{p}_{12}+\bar{\mathbbmss{p}}_{12}$ and $\bar{\mathbbmss{v}}=\mathbbmss{p}_{12}-\bar{\mathbbmss{p}}_{12}$, as suggested in Section \ref{ssec::2BS_calc}. Because the integral on $\bar{\mathbbmss{v}}$ is the most sensitive, it is recommended to assign the greatest number of points to it.
    \item Accuracies being chosen, identify the different $x$ values and $\mathbbmss{p}_{12}$ values at which an evaluation of the function $\Theta$ will be required and perform all the necessary evaluations. Gauss-Legendre quadrature works very well for the integral on $u$. The result is stored for further use.
    \item For each $x$ value, compute the two-body ME in the same way it has been done in Section \ref{sec:2GB}. Thanks to the previous step, two-body wave functions $\Theta$ have already been tabulated. Again, the result is stored for further use.
    \item At this stage, only the integral on $x$ remains. Perform this last integral by using the evaluations of two-body ME stored during the last step. As already mentioned, a Gauss-Laguerre quadrature is the most adapted.
\end{itemize} 
For each value of $j_{12}$ and $\lambda_{12}$, that is for each term in the sum in equation \eqref{eq::3GB_spec_integral}, four integrals are evaluated numerically. Since the sum extends to $j_{12}\to \infty$, the number of required integrals would be infinite. In practice, the sum on $j_{12}$ is truncated at a given level, denoted $j_{\text{max}}$. Plugging the computed integrals into equation \eqref{eq::3GB_spec_integral} allows for an approximate evaluation of two-body potential ME on unsymmetrical trial states. Because of the multiple quadratures and because of this truncation, the exactness of the evaluation cannot be guaranteed. The ME on unsymmetrical states can be combined following expressions \eqref{eq::3GB_ME_VA'0A'0} to \eqref{eq::3GB_ME_VA'0A0''} in order to compute potential ME on symmetrical trial states. To get the full potential energy, the factor $3$ from equation \eqref{eq::3GB_spec_v=3v} is not to be forgotten. Once potential ME on symmetrical states have been computed, the contribution from kinetic energy is added and Hamiltonian ME are obtained. This calculation is repeated for many different couples $(\mathbbmss{a},\mathbbmss{b})$ and only the minimum value is retained as an approximation for the true spectrum. 

\begin{figure}
\centering
\includegraphics[scale=0.4]{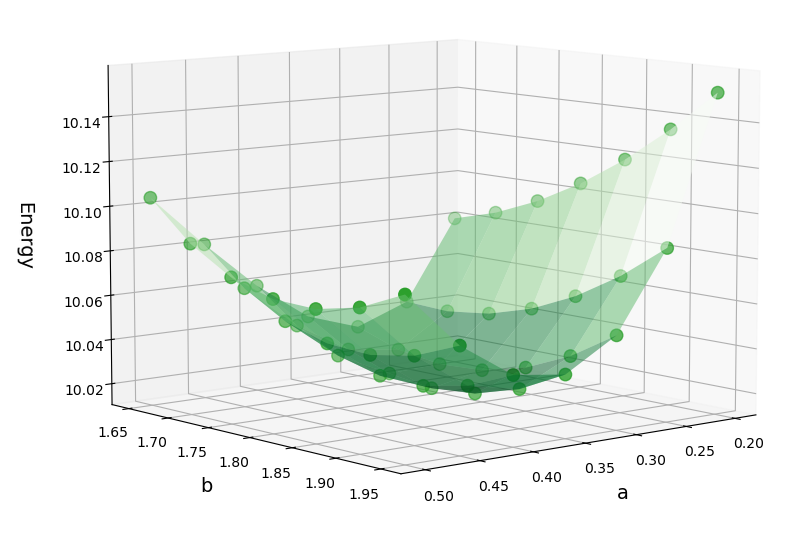}
\includegraphics[scale=0.4]{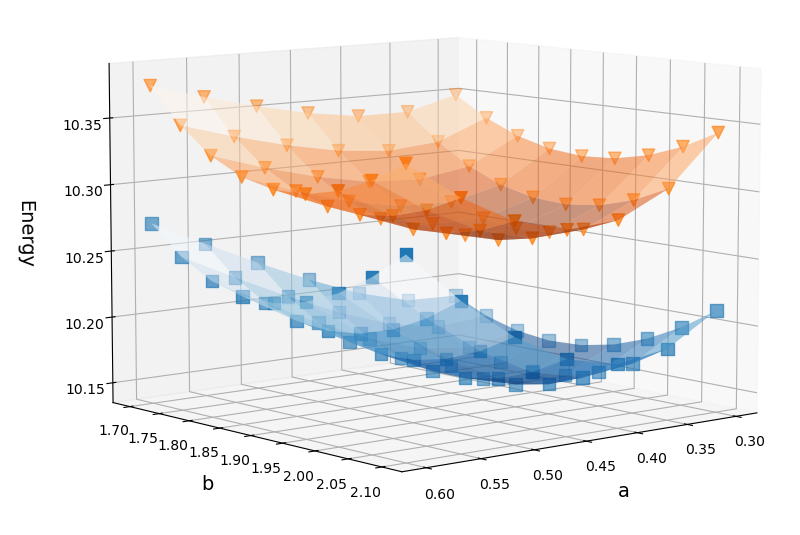}
\caption{Plot of Hamiltonian ME for a single trial state depending on the non-linear variational parameters $\mathbbmss{a}$ and $\mathbbmss{b}$. Left : a symmetric state $\ket{\Psi_{a,b};A'_2;0;1^{\pm-}}$ is considered. Right : symmetric states $\ket{\Psi_{a,b};A''_2;0;1^{\pm-}}$ are considered (even with orange triangles and odd with blue squares). Quadrature with $30$ points have always been used except for $\bar{\mathbbmss{v}}$ where $100$ points have been uses. A cut-off $j_{\text{max}}=20$ has been chosen.}
\label{fig::3GB_spec_Ap0&App0}
\end{figure}

\begin{figure}
\centering
\includegraphics[scale=0.35]{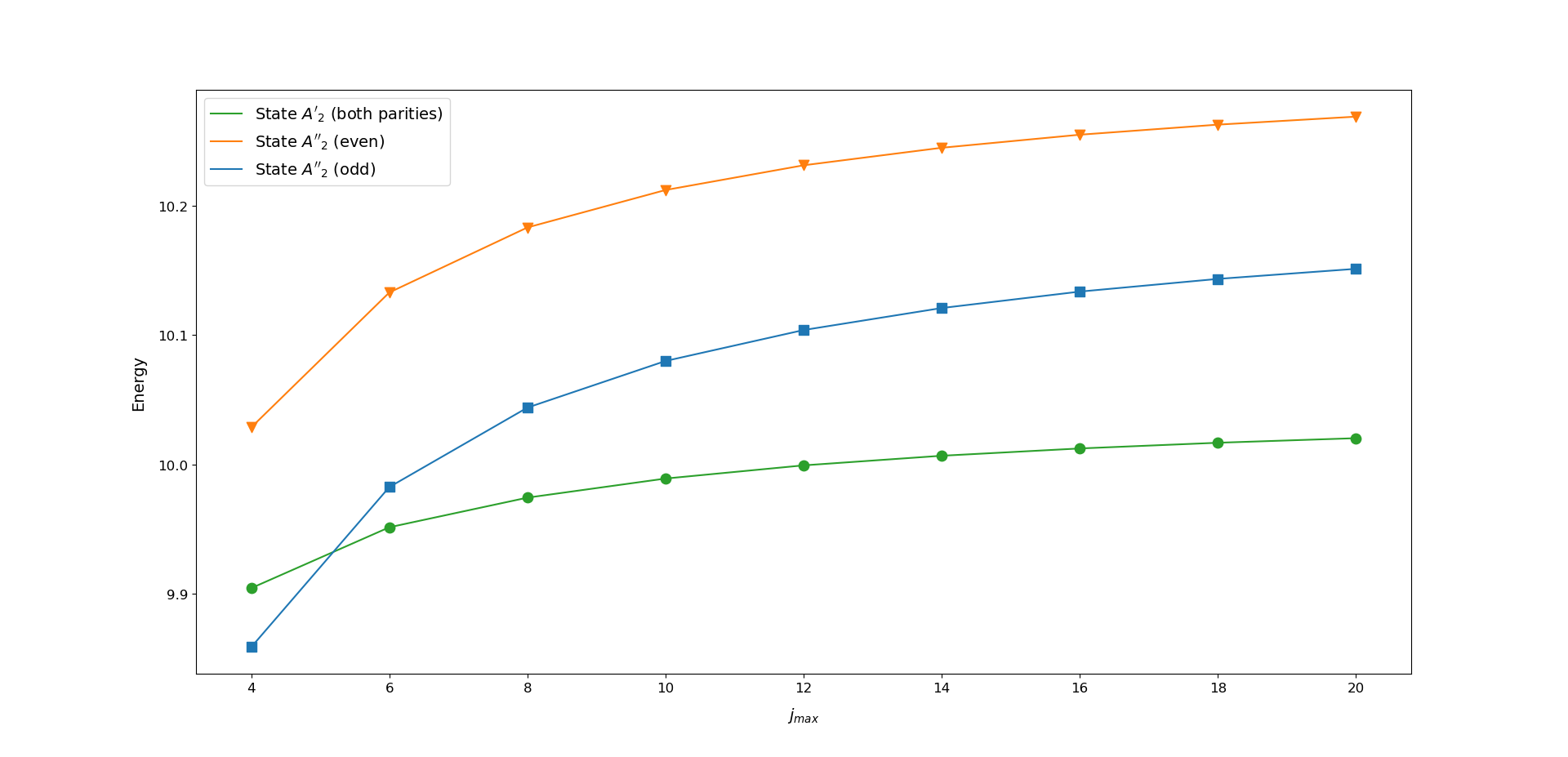}
\caption{Evolution of the lowest three-gluon glueball masses depending on the cut-off for the sum in $j_{12}$. Results for $\ket{\Psi_{a,b};A'_2;0;1^{\pm-}}$ use $(\mathbbmss{a},\mathbbmss{b})=(0.35,1.8)$ and are displayed with green circles. Results for$\ket{\Psi_{a,b};A''_2;0;1^{+-}}$ use $(\mathbbmss{a},\mathbbmss{b})=(0.45,1.95)$ and are displayed with orange triangles. Results for $\ket{\Psi_{a,b};A''_2;0;1^{--}}$ use $(\mathbbmss{a},\mathbbmss{b})= (0.35,1.9)$ and are displayed with blue squares.}
\label{fig::3GB_j12conv}
\end{figure}

Since these states are the simplest to handle, convergence properties are illustrated using $\ket{\Psi_{a,b};A'_2;0;1^{\pm-}}$ and $\ket{\Psi_{a,b};A''_2;0;1^{\pm-}}$. All the following evaluations will be performed with $\alpha_s=0.450$ \cite{math08}. To start with, calculations are performed for various numbers of points used in the different quadratures. All these computations employs for now the same cutoff at $j_{12}=20$. The minimisation process is executed for each configurations. The results are summarized in Table~\ref{tab::3GB_spec_acc}. As expected, the integrals over $u$ and $x$ converge with as few as $30$ points. While slightly more sensitive, the integral on $\mathbbmss{v}$ does not require more than $50$ points for sufficient accuracy in the current context. The key parameter for ensuring precision is the number of points used in the integral over $\mathbbmss{v}$. Using $100$ points, the energy stabilizes to two significant digits. The search for a minimum in the variational parameters can also be illustrated. Fig.~\ref{fig::3GB_spec_Ap0&App0} shows energy evaluations over a range of $(\mathbbmss{a},\mathbbmss{b})$ pairs. Clear minima are observed for each state. Lastly, the convergence of the sum over $j_{12}$ can be examined by repeating evaluations with various cutoff values. The result is displayed in Fig.~\ref{fig::3GB_j12conv}. Although the convergence behavior differs slightly between states, a reasonable convergence is achieved in all cases. Notice that the $j_{12}$ convergence can also be assessed by calculating normalisation by means of equation \eqref{eq::3GB_spec_integral}, substituting $\mathbbmss{V}(r_{12})$ with the identity operator. 

Before to move on to the determination of a complete spectrum, the mixing between different symmetric states is to be investigated. This issue will be illustrated using $\ket{\Psi_{a,b};A'_2;0;1^{+-}}$ and $\ket{\Psi_{a,b};A''_2;0;1^{+-}}$. Without allowing for mixing, approximate energies are simply obtained by evaluating the Hamiltonian on each state separately,
\begin{align}
    &E_{A'_2;0} = 10.020, &&E_{A''_2;0} = 10.269.
\end{align}
Introducing mixing between these state requires the evaluation of off-diagonal Hamiltonian ME and solving the corresponding eigenvalue problem. The off-diagonal kinetic ME cancel, while the off-diagonal potential ME are calculated using formula \eqref{eq::3GB_ME_VA'0A0''}. The splitting is found to be maximal when both states share the same variational parameters. The resulting optimal eigenvalues of the Hamiltonian matrix are
\begin{align}
    &E_{1} = 9.9902, &&E_{2} = 10.3319.
\end{align}
As shown, the mixing slightly affects the energies, but this effect does not exceed $1\%$ of the original value. Mixing with states of higher mass is expected to have an even smaller impact due to the increasing energy gap. Consequently, the following calculations will be performed using pure symmetric states.


\begin{table}
\centering
\begin{tabular}{l r}
\multicolumn{2}{l}{For $\ket{\Psi;A'_2;0;1^{\pm-}}$:\phantom{abcd}} \\
\hline\hline
$(N_{u},N_{\mathbbmss{v}},N_{\bar{\mathbbmss{v}}},N_{x})$ \hspace{0.4cm}  & Energy \\
\hline
$(30,30,30,30)$ & $9.9499$\\ \\[-2mm]
$(50,30,30,30)$ & $9.9499$\\
$(100,30,30,30)$ & $9.9499$\\ \\[-2mm]
$(30,50,30,30)$ & $9.9536$\\
$(30,100,30,30)$ & $9.9536$\\ \\[-2mm]
$(30,30,50,30)$ & $9.9895$\\
$(30,30,100,30)$ & $10.0202$\\ \\[-2mm]
$(30,30,30,50)$ & $9.9499$\\
$(30,30,30,100)$ & $9.9499$\\
\hline
\end{tabular}\hfill
\begin{tabular}{l r}
\multicolumn{2}{l}{For $\ket{\Psi;A''_2;0;1^{+-}}$:\phantom{abcd}} \\
\hline\hline
$(N_{u},N_{\mathbbmss{v}},N_{\bar{\mathbbmss{v}}},N_{x})$ \hspace{0.4cm} & Energy \\
\hline
$(30,30,30,30)$ & $10.1793$\\ \\[-2mm]
$(50,30,30,30)$ & $10.1793$\\
$(100,30,30,30)$ & $10.1793$\\ \\[-2mm]
$(30,50,30,30)$ & $10.1817$\\
$(30,100,30,30)$ & $10.1817$\\ \\[-2mm]
$(30,30,50,30)$ & $10.2297$\\
$(30,30,100,30)$ & $10.2689$\\ \\[-2mm]
$(30,30,30,50)$ & $10.1793$\\
$(30,30,30,100)$ & $10.1793$\\
\hline
\end{tabular}\hfill
\begin{tabular}{l r}
\multicolumn{2}{l}{For $\ket{\Psi;A''_2;0;1^{--}}$:\phantom{abcd}} \\
\hline\hline
$(N_{u},N_{\mathbbmss{v}},N_{\bar{\mathbbmss{v}}},N_{x})$ \hspace{0.4cm} & Energy \\
\hline
$(30,30,30,30)$ & $10.0566$\\ \\[-2mm]
$(50,30,30,30)$ & $10.0566$\\
$(100,30,30,30)$ & $10.0566$\\ \\[-2mm]
$(30,50,30,30)$ & $10.0616$\\
$(30,100,30,30)$ & $10.0616$\\ \\[-2mm]
$(30,30,50,30)$ & $10.1097$\\
$(30,30,100,30)$ & $10.1512$\\ \\[-2mm]
$(30,30,30,50)$ & $10.0566$\\
$(30,30,30,100)$ & $10.0566$\\
\hline
\end{tabular}
\caption{Evaluations of the glueball masses in unit of $\sqrt{\sigma}$ for different numbers of points for the quadratures. For $\ket{\Psi;A'_2;0;1^{\pm-}}$, energies are given for $(\mathbbmss{a}, \mathbbmss{b})=(0.35,1.8)$. For $\ket{\Psi;A'_2;0;1^{+-}}$, energies are given for $(\mathbbmss{a}, \mathbbmss{b})=(0.45,1.95)$. For $\ket{\Psi;A'_2;0;1^{--}}$, energies are given for $(\mathbbmss{a}, \mathbbmss{b})=(0.35,1.9)$. A cut-off at $j_{12}=20$ has been used. }\label{tab::3GB_spec_acc}
\end{table}


\begin{figure}
\centering
\includegraphics[scale=0.4]{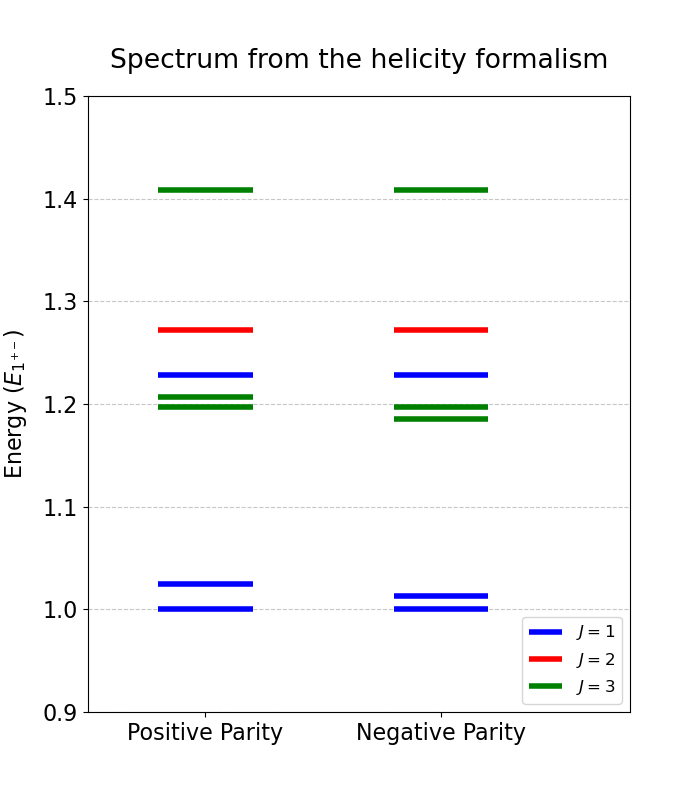}
\includegraphics[scale=0.4]{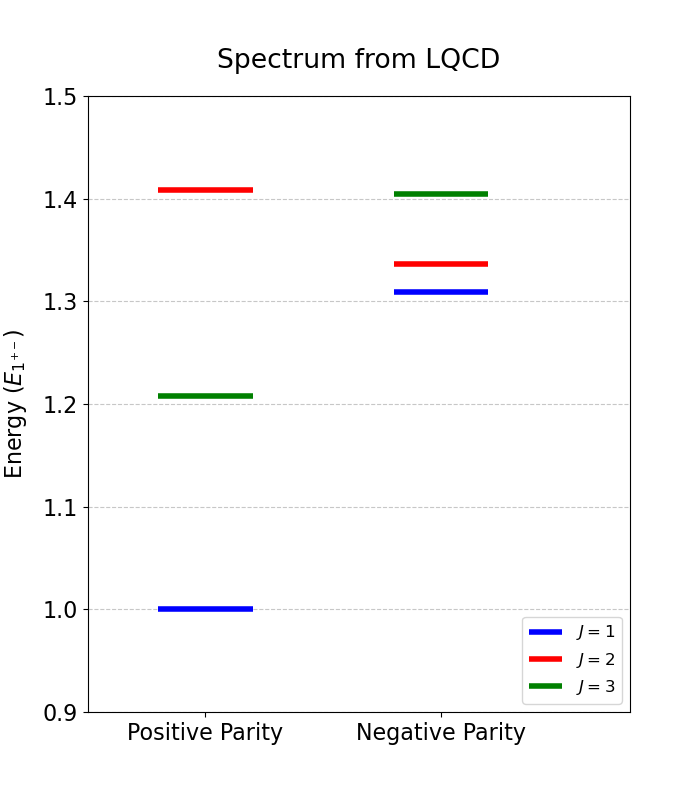}
\caption{Comparison of glueball spectra obtained using the helicity formalism (left, this work) and LQCD (right, references \cite{chen06,morn99}). In both cases, spectra are provided in unit of the lowest mass. Concerning calculations with the helicity formalism, a strong coupling constant $\alpha_s=0.450$ has been used.}
\label{fig::3GB_spec_energies}
\end{figure}

A concrete three-gluon glueball spectrum can now be computed. Again, $\alpha_s=0.450$ is used. Calculations are performed with a cut-off at $j_{12}=20$ for the summation and with $30$ points for each quadrature, except for $\bar{\mathbbmss{v}}$ where $100$ points are used. Full optimisations of the variational parameters $\mathbbmss{a}$ and $\mathbbmss{b}$ are performed. The states $\ket{\Psi_{a,b};A'_2;0;J^{\pm-}}$, $\ket{\Psi_{a,b};A''_2;0;J^{\pm-}}$ and
$\ket{\Psi_{a,b};A'_2;1;J^{\pm-}}$ are investigated up to $J=3$. However, states $\ket{\Psi_{a,b};A''_2;1;J^{\pm-}}$ are not considered, as they involve significantly more ME computations on unsymmetrical states, leading to a substantial increase in computational cost. The resulting three-gluon glueball masses, expressed in unit of $\sqrt{\sigma}$, are presented in Table \ref{tab::3GB_spec_energies}. For comparison, Figure \ref{fig::3GB_spec_energies} displays these masses alongside results from LQCD. In Figure \ref{fig::3GB_spec_energies}, masses are normalized to the lowest state, $1^{+-}$. Several observations support a relatively good agreement between the current calculation and LQCD results \cite{chen06,morn99}.
\begin{itemize}
\item The lowest $1^{+-}$ and $3^{+-}$ three-gluon glueballs obtained using the helicity formalism appear at the correct energy. However, the helicity formalism predicts a pair of states, whereas LQCD predicts only a single state \cite{chen06,morn99}.
\item The lowest $2^{+-}$ glueball is found with the same hierarchy in both approaches. However, its energy is $10\%$ lower using the helicity formalism. Unlike the lowest $1^{+-}$ and $3^{+-}$, the $2^{+-}$ does not exhibit a nearby secondary state, but this is an artifact resulting from the omission of $\ket{\Psi_{a,b};A''_2;1;J^{\pm-}}$ states in the analysis. The $10\%$ discrepancy may be explained by this omission : including $\ket{\Psi_{a,b};A''_2;1;2^{+-}}$ could slightly shift the $2^{+-}$ average energy. It may also stem from the simplicity of the Hamiltonian used, which does not account for spin interactions.
\item For negative parity, the first excited $1^{--}$ state predicted by the helicity formalism appears similar to the lowest $1^{--}$ observed in LQCD. As for $2^{+-}$, the relative energy is around $5\%$ too low. This discrepancy can likely be explained with the same arguments as for the $2^{+-}$ state.
\item Finally, a $3^{--}$ state analogous to the one observed in LQCD is also predicted. In this case, energies from the helicity formalism and LQCD are in good agreement.
\end{itemize}
On the other hand, the spectrum depicted by the helicity formalism does not fully align  with the LQCD results. In particular, the helicity formalism predicts states that are not observed in LQCD. Overall, the helicity spectrum appears to exhibit near- parity degeneracy, whereas the LQCD spectrum clearly separates into two distinct columns. 
\begin{itemize}
    \item Low-lying $1^{--}$ and $3^{--}$ three-gluon glueballs are predicted by the helicity formalism. These states appear, roughly speaking, at the same energy than their positive parity counterparts. According to \cite{chen06,morn99}, in LQCD, the lowest $1^{--}$ and $3^{--}$ lie well above the lowest $1^{+-}$ and $3^{+-}$.
    \item On the positive parity side, the helicity formalism predicts $1^{+-}$ and $3^{+-}$ states in the vicinity of the lowest $2^{+-}$ glueball. Again, such states are absent in references \cite{chen06,morn99}.
\end{itemize}
Several explanations can be proposed for this discrepancy. On the one hand, constituent approaches are known to sometimes overestimate the number of resonances. However, this effect typically impacts excited states rather than low-lying ones. On the other hand, it is possible that LQCD methods have missed a few states due to the challenges in interpreting the signal and in performing the continuum limit. For example, it seems that a more recent study of LQCD results observes an excited $1^{+-}$ glueball state whose energy is consistent with the one obtained using the helicity formalism \cite{athe20}.

\begin{figure}
    \centering
    \includegraphics[scale=0.42]{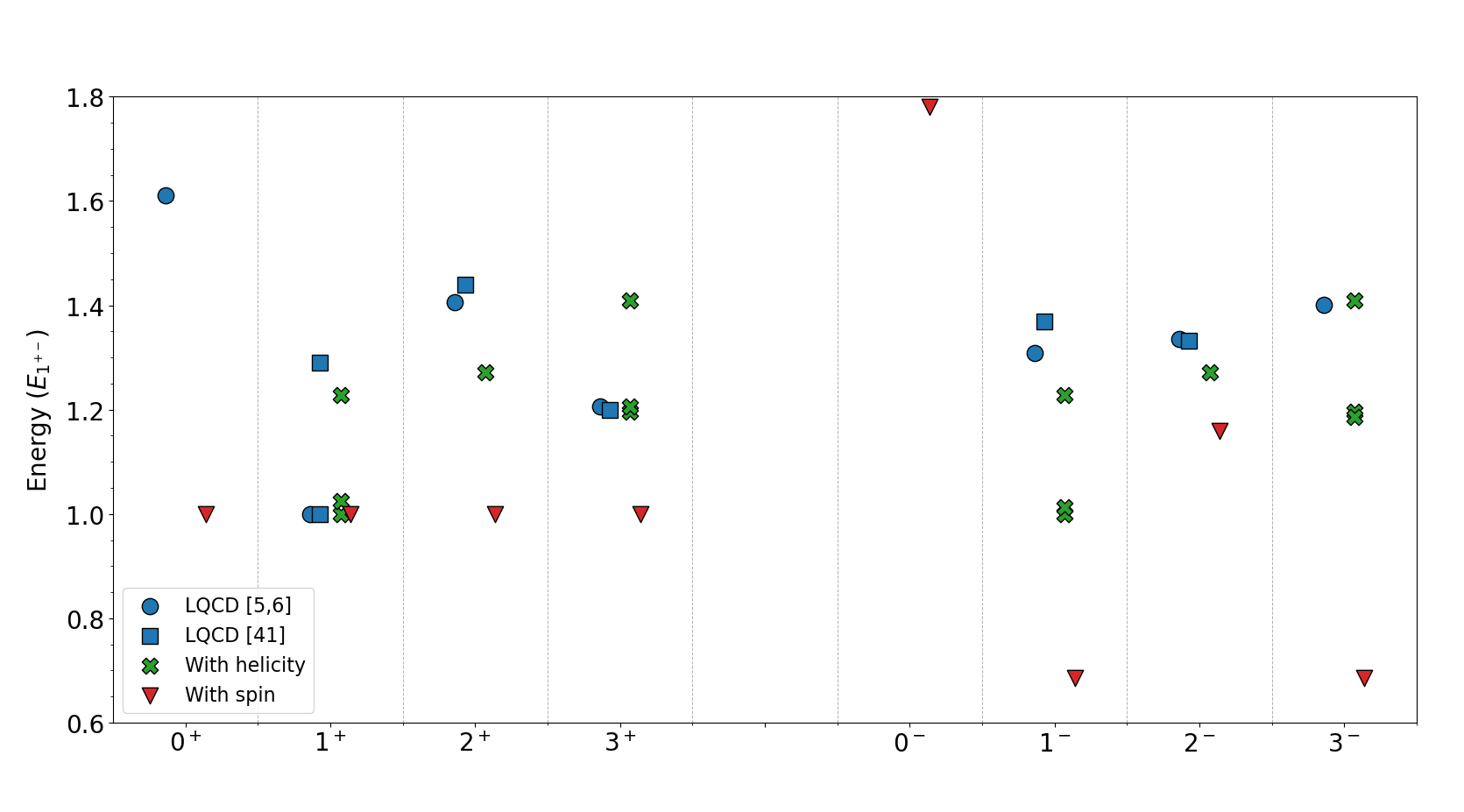}
    \caption{Comparison of low-lying negative charge conjugation glueball spectra computed using different approaches. Blue dots and squares represent the results from the LQCD calculations from references \cite{morn99,chen06} and \cite{athe20}, respectively. Red triangles correspond to results obtained by considering three spin-$1$ particles governed by the Hamiltonian \eqref{eq::3GB_ham_dimless}. Finally, green crosses are obtained by considering helicity degrees of freedoms for the constituent gluons. Masses are normalized to the $1^{+-}$ state.}
    \label{fig::3GB_hel_vs_spin}
\end{figure}

For two-gluon glueballs, which lies in the $C=+$ part of the spectrum, it is already know that helicity degrees of freedom are necessary to reproduce the lattice spectrum \cite{math08}. It would be interesting to compare the $C = -$ glueball spectrum obtained with helicity degrees of freedom in the present model to the $C = -$ spectrum obtained when longitudinal spin components for the gluon are included. Fortunately, such results have already been published \cite{llan06, math08b}. However, the Hamiltonian considered in these studies includes spin interactions. To ensure a fair comparison, eigenenergies from the dimensionless Hamiltonian \eqref{eq::3GB_ham_dimless} are computed considering three spin-$1$ particles and using oscillator bases expansions \cite{silv20,chev24}. Figure \ref{fig::3GB_hel_vs_spin} illustrates the comparison. Energies are again given in unit of the lowest $1^{+-}$ state. LQCD results from reference \cite{athe20} are also included. It is immediately apparent that the spectrum computed with spin degrees of freedom exhibits several flaws.
\begin{itemize}
    \item Positive parity states are all degenerate, whereas LQCD predicts a definite hierarchy for the lowest $J^{+-}$ glueball states. In particular, spin degrees of freedom predict a low-lying $0^{+-}$ glueball, which LQCD calculations place at a higher energy.
    \item On the negative parity side, the $1^{--}$ and $3^{--}$ spin eigenstates share the same energy, while the $2^{--}$ state lies significantly higher. In contrast, LQCD calculations predict nearly degenerated states with a soft ordering in ascending total angular momentum.
\end{itemize}
On the other hand, using helicity degrees of freedom better reproduces the overall spectrum, apart from the aforementioned inaccuracies and extra states. As with two-gluon glueballs \cite{math08}, it appears that constituent gluons should be treated as particles endowed with massless helicity quantum numbers.

\begin{figure}
\centering
\includegraphics[scale=0.4]{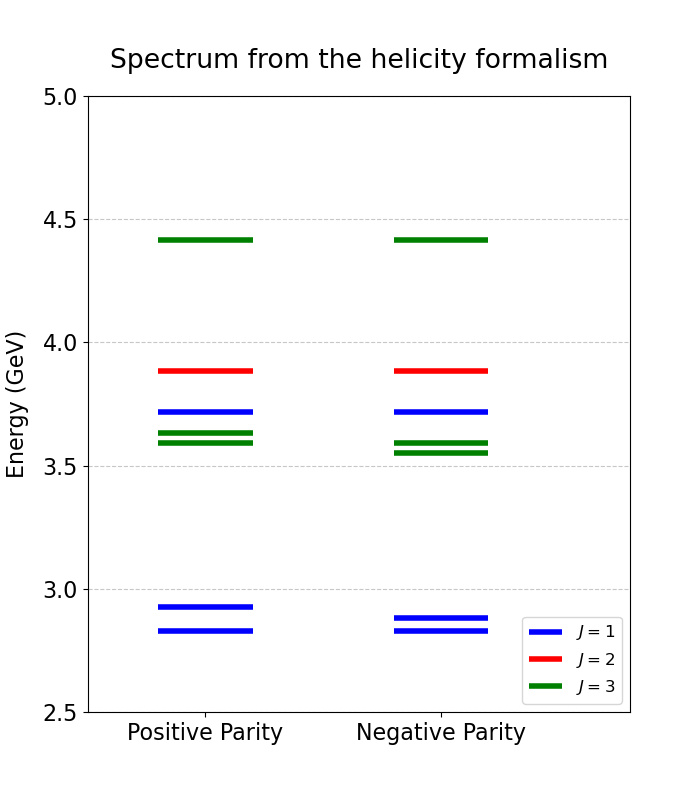}
\includegraphics[scale=0.4]{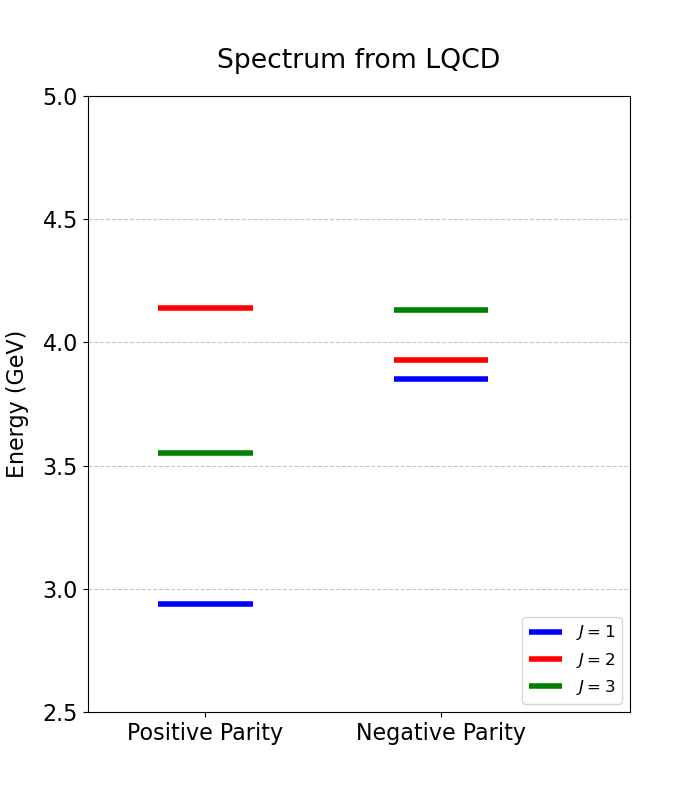}
\caption{Comparison of glueball spectra obtained using the helicity formalism with the modified potential \eqref{eq::3GB_spec_ham_pot_mod} (left, this work) and quenched lattice QCD (right, references \cite{chen06,morn99}). In both cases, spectra are provided in GeV.}
\label{fig::3GB_spec_energies_GeV}
\end{figure}

The previous analysis focused on the relative spectrum. To obtain a genuine energy spectrum requires to fix the mesonic string tension $\sigma$. The best agreement with LQCD results is achieved for a value of $\sigma = 0.086$GeV$^2$, which seems a factor 2 smaller compared to the values found in the literature (see for instance \cite{math08,theu01,gian19,caps86}). This issue can be addressed by reasonably modifying the Hamiltonian \eqref{eq::3GB_spec_ham}. In baryon studies, it is common to complement the potential with a constant term, particularly when dealing with light quarks \cite{caps86}.  Replacing the potential \eqref{eq::3GB_spec_ham_pot} with 
\begin{equation}
    V(r) = \sigma r - \frac{3\alpha_s}{2r}+C \label{eq::3GB_spec_ham_pot_mod}
\end{equation}
were $\sigma = 0.150$GeV$^2$, $\alpha_s = 0.450$ and $C = - 0.375$GeV results in the spectrum displayed in Figure \ref{fig::3GB_spec_energies_GeV}. The agreement between  absolute spectra is as good as that for relative spectra, and the parameters used are more consistent with values typically found in the literature. It is also important to notice that the methodology employed in this work inherently provides approximate results. This approximate character may also explain the discrepancies observed in both spectra.


\begin{table}
\centering
\begin{tabular}{l r r r r}
\hline \hline
State \hspace{1.75cm} & Opt. $\mathbbmss{b}$ & \hspace{0.5cm} Opt. $\mathbbmss{a}$ & \hspace{1cm} $E$ \\
\hline
$\ket{\Psi;A'_2;0;1^{\pm-}}$ & $1.80$ & $0.35$ & $10.020$ \\
$\ket{\Psi;A''_2;0;1^{+-}}$ & $1.95$ & $0.45$ & $10.269$ \\
$\ket{\Psi;A''_2;0;1^{--}}$ & $1.90$ & $0.35$ & $10.151$ \\[-0.2cm] \\
$\ket{\Psi;A'_2;0;3^{\pm-}}$ & $2.25$ & $0.55$ & $11.990$ \\
$\ket{\Psi;A''_2;0;3^{+-}}$ & $2.30$ & $0.55$ & $12.090$ \\
$\ket{\Psi;A''_2;0;3^{--}}$ & $2.25$ & $0.50$ & $11.881$ \\[-0.2cm] \\
$\ket{\Psi;A'_2;1;1^{\mp-}}$ & $2.40$ & $0.6$ & $12.305$ \\[-0.2cm] \\
$\ket{\Psi;A'_2;1;2^{\mp-}}$ & $2.40$ & $0.6$ & $12.742$ \\[-0.2cm] \\
$\ket{\Psi;A'_2;1;3^{\mp-}}$ & $2.65$ & $0.95$ & $14.113$\\
\hline\hline    
\end{tabular}
\caption{Display of the dimensionless three-gluon glueball spectrum obtained by using the three-body helicity formalism. Energies are given in unit of $\sqrt{\sigma}$. Only a single trial state is considered and the optimisation on variational parameters $\mathbbmss{a}$ and $\mathbbmss{b}$ is performed. Calculations are performed until $J=3$ and $|\mu|=1$. Only one of the two states with $|\mu|=1$ is considered (namely, $A'_0$). All integrals are computed with $30$ point, except for those on $\bar v$ which use $100$ points. The sum in $j_{12}$ is truncated at $j_{12}=20$.}
\label{tab::3GB_spec_energies}
\end{table}


\section{Conclusion}
\label{sec:conclu}

In the framework of constituent approaches, the current work exploited the two- and three-body helicity formalism to acquire two- and three-gluon glueball spectra. It required to sum up different properties from the existing literature as well as to complement them with innovative results. A definite methodology able to infer a spectrum for helicity states and based on the variational theorem has been set up for both two- and three-body systems. 

Concerning two-gluon glueballs, the spectrum obtained using the helicity formalism is compatible with LQCD calculations. This agreement gives credit in the methodology implemented and motivates the generalisation to three-gluon glueballs. Although similar calculations were already present in the literature \cite{math08,szcz03,szcz96}, the current work proposed a new perspective on this spectrum.

The situation concerning three-gluon glueballs is less conclusive. In summary, in units of the meson string tension, the helicity formalism proves able to reproduce the spectrum from LQCD with a satisfying accuracy but predicts supplementary states whose origin remains elusive. The comparison with calculations involving three spin-$1$ particles favored the use of helicity degrees of freedom. When comparing spectra in physical units, a mesonic string tension too low by a factor of $2$ compared to its standard value is to be used. This flaw can be overcome by adding a constant term in the potential in agreement with the potential used in baryons \cite{caps86}. Although it allows to use parameter values in their physical range, it slightly deteriorates the agreement with LQCD. In addition, this parameter was not necessary to describe two-gluon glueballs.

Additional research may help unveiling the origin of these discrepancies. Improvements about the accuracy of the resolution method in the case of three-gluon glueballs may give credit to the spectrum obtained using the helicity formalism. Including more trial states in the resolution to test the convergence, truly implementing the possibility of mixing between the different symmetrical states, increasing the number of points in the quadrature or increasing the cutt-off in $j_{12}$ are all trails to reach this goal. Hamiltonian \eqref{eq::3GB_spec_ham} may also be refined and complemented, for instance, with spin-based interactions. This may help to decide whether the extra states predicted by the current methodology are artifacts of the resolution method or not. 

Setting aside the question of the extra states, the solutions obtained using the helicity formalism may also yield additional insights into glueball properties. For instance, effort could be directed toward evaluating observables beyond the energy, such as decay rates.


\section*{Acknowledgements} 

C. C. would like to thank the Fonds de la Recherche Scientifique - FNRS for the financial support. V.M. is a Serra Húnter fellow and acknowledges support from the Spanish national Grants PID2023-147112NB-C21 and CNS2022-136085. The authors thank C. Semay and F. Buisseret for its advice and a careful reading of the manuscript.


\appendix

\renewcommand{\thesection}{Appendix \Alph{section}}

\renewcommand{\theequation}{\Alph{section}.\arabic{equation}}

\section{Helicity states for massless particles}
\label{app:massless}

In Section \ref{sec:1BS2BS}, massive particle states were discussed. For massless ones, the Pauli-Lubanski Casimir is identically zero while helicity proves to be Lorentz invariant. As a result, a general massless states $\ket{\phi;0;\lambda}$ is by definition an eigenstate of the squared-momentum operator with a null eigenvalue and of the helicity operator,
\begin{align}
P^2\ket{\psi;0;\lambda}=0\ket{\psi;0;\lambda}, && \Lambda \ket{\psi;0;\lambda} = \lambda \ket{\psi;0;\lambda}.
\end{align}
If helicity is truly Lorentz invariant, it is nevertheless reversed by a parity transformation, meaning that massless states always come by pairs, one with helicity $\lambda$ and the other one with helicity $-\lambda$. This feature justifies the aforementioned simplification of reality: although spin is not formally defined in the massless case, these two states with same nature but opposed helicities are often considered as the two facets of a spin $\lambda$ particle whose intermediary projections are forbidden. This is the case when someone consider left-handed and right-handed photons as a unique spin $1$ particle whose $0$ projection is forbidden. 

To describe the actual state of a massless particle, complete orthonormal sets are again necessary. Since massless particles have no rest frame, the corresponding states are defined as eigenstates of the helicity operator, and only the $p$-helicity states are suitable to describe them. These are eigenstates of the exact same operators than massive ones but with zero as eigenvalue for both  Casimir operators,
\begin{align}
    &P^2 \ket{0;p\theta\phi;s\lambda} = 0 \ket{0;p\theta\phi;s\lambda}, &&W^2 \ket{0;p\theta\phi;s\lambda} = 0 \ket{0;p\theta\phi;s\lambda}.\label{eq::1BS_phelMS_massless}
\end{align}
Above, $s$ rather specifies which values $\lambda$ can take than is a true quantum number. Massless $p$-helicity states significantly differ from massive ones in the sense that, because massless particles cannot be taken at rest, relation \eqref{eq::1BS_phelicity} cannot remain valid for these. A similar relation can nevertheless be obtained by turning the rest state into another reference state having a more suitable four-momentum for massless particles. Choosing this reference four-momentum as $k=(1,0,0,1)$ and denoting the associated reference state $\ket{0;s\lambda}$, the adapted relation reads
\begin{equation}
\begin{aligned}
    \ket{0;p\theta\phi;s\lambda} &= U(L_h(0,p,\theta,\phi))\ket{0;s\lambda} \\
    &= U(R(\phi,\theta,0) L_z(0,p))\ket{0;s\lambda}.
\end{aligned} \label{eq::1BS_phelicity_massless}
\end{equation}
Above, $L_z(0,p)$ refers to a boost that provides a momentum of modulus $p$ along the $z$-axis to a massless particle having initially the dimensionless reference four-momentum $k$. Notice that, because $\ket{0;s\lambda}$ has a unit spatial momentum along the $z$ axis, this state proves to be eigenstate of both $W_3$ and $\Lambda$, these two operators being equal in the current circumstances. 

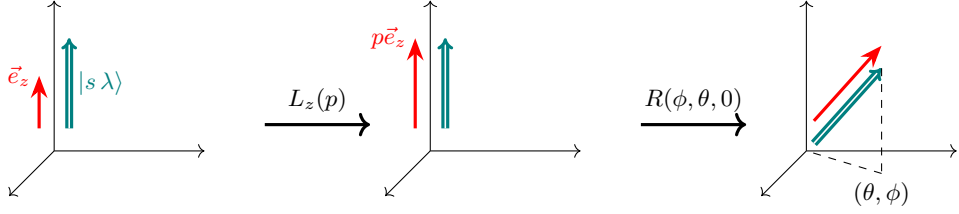
\begin{figure}
\begin{center}
\begin{tikzpicture}
\draw[->] (0,0) - - (0,2);
\draw[->] (0,0) - - (2,0);
\draw[->] (0,0) - - (-0.6,-0.6);
\draw[-{Implies},double,very thick,teal] (0.2,0.3) - - (0.2,1.5);
\node[right, teal](s) at (0.2,0.9) {$\ket{s\,\lambda}$};
\draw[-{Stealth},very thick,red] (-0.2,0.3) - - (-0.2,1);
\node[left, red](s) at (-0.2,1) {$\vec{e}_z$};
\draw[->,very thick] (2.8,0.35) - - (4.2,0.35);
\node[above](L) at (3.5,0.4) {$L_z(p)$};

\draw[->] (5,0) - - (5,2);
\draw[->] (5,0) - - (7,0);
\draw[->] (5,0) - - (4.4,-0.6);
\draw[-{Implies},double,very thick,teal] (5.2,0.3) - - (5.2,1.5);
\draw[-{Stealth},very thick,red] (4.8,0.3) - - (4.8,1.5);
\node[left, red](s) at (4.8,1.5) {$p\vec{e}_z$};
\draw[->,very thick] (7.8,0.35) - - (9.2,0.35);
\node[above](R) at (8.5,0.4) {$R(\phi,\theta,0)$};

\draw[->] (10,0) - - (10,2);
\draw[->] (10,0) - - (12,0);
\draw[->] (10,0) - - (9.4,-0.6);
\draw[-{Stealth},very thick,red] (10.1,0.4) - - (11,1.4);
\draw[-{Implies},double,very thick,teal] (10.1,0.1) - - (11,1.1);
\draw[dashed] (11,1.1) - - (11,-0.3);
\draw[dashed] (11,-0.3) - - (10,0);
\node[below](angles) at (11,-0.3) {$(\theta,\phi)$};
\end{tikzpicture}
\caption{Graphic illustration for the definition \eqref{eq::1BS_phelicity_massless} of massless one-body $p$-helicity states.\label{fig::1BS_phelicity_massless}}
\end{center}
\end{figure}

Because of this adaptation, the graphic interpretation of massive $p$-helicity states must slightly be adapted, the initial state having a non-zero spatial momentum. The new diagram is shown in Fig.~\ref{fig::1BS_phelicity_massless}. Apart from this modification in the initial state, both diagrams are to be understood similarly.

Let us now focus on the adaptation of properties \eqref{eq::1BS_parityProp} and \eqref{eq::1BS_canProp} to massless particles. First, the definition of intrinsic parity \eqref{eq::1BS_parityProp} making use of rest states, it must be adapted to a massless reference state. Applying parity on this state reverses its non-zero spatial momentum and, as already mentioned, opposes the helicity quantum number. It results in the following relation for the action of parity on a reference state,
\begin{equation}
    \Pi\ket{0;s\lambda} = \eta\, U(R(0,\pi,0))\ket{0;s-\lambda}
\end{equation}
where $\eta$ is defined as the intrinsic parity of the massless particle. This modification considered, one can roughly follow the same calculation steps than in the massive case to demonstrate that
\begin{equation}
    \Pi\ket{0;p\theta\phi;s\lambda} = \eta(-1)^{\lambda} \ket{0;p(\pi-\theta)(\pi+\phi);s-\lambda}. \label{eq::1BS_parity_massless}
\end{equation}
Formula \eqref{eq::1BS_parity_massless} replaces \eqref{eq::1BS_parity} in presence of massless particles. The discussion of relation \eqref{eq::1BS_canProp} must also partially be revised. In the massive case, the Wigner rotations $R_W$ defined in relation \eqref{eq::1BS_wignRot} is shown to belong to $SO(3)$, namely the massive little group. Setting $m=0$, the combination \eqref{eq::1BS_wignRot} turns out to no longer belong to the massive little group but to the massless one, $ISO(2)$. Therefore, the transformation is no longer parameterised by Euler angles but by three $ISO(2)$ parameters, such as the $\alpha$, $\beta$ and $\theta$ parameters suggested in reference \cite{wein95}, and the action of $R_W$ on physical states is no longer provided by $SU(2)$ Wigner $D$ matrices but instead reduces to \cite{lind03}
\begin{equation}
    D^s_{\lambda'\lambda}(\alpha_W,\beta_W,\theta_W)= e^{i\theta_W\lambda}\delta_{\lambda'\lambda}. \label{eq::1BS_D_ISO2}
\end{equation}
The $\delta_{\lambda'\lambda}$ factor ensures that, as expected for massless states, the helicity quantum number remains invariant while boosting the state. This replacement done, relation \eqref{eq::1BS_canProp} can be freely used even while dealing with massless particles. Although $R_W$ is no longer a true spatial rotation, this transformation is still called Wigner rotation through misuse of language.

As already mentioned in the main text, helicity states for two massless bodies behave quite similarly to those of massive ones. The demonstrations are very similar to their massive counterparts, with the main differences generally arising from the phase factor in definition \eqref{eq::2BS_pdef}, which must be adapted by replacing $s_2$ with $|\lambda_2|$. The one-body parity transformation rule \eqref{eq::1BS_parity} must also be replaced by the massless version \eqref{eq::1BS_parity_massless}. The rest of the demonstrations remain analogous to the massive case, taking advantage of the fact that the reference helicity states for the massless particles, $\ket{0;s\lambda}$, are also $J_3$ eigenstates.


\section{Examples of two-body calculations in momentum space}
\label{app:exem_momspace}

To simplify, evaluations will be performed on pure two-body canonical states. Three different two-body systems will be considered,
\begin{align}
    \mathcal{H}_{\text{Coul}}(p,r)&= \frac{p^2}{2\mu} - \frac{g}{r}, \label{eq::app_exmom_coul}\\
    \mathcal{H}_{\text{lin}}(p,r)&= \frac{p^2}{2\mu} + \lambda r,\label{eq::app_exmom_lin}\\
    \mathcal{H}_{\text{Ful}}(p,r)&=  \sqrt{m_1^2+p^2} + \sqrt{m_2^2+p^2} + A r - \frac{\kappa}{r},
\end{align}
with $\mu = m_1m_2/(m_1+m_2)$ denoting the reduced mass. For convenience, in all cases, both particles are chosen with equal non-zero masses, $m_1=m_2=m$ and $\mu = m/2$. Natural units are used. Hamiltonian ME will be computed for a Gaussian-like $\Xi$ helicity wave function with one parameter,
\begin{align}
    &\Xi_a(p)=2\left(\frac{(2a)^3}{\pi}\right)^{1/4}\,p\,e^{-ap^2}.
\end{align}
Above, the constant has been fixed to ensure that the condition \eqref{eq::2BS_norm_xi} holds. The variational theorem ensures that, for any angular momentum $\ell$ and for all value of the parameter $a$, this ME consist of an upper-bound for the ground state energy with angular momentum $\ell$ of the corresponding Hamiltonian. This approximated eigenenergy is the one referred as SGA in Section \ref{sec:2GB}. 

Concerning kinetic energies, the integral from relation \eqref{eq::2BS_kinEnergy} can be solved analytically for both non- and semi-relativistic kinematics,
\begin{align}
    |A|^2 \int \diff p\, p^2e^{-2ap^2} \frac{p^2}{2\mu} &= \frac{3}{8}\frac{1}{a \mu},\\
    |A|^2 \int \diff p\, p^2e^{-2ap^2} 2\sqrt{m^2+p^2} &= \sqrt{\frac{8a}{\pi}}\,m^2 e^{a m^2} K_1\left(a m^2\right).
\end{align}
Above $K_n$ refers to a modified Bessel function of second kind. Potential ME are managed using relations \eqref{eq::2BS_coulomb} and \eqref{eq::2BS_HersFormula_uv} from previous sections. As already mentioned, numerical integration methods based on Gauss-Legendre quadrature are used.\footnote{For both potentials, to compute ME using \eqref{eq::2BS_coulomb_uv} and \eqref{eq::2BS_HersFormula_uv}, Gauss-Legendre quadratures with $1000$ points are used for integrals on $\bar v$ while integrals on $v$ are handled with a change of variables $v=v_t/(1-v_t)$ and only $300$ points. Integrals on $v$ turn out to be clearly less sensitive than the ones on $\bar v$.} Results for the potential and the kinetic energy are then gathered to compute Hamiltonian matrix-elements and the obtained upper bounds are minimised on the variational parameter $a$. With a single trial state, this method only provides approximations for the lowest state of a given angular momentum $\ell$.

Let us now focus on the results of the three tests. The Table~\ref{tab::2BS_CoulombPot} displays the upper bounds obtained for the Hamiltonian $\mathcal{H}_{\text{Coul}}$. For this test, the mass $m$ of the particles has been fixed to $2$ while the constant of the Coulomb potential $g$ has been fixed to $1$, both in arbitrary units. One will notice that the obtained upper bounds are far from accurate, especially for $\ell=3$. This is a regular feature when using a Gaussian trial wave function to solve a divergent potential. One can check that turning the trial wave function into the exact one for the $\ell=0$ ground state \cite{flug94},
\begin{equation}
    \Xi^{\text{Coul}}(p) \propto \frac{p}{(1+p^2)^2},
\end{equation}
allows to exactly reproduce its eigenenergy, as expected.
\begin{table}
    \centering
    \begin{tabular}{l r r r}
    \hline
    $\ell$\hspace{2mm} & \hspace{2mm} $E_{\text{exact}}$ & \hspace{2mm} $E_{\text{SGA}}$ & \hspace{2mm} $a_{\text{opt}}$\\
    \hline \hline
    $0$ & $-0.50$ & $-0.42$ & $0.89$ \\
    $1$ & $-0.25$ & $-0.12$ & $3.16$ \\
    $2$ & $-0.125$ & $-0.05$ & $6.99$ \\
    \hline
    \end{tabular}
    \caption{Spectra comparison for the Coulomb Hamiltonian \eqref{eq::app_exmom_coul}. Upper bounds obtained with the SGA, ($E_{\text{SGA}}$), are compared to the analytical spectrum from \cite{yane94}, $E_{\text{exact}}$, for various angular momenta $\ell$. Optimised values of the variational parameter, $a_{\text{opt}}$ are also displayed in the fourth column. Arbitrary units are used: $m = 2$ and $g = 1$.}
    \label{tab::2BS_CoulombPot}
\end{table}
Concerning the pure linear Hamlitonian $\mathcal{H}_{\text{lin}}$, results in arbitrary units are displayed in Table~\ref{tab::2BS_LinPot} for unit masses and linear potential constant. An analytical spectrum for $\ell=0$ is provided in \cite{luch91}. For $\ell>0$, the SGA is compared to the spectrum provided by another approximation method, the Lagrange-mesh method \cite{sema01}. One can see that a single Gaussian provides a quite accurate upper bound for the $\ell=0$ ground state. For $\ell>0$, as expected, the very accurate results provided by the Lagrange-mesh method lies below those obtained with the SGA.
\begin{table}
    \centering
    \begin{tabular}{l r r r r}
    \hline
    $\ell$\hspace{2mm} & \hspace{2mm} $E_{\text{exact}}$ & \hspace{2mm} $E_{\text{LM}}$ & \hspace{2mm} $E_{\text{SGA}}$ & \hspace{2mm} $a_{\text{opt}}$\\
    \hline \hline
    $0$ & $2.338$ & $2.338$ & $2.345$ & $0.960$ \\
    $1$ & N.A. & $3.361$ & $3.472$ & $0.648$ \\
    $2$ & N.A. & $4.248$ & $4.556$ & $0.494$ \\
    \hline
    \end{tabular}
    \caption{Spectra comparison for the linear Hamiltonian \eqref{eq::app_exmom_lin}. Upper bounds obtained with the SGA, ($E_{\text{SGA}}$), are compared to the analytical spectrum from \cite{luch91} for $\ell=0$ ($E_{\text{exact}}$) and to the spectrum from the Lagrange mesh method \cite{sema01} for various angular momenta, ($E_{\text{LM}}$). Optimised values of the variational parameter, $a_{\text{opt}}$ are also displayed in the fourth column. Arbitrary units are used : $m=1$ and $\lambda=1$.}
    \label{tab::2BS_LinPot}
\end{table}
Finally, the Hamiltonian $\mathcal{H_{\text{Ful}}}$ is investigated. This Hamiltonian has been used in \cite{fulc94} to model bottomonium and charmonium systems. Table~\ref{tab::2BS_FulcherPot} compares the results obtained with the SGA to the spectrum provided in the reference. In both cases and for each $\ell$ values, the single Gaussian upper-bound achieves an accuracy below $2$\%.
\begin{table}
    \hfill
    \begin{tabular}{l r r r r}
    \hline
    $\ell$\hspace{2mm} & \hspace{4mm} $E_{\text{exact}}$ & \hspace{4mm} $E_{\text{SGA}}$ & \hspace{5mm} [$\varepsilon$] & \hspace{4mm} $a_{\text{opt}}$\\
    \hline \hline
    $0$ & $9.448$ & $9.499$ & [$0.5\%$] & $0.391$ \\
        $1$ & $9.900$ & $9.920$ & [$0.2\%$] & $0.450$ \\
    $2$ & $10.150$ & $10.204$ & [$0.5\%$] & $0.397$ \\
    \hline
    \end{tabular}\hfill
    \begin{tabular}{l r r r r}
    \hline
    $\ell$\hspace{2mm} & \hspace{4mm} $E_{\text{exact}}$ & \hspace{4mm} $E_{\text{SGA}}$ & \hspace{5mm} [$\varepsilon$] & \hspace{4mm} $a_{\text{opt}}$\\
    \hline \hline
    $0$ & $3.067$ & $3.094$ &[$0.9\%$]& $1.383$ \\
    $1$ & $3.504$ & $3.531$ &[$0.8\%$] & $1.141$ \\
    $2$ & $3.811$ & $3.886$ &[$2\%$] & $0.903$ \\
    \hline
    \end{tabular} \hfill \hfill
    
    \caption{Spectra comparison for Fulcher's Hamiltonian $\mathcal{H_{\text{Ful}}}$ \cite{fulc94} (bottomonium at left, charmonium at right). Upper bounds obtained with the SGA, ($E_{\text{SGA}}$), are compared to the spectrum from \cite{luch91} for various angular momenta $\ell$. Optimised values of the variational parameter, $a_{\text{opt}}$ are also displayed in the fourth column. Energy results and  $a$ values are respectively provided in GeV and GeV$^{-1}$. Relatives errors $\varepsilon$ are indicated in square brackets.}
    \label{tab::2BS_FulcherPot}
\end{table}


\section{Properties of Berman's States}
\label{app:prop_berm}

This Appendix is devoted to the demonstration of some properties about Berman's states missing in the literature. Let us start with property \eqref{eq::BD_changeRefState}, which is about the modification of the reference plane in the definition of Berman's $J$ helicity states. For the sake of conciseness, in the following, $R(\alpha,\beta,\gamma)$ will be denoted $R$ and $\diff\alpha\diff\!\cos\!\beta\diff\gamma$ will be denoted $\diff R$. First, an arbitrary rotation $\bar R$ is artificially introduced in \eqref{eq::BD_Jdef}.
The $\bar R$ rotation is applied on the reference state $\ket{w_1 w_2 w_3 ; \lambda_1 \lambda_2 \lambda_3}$ while its inverse is composed with the rotation $R$,
\begin{equation}
\begin{aligned}
\ket{JM\mu ; w_1 w_2 w_3; \lambda_1 \lambda_2 \lambda_3} = \sqrt{\frac{2J+1}{8\pi^2}} \int \diff R\, D^{J*}_{M\mu}(R)\,U\left(R\bar{R}^{-1}\right) \left(U(\bar{R}) \ket{w_1 w_2 w_3 ; \lambda_1 \lambda_2 \lambda_3}\right).
\end{aligned}
\end{equation}
Turning the integration on $R$ Euler angles into an integration on $R\bar R^{-1}$ Euler angles, one gets
\begin{equation}
\begin{aligned}
\ket{JM\mu ; w_1 w_2 w_3; \lambda_1 \lambda_2 \lambda_3}
= \sum_{\mu'=-J}^J D^{J*}_{\mu'\mu}(\bar R)\sqrt{\frac{2J+1}{8\pi^2}} \int \diff (R\bar R^{-1})\,  D^{J*}_{M\mu'}(R\bar R^{-1}) U\left(R\bar{R}^{-1}\right)&\\
\left(U(\bar{R}) \ket{w_1 w_2 w_3 ; \lambda_1 \lambda_2 \lambda_3}\right)&.
\end{aligned}
\end{equation}
In the right-hand side of this equation, one recognize the definition of a Berman's $J$ helicity state whose reference state would have been tilted with the rotation $\bar R$. Denoting this state $\ket{JM\mu'; w_1 w_2 w_3; \lambda_1 \lambda_2 \lambda_3}_{\bar R}$, one finally gets formula \eqref{eq::BD_changeRefState}, 
\begin{equation}
\begin{aligned}
\ket{JM\mu ; w_1 w_2 w_3; \lambda_1 \lambda_2 \lambda_3} = \sum_{\mu'=-J}^J D^{J}_{\mu\mu'}(\bar R^{-1}) \ket{JM\mu'; w_1 w_2 w_3; \lambda_1 \lambda_2 \lambda_3}_{\bar R}.\label{eq::Aconv_final}
\end{aligned}
\end{equation}
This property can be used to relate different conventions for Berman's states. In reference \cite{berm65}, Berman's reference states are defined considering the momentum of the third particle as opposite to the $x$ axis in \cite{berm65} while it is opposite to the $y$ axis in the current work. In light of the previous developments, these two definitions for Berman's $J$-helicity states have a chance to differ each-other. The rotation $\bar R$ that conveys from one convention to the other one brings the $y$ axis along the $x$ one, meaning
\begin{equation}
    \bar R = R(0,0,-\pi/2).
\end{equation}
Applying the aforementioned relation to this case, one gets
\begin{equation}
\begin{aligned}
\ket{JM\mu ; w_1 w_2 w_3; \lambda_1 \lambda_2 \lambda_3} &= \sum_{\mu'=-J}^J D^{J}_{\mu\mu'}(0,0,\pi/2) \ket{JM\mu'; w_1 w_2 w_3; \lambda_1 \lambda_2 \lambda_3}_{\text{\cite{berm65}}}\\
& = i^{-\mu} \ket{JM\mu; w_1 w_2 w_3; \lambda_1 \lambda_2 \lambda_3}_{\text{\cite{berm65}}},
\end{aligned}
\end{equation}
where the notation $\ket{...}_{\text{\cite{berm65}}}$ refers to an helicity state that uses conventions from \cite{berm65}. Although both states have the same physical meaning, they differ by their phase conventions.

\begin{figure}
\begin{center}
\begin{tikzpicture}
\tikzset{
    partial ellipse/.style args={#1:#2:#3}{
        insert path={+ (#1:#3) arc (#1:#2:#3)}}}

\draw[->] (0,0) - - (0,2);
\node[left](z) at (0,2) {$y$};
\draw[->] (0,0) - - (2,0);
\node[right](y) at (2,0) {$x$};

\draw[-{Stealth},very thick,atomictangerine] (0,0) - - (0,-1.5);
\node[below, atomictangerine](s) at (0,-1.5) {$\vec{p}_3$};

\draw[-{Stealth},very thick,red] (0,0) - - (1.3,0.75);
\node[above right, red](s) at (1.3,0.75) {$\vec{p}_1$};

\draw[-{Stealth},very thick,burntorange] (0,0) - - (-1.3,0.75);
\node[above left, burntorange](s) at (-1.3,0.75) {$\vec{p}_2$};

\filldraw[white] (0,0) circle (0.13);
\draw (0,0) circle (0.13);
\filldraw (0,0) circle (0.04);
\node[below left](s) at (0,0) {$z$};

\draw[-{Stealth},very thick] (2.6,0) - - (4.6,0);
\node[above](s) at (3.5,0) {\small$R(0,-\pi,0)$};

\draw[->] (6,0) - - (6,2);
\draw[->] (6,0) - - (8,0);

\draw[-{Stealth},very thick,atomictangerine] (6,0) - - (6,-1.5);

\draw[-{Stealth},very thick,burntorange] (6,0) - - (7.3,0.75);

\draw[-{Stealth},very thick,red] (6,0) - - (4.7,0.75);

\draw[-{Straight Barb[length=0.5mm,width=0.8mm]},thick] (6,0.65) [partial ellipse=0:340:0.4cm and 0.2cm] ;

\filldraw[white] (6,0) circle (0.13);
\draw (6,0) circle (0.13);
\filldraw (6,0) circle (0.04);

\draw[-{Stealth},very thick] (8.6,0) - - (10.6,0);
\node[above](s) at (9.5,0) {\small$R(\varphi_{13},0,0)$};

\draw[->] (12,0) - - (12,2);
\draw[->] (12,0) - - (14,0);

\draw[-{Stealth},very thick,red] (12,0) - - (12,-1.5);
\node[below, red](s) at (12,-1.5) {$\vec{p}_1$};

\draw[-{Stealth},very thick,atomictangerine] (12,0) - - (13.3,0.75);
\node[above right, atomictangerine](s) at (13.3,0.75) {$\vec{p}_3$};

\draw[-{Stealth},very thick,burntorange] (12,0) - - (10.7,0.75);
\node[above left, burntorange](s) at (10.7,0.75) {$\vec{p}_2$};

\draw[-{Straight Barb[length=0.5mm,width=0.8mm]},thick] (12,0) [partial ellipse=40:140:0.7cm and 0.7cm] ;
\draw[-{Straight Barb[length=0.5mm,width=0.8mm]},thick] (12,0) [partial ellipse=160:260:0.7cm and 0.7cm] ;
\draw[-{Straight Barb[length=0.5mm,width=0.8mm]},thick] (12,0) [partial ellipse=280:380:0.7cm and 0.7cm] ;

\filldraw[white] (12,0) circle (0.13);
\draw (12,0) circle (0.13);
\filldraw (12,0) circle (0.04);

\end{tikzpicture}
\end{center}
\caption{Diagram justifying the role of $R(\varphi_{13},\pi/2,0)$ in the application of $\mathbb{P}_{13}$ on Berman's states.}
\label{AP13_fig1}
\end{figure}
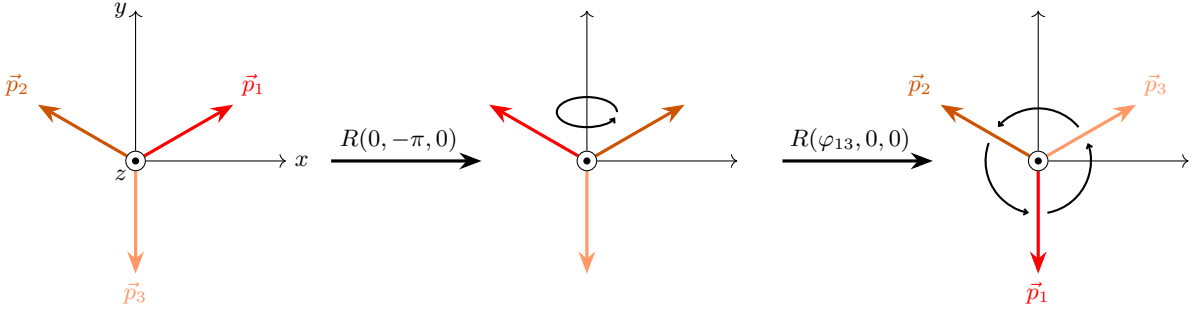

The action of $\mathbb{P}_{13}$ on Berman's $J$-helicity states also deserves a few explanations. The interest of symmetry being limited to the study of identical particles, spins and masses of the three particles will be supposed equal, $s_1=s_2=s_3=s$ and $m_1=m_2=m_3=m$. Calculations have first to be performed at the $p$-helicity states level. The permutation operator exchanges the states of particle $1$ and $3$,
\begin{align}
\mathbb{P}_{13}\ket{\alpha\beta\gamma;w_1w_2w_3;\lambda_1\lambda_2\lambda_3}= U(R(\alpha,\beta,\gamma))\big(&U(R(\phi_3,\pi/2,0)L_z(p_3))\ket{s\lambda_3} \nonumber \\
\otimes\ &U(R(\phi_2,\pi/2,0)L_z(p_2))\ket{s\lambda_2}  \label{AP13_init}\\
\otimes\ &U(R(\phi_1,\pi/2,0)L_z(p_1))\ket{s\lambda_1} \nonumber
\big).
\end{align}
It is not clear if the left-hand side of this equation fits with the structure of a Berman's $p$-helicity states because for now relations \eqref{eq::BD_pdef2_1} and \eqref{eq::BD_pdef2_2} are not verified. This difficulty can be overcame by the insertion of a rotation $R(\varphi_{13},-\pi,0)$. The angles are chosen to mimic the action of $\mathbb{P}_{13}$ on the momenta, as illustrated on Fig.~\ref{AP13_fig1}. Algebraically, one gets,
\begin{equation}
\begin{aligned}
\mathbb{P}_{13}\ket{\alpha\beta\gamma;w_1w_2w_3;\lambda_1\lambda_2\lambda_3}= U(R(\alpha,\beta,\gamma)R(\varphi_{13},-\pi,0)) \big(U(R(0,\pi,-\varphi_{13})R(\phi_3,\pi/2,0)L_z(p_3))\ket{s\lambda_3} &\\ 
\otimes \ U(R(0,\pi,-\varphi_{13})R(\phi_2,\pi/2,0)L_z(p_2))\ket{s\lambda_2} &\\
\otimes \ U(R(0,\pi,-\varphi_{13})R(\phi_1,\pi/2,0)L_z(p_1))\ket{s\lambda_1} & \big). \\
\end{aligned}   
\end{equation}
The $SO(3)$ multiplication law can be used to reduce the previous expression
\begin{align}
  R(\alpha,\beta,\gamma)R(\varphi_{13},-\pi,0)&=R(\pi+\alpha,\pi-\beta,-(\pi+\gamma+\varphi_{13})),\\
   R(0,\pi,-\varphi_{13})R(\phi,\pi/2,0)&=R(\pi+\varphi_{13}-\phi,\pi/2,\pi)\hspace{4mm} (\forall\, \phi\in \mathbb{R}).
\end{align}
With these relations the action of $\mathbb{P}_{13}$ on $\ket{\alpha\beta\gamma;Ww_1w_2;\lambda_1\lambda_2\lambda_3}$ becomes,
\begin{equation}
\begin{aligned}
\mathbb{P}_{13}\ket{\alpha\beta\gamma;w_1w_2w_3;\lambda_1\lambda_2\lambda_3}= U(R(\pi+\alpha,\pi-\beta,-(\pi+\gamma+\varphi_{13})) \big(U(R(\pi+\varphi_{13}-\phi_3,\pi/2,\pi)L_z(p_3))\ket{s\lambda_3}& \\ 
\otimes \ U(R(\pi+\varphi_{13}-\phi_2,\pi/2,\pi)L_z(p_2))\ket{s\lambda_2}& \\
\otimes \ U(R(\pi+\varphi_{13}-\phi_1,\pi/2,\pi)L_z(p_1))\ket{s\lambda_1}& \big). \\
\end{aligned}   
\end{equation}
This relation can even further be simplified using expression \eqref{eq::BD_pdef2_2} for $\phi_1$, $\phi_2$ and $\phi_3$,
\begin{equation}
\begin{aligned}
\mathbb{P}_{13}\ket{\alpha\beta\gamma;w_1w_2w_3;\lambda_1\lambda_2\lambda_3}&\, = U\big(R(\pi+\alpha,\pi-\beta,-(\pi+\gamma+\varphi_{13}))\big)\big(U(R(\varphi_{13}-\pi/2,\pi/2,\pi)L_z(p_3))\ket{s\lambda_3} \\ 
&\hspace{5.67cm}\otimes \ U(R(3\pi/2-\varphi_{12},\pi/2,\pi)L_z(p_2))\ket{s\lambda_2} \\
&\hspace{5.67cm} \otimes \ U(R(3\pi/2,\pi/2,\pi)L_z(p_1))\ket{s\lambda_1} \big) \\
= &\ U\big(R(\pi+\alpha,\pi-\beta,-(\pi+\gamma+\varphi_{13}))\big) \big(U(R(\varphi_{13}-\pi/2,\pi/2,\pi)L_z(p_3))\ket{s\lambda_3} \\ 
&\hspace{5.25cm}\otimes \ U(R(\varphi_{13}+\varphi_{23}-\pi/2,\pi/2,\pi)L_z(p_2))\ket{s\lambda_2} \\
&\hspace{5.25cm} \otimes \ U(R(3\pi/2,\pi/2,\pi)L_z(p_1))\ket{s\lambda_1} \big). \\
\end{aligned}   
\end{equation}
Last equality makes use of $\varphi_{12}+\varphi_{23}+\varphi_{13}=2\pi$. The three angles in the reference state now satisfy relations \eqref{eq::BD_pdef2_1} and \eqref{eq::BD_pdef2_2} with $w_1$ and $w_3$ exchanged. To definitely obtain a true $p$-helicity state, the $\pi$ rotations around the $z$ axis have to be absorbed in the $\ket{s\lambda_i}$. This operation produces three $(-1)^{-\lambda_i}$ phase factors,
\begin{equation}
\begin{aligned}
\mathbb{P}_{13}\ket{\alpha\beta\gamma;w_1w_2w_3;\lambda_1\lambda_2\lambda_3}= (-1)^{-\lambda_1-\lambda_2-\lambda_3} \ket{(\pi+\alpha)\,(\pi-\beta)\,(-\pi-\gamma-\varphi_{13});w_3w_2w_1;\lambda_3\lambda_2\lambda_1}.
\end{aligned}
\end{equation}

Now that the relation has been written for Berman's $p$-helicity states, the corresponding relation for Berman's $J$-helicity states can be obtained,
\begin{equation}
    \begin{aligned}
    &\mathbb{P}_{13} \ket{JM\mu;w_1w_2w_3;\lambda_1\lambda_2\lambda_3} =  \sqrt{\frac{2J+1}{8\pi^2}} \int \diff\alpha \diff\!\cos\!\beta \diff\gamma \, D^{J*}_{M\mu}(\alpha,\beta,\gamma) \mathbb{P}_{13}\ket{\alpha\beta\gamma ; w_1 w_2 w_3; \lambda_1 \lambda_2 \lambda_3} \\
    &\hspace{2.05cm} = (-1)^{-\lambda_1-\lambda_2-\lambda_3}  \sqrt{\frac{2J+1}{8\pi^2}} \int \diff\alpha \diff\!\cos\!\beta \diff\gamma \, D^{J*}_{M\mu}(\alpha,\beta,\gamma) \\
    &\hspace{7.1cm} \ket{(\pi+\alpha)\,(\pi-\beta)\,(-\pi-\gamma-\varphi_{13}) ; w_3 w_2 w_1; \lambda_3 \lambda_2 \lambda_1}.
    \end{aligned}
\end{equation}
The integration variables can be changed to fit with the new angles in the $p$-helicity state,
\begin{equation}
    \begin{aligned}
    \mathbb{P}_{13} \ket{JM\mu;w_1w_2w_3;\lambda_1\lambda_2\lambda_3} = (-1)^{-\lambda_1-\lambda_2-\lambda_3} \sqrt{\frac{2J+1}{8\pi^2}} \int \diff\alpha'\diff\!\cos\!\beta'\diff\gamma'\, D^{J*}_{M\mu}(\alpha'-\pi,\pi-\beta',-\pi-\gamma'-\varphi_{13})& \\
    \hspace{1.8cm} \ket{\alpha'\,\beta'\,\gamma'; w_3 w_2 w_1 ; \lambda_3 \lambda_2 \lambda_1}&. 
    \end{aligned}
\end{equation}
Making use of Wigner $D$ matrices properties \cite{vars88}, the aforementioned relation can be reduced,
\begin{equation}
    \begin{aligned}
    \mathbb{P}_{13} \ket{JM\mu;w_1w_2w_3;\lambda_1\lambda_2\lambda_3} = (-1)^{-\lambda_1-\lambda_2-\lambda_3} \sqrt{\frac{2J+1}{8\pi^2}} \int \diff\alpha'\diff\!\cos\!\beta' \diff\gamma'\,\left((-1)^{J-\mu} e^{-i\varphi_{13}\mu} D^{J*}_{M-\mu}(\alpha',\beta',\gamma')\right)& \\
    \hspace{0cm} \ket{\alpha'\,\beta'\,\gamma'; w_3 w_2 w_1 ; \lambda_3 \lambda_2 \lambda_1}&\\
    = (-1)^{J-\mu-\lambda_1-\lambda_2-\lambda_3} e^{-i\varphi_{13}\mu} \ket{J M-\mu;w_3w_2w_1;\lambda_3\lambda_2\lambda_1}.\hspace{3.35cm}&
    \end{aligned}
\end{equation}
This closes the demonstration of the property. The action of $\mathbb{P}_{23}$ is demonstrated in a similar way. 


\section{Demonstration of the change of basis formula from Berman's to Wick's states}
\label{app:BtoW}

This Appendix is dedicated to the derivation of relation \eqref{eq::BtoW_fin}. It is divided in three parts. First, Berman's $p$-helicity states are restructured to include a state for particle $1$ and $2$ in their CoMF. Then, this rewriting is used to prove the change of basis formula. Finally, a consistency check concerning normalisation of the states is suggested.

\subsubsection*{Rewriting of $p$-helicity states}

To fit with Wick's structure, particle $1$ and $2$ in Berman's states are to be brought in their own CoMF. This operation is performed on the reference states \eqref{eq::BD_refdef} at first. The boost that transitions between the CoMF of the entire system and the $12$-CoMF is the one that imparts a momentum $\bm{p_3}$ to a particle of mass $m_{12}$ initially at rest. It writes down as follows
\begin{equation}
L_3=R(3\pi/2,\pi/2,0)L_z(m_{12},p_3)R^{-1}(3\pi/2,\pi/2,0).\label{eq::BtoW_L3}
\end{equation}
This $L_3$ boost can be artificially inserted into the definition of Berman's reference states \eqref{eq::BD_refdef},
\begin{equation}
\begin{aligned}
\ket{w_1w_2w_3;\lambda_1\lambda_2\lambda_3}
=\ U(L_3^{-1} L_3)  \Big[\, U(R(\phi_1,\pi/2,0)L_z(p_1))\ket{s_1\lambda_1}\,\otimes\,U(R(\phi_2,\pi/2,0)L_z(p_2))\ket{s_2\lambda_2}&\Big]\\ 
\otimes\, U(R(\phi_3,\pi/2,0)L_z(p_3))\ket{s_3\lambda_3}&\\
= U(L_3)^{-1} \Big[\, U(L_3)U(R(\phi_1,\pi/2,0)L_z(p_1))\ket{s_1\lambda_1}\,\otimes\,U(L_3)U(R(\phi_2,\pi/2,0)L_z(p_2))\ket{s_2\lambda_2}&\Big]\\
\otimes\,U(R(\phi_3,\pi/2,0)L_z(p_3))\ket{s_3\lambda_3}&.
\end{aligned}
\end{equation}
Inside the brackets, both particle $1$ and particle $2$ are subjected to the $L_3$ boost. The subsequent development focuses on the first particle but the same applies to the second one. The expression to simplify results from applying a Lorentz boost on a one-body helicity state \eqref{eq::1BS_phelicity},
\begin{align}
U(L_3)U(R(\phi_1,\pi/2,0)L_z(p_1))\ket{s_1\lambda_1} = U(L_3)\ket{p_1\pi/2\,\phi_1;s_1\lambda_1}_0.
\end{align}
For further clarification, the third angle convention is written as an index. The situation is the one handled by property \eqref{eq::1BS_canProp},
\begin{align}
U(L_3)\ket{p_1\pi/2\,\phi_1;s_1\lambda_1}_0 = \sum_{\lambda_1^\prime} D^{s_1}_{\lambda_1^\prime\,\lambda_1}(R_W^1)\ket{p_{12}\theta_{12}\phi_{12};s_1\lambda_1^\prime}_{0}. \label{eq::BtoW_Wr1}
\end{align}
By construction, after applying the $L_3$ boost, the momentum ends in the $12$-CoMF. Corresponding coordinates have been denoted $p_{12}$, $\theta_{12}$ and $\phi_{12}$, in agreement with notations from Wick's definition \eqref{eq::WD_sub2BS}. Because, in the reference state, momenta are chosen as lying in the $xy$ plane and because $L_3$ is a boost along the $y$ axis, boosted momenta stays in that same plane. As a consequence, the polar angle $\theta_{12}$ is shown to be identically equal to $\pi/2$. Concerning expressions of $p_{12}$ and $\phi_{12}$ in terms of the energies $w_1$, $w_2$ and $w_3$, these are obtained by evaluating different Lorentz invariant combinations of four-momenta\footnote{Namely $(\bm{P_1}+\bm{P_2})^2$, $(\bm{P_1}+\bm{P_3})^2$ and $(\bm{P_1}+\bm{P_2}+\bm{P_3})^2$.} in both the $12$-CoMF and the ECoMF. This allows to show that 
\begin{align}
    &2p_{12} = \sqrt{(w_1+w_2)^2-w_3^2},&&\cos(\pi/2-\phi_{12}) = \frac{w_1-w_2}{w_3}.
\end{align}
The Wigner rotation $R_W^1$ is obtained by specifying expression \eqref{eq::1BS_wignRot} to the current situation,
\begin{equation}
R_W^1=(R(\phi_{12},\pi/2,0) L_z(m_1,p_{12}))^{-1}\, L_3\, (R(\phi_1,\pi/2,0) L_z(m_1,p_1)). \label{eq::BtoW_wignerRot_1}
\end{equation}
This expression for $R_W^1$ and the knowledge of each particle masses and energies theoretically allow to determine the parameters for the Wigner rotation and to deduce the corresponding $D$ matrix. The same calculation is applied to the second particle. By collecting the results, the following intermediate expression is obtained for the three-body reference state,
\begin{equation}
\begin{aligned}
\ket{w_1w_2w_3;\lambda_1\lambda_2\lambda_3} \hspace{13cm} &\\ 
= \sum_{\lambda_1',\,\lambda_2'} D^{s_1}_{\lambda_1^\prime\,\lambda_1}(R_W^1) D^{s_2}_{\lambda_2^\prime\,\lambda_2}(R_W^2)
\Big(U(L_3)^{-1} \Big[\ket{p_{12}(\pi/2)(\phi_{12});s_1\lambda_1'}_{0}\,\otimes\,\ket{p_{12}(\pi/2)(\pi+\phi_{12});s_2\lambda_2'}_{0} \Big]&\\
\otimes\,\ket{p_3(\pi/2)\phi_3;s_3\lambda_3}_0\Big)&.
\end{aligned}
\end{equation}
Above, the definition of one-body helicity state \eqref{eq::1BS_phelicity} has also been used for the last particle. By using the definition of $L_3$, its invert can be written as
\begin{equation}
\begin{aligned}
L_3^{-1} = R(5\pi/2,\pi/2,0)L_{z}(m_{12},p_3)R^{-1}(5\pi/2,\pi/2,0).\label{eq::BtoW_L3-1}
\end{aligned}    
\end{equation} 
The boost $L_3^{-1}$ proves to fit the definition of a canonical boost \eqref{eq::1BS_pcanonical}. An helicity boost would be more convenient to subsequently construct helicity states. To perform the conversion, the rotation on the right can be absorbed into both one-body helicity states, using the invariance of helicity under rotations \eqref{eq::1BS_rotProp},
\begin{align}
R^{-1}(5\pi/2,\pi/2,0)\ket{p_{12}(\pi/2)(\phi_{12});s_1\lambda_1'}_{0} =\ & e^{i\xi_1\lambda_1'}\ket{p_{12}(\pi/2-\phi_{12})(3\pi/2);s_1\lambda_1'}_{0},\\
R^{-1}(5\pi/2,\pi/2,0)\ket{p_{12}(\pi/2)(\pi+\phi_{12});s_2\lambda_2'}_{0} =\ & e^{i\xi_2\lambda_2'}\ket{p_{12}(\pi/2+\phi_{12})(5\pi/2);s_2\lambda_2'}_{\pi}.
\end{align}
Because in definition \eqref{eq::WD_sub2BS} the second particle is considered as opposed to the first one, it is brought in the $\pi$ convention instead of the $0$ one. Above, the expressions for the polar and azimuth angles from the right-hand side state are obtained by applying $R^{-1}(5\pi/2,\pi/2,0)$ on both particle momenta. Concerning $\xi_1$ and $\xi_2$, their values are obtained by explicitly multiplying both rotations that provide their angles to the one-body helicity states by $R^{-1}(5\pi/2,\pi/2,0)$,
\begin{equation}
\begin{aligned}
R^{-1}(5\pi/2,\pi/2,0)\,R(\phi_{12},\pi/2,0)&=R(3\pi/2,\pi/2-\phi_{12},\pi/2)\\
R^{-1}(5\pi/2,\pi/2,0)\,R(\pi+\phi_{12},\pi/2,0)&=R(5\pi/2,\pi/2+\phi_{12},-\pi/2).\label{eq::BtoW_xi}
\end{aligned}
\end{equation}
Noticing that $\phi_{12}$ itself lies in between $-\pi/2$ and $\pi/2$, these two formulas have been tuned so that the second Euler angle of both combined rotations lies in between $0$ and $\pi$. The combined rotations are neither in the $0$ nor in the $\pi$ convention. This issue is resolved by adding the suitable rotation around the $z$ axis. After acting on the helicity state, this results in a simple phase factor,
\begin{align}
    &e^{i\lambda_1'\xi_1} = e^{-i\lambda_1'\pi/2}, && e^{i\lambda_2'\xi_2} = e^{3i\lambda_2'\pi/2}.
\end{align}
At this stage, the reference state has been restructured as follows, 
\begin{equation}
\begin{aligned}
&\ket{w_1w_2w_3;\lambda_1\lambda_2\lambda_3} \\
&\hspace{1cm} = \sum_{\lambda_1',\lambda_2'} D^{s_1}_{\lambda_1^\prime\,\lambda_1}(R_W^1) D^{s_2}_{\lambda_2^\prime\,\lambda_2}(R_W^2)\, e^{-i(\lambda_1'-3\lambda_2')\pi/2} \bigg(U(R(5\pi/2,\pi/2,0)L_{z}(m_{12},p_3)) \\
&\hspace{2cm}\Big[\ket{p_{12}(\pi/2-\phi_{12})(3\pi/2);s_1\lambda_1'}_{0}\,\otimes\,\ket{p_{12}(\pi/2+\phi_{12})(5\pi/2);s_2\lambda_2'}_{\pi} \Big]\otimes\,\ket{p_3(\pi/2)(3\pi/2);s_3\lambda_3}_0\bigg).
\end{aligned}
\end{equation}
Two small modifications remain to be performed. First, the third particle is for now in the $0$ convention but, because the momentum of particle $3$ will be considered as opposed, its third angle convention should be changed to the $\pi$ one, thereby adding a $e^{i\lambda_3\pi}$ phase factor. Lastly, the state inside the brackets can be rewritten as a two-body helicity state \eqref{eq::2BS_pdef_2}. This only requires an additional phase $(-1)^{s_2-\lambda_2'}$. The final expression for the reference state is
\begin{equation}
\begin{aligned}
\ket{w_1w_2w_3;\lambda_1\lambda_2\lambda_3} &=\sum_{\lambda_1',\lambda_2'} D^{s_1}_{\lambda_1^\prime\,\lambda_1}(R_W^1) D^{s_2}_{\lambda_2^\prime\,\lambda_2}(R_W^2)\, e^{i(2s_2+\lambda_2'-\lambda_1'+2\lambda_3)\pi/2}\\
&\hspace{1.5cm}\Big(U(R(5\pi/2,\pi/2,0)L_{z}(m_{12},p_3)) \ket{p_{12}(\pi/2-\phi_{12})(3\pi/2);s_1\lambda_1's_2\lambda_2'}\\
&\hspace{7.7cm}\otimes\,\ket{p_3(\pi/2)(3\pi/2);s_3\lambda_3}_{\pi}\Big).
\end{aligned}\label{eq::BtoW_phs}
\end{equation}
This expression has the desired structure: particles $1$ and $2$ are coupled in their own CoMF, and this subsystem is boosted towards the ECoMF to be coupled with the third particle.

\subsubsection*{Rewriting of $J$-helicity states} \label{sec::BtoW_RJHS}

Now that an intermediary coupling has been added in Berman's $p$-helicity states, the development of Berman's $J$-helicity states in Wick's $J$-helicity states can be performed. By inserting relation \eqref{eq::BtoW_phs} into the definition \eqref{eq::BD_Jdef}, the following expression is obtained,
\begin{align}
\begin{aligned}
\ket{JM\mu ; w_1 w_2 w_3 ; \lambda_1 \lambda_2 \lambda_3}
= \sqrt{\frac{2J+1}{8\pi^2}} \sum_{\lambda_1',\lambda_2'} 
D^{s_1}_{\lambda_1^\prime\,\lambda_1}(R_W^1) D^{s_2}_{\lambda_2^\prime\,\lambda_2}(R_W^2) e^{i(2s_2+\lambda_2'-\lambda_1'+2\lambda_3)\pi/2}\hspace{2.4cm}&
\\
\phantom{\sum_{\lambda_1',\lambda_2'}} \int \diff\alpha \, \diff\!\cos\!\beta \, \diff\gamma \, D^{J*}_{M\mu}(\alpha,\beta,\gamma) 
U\left(R\left(\alpha,\beta,\gamma\right)\right)\Big(U(R(5\pi/2,\pi/2,0) L_{z}(m_{12},p_3)) &\\
\ket{p_{12}(\pi/2-\phi_{12})(3\pi/2);s_1\lambda_1's_2\lambda_2'} \otimes\,\ket{p_3(\pi/2)(3\pi/2);s_3\lambda_3}_{\pi}&\Big).
\end{aligned}\label{eq::BtoW_JH1}
\end{align}
To fit with Wick's definition, eigenstates of the total angular momentum relative to particle $1$ and $2$ would be preferable to the current $\ket{p_{12}(\pi/2-\phi_{12})(3\pi/2);s_1\lambda_1's_2\lambda_2'}$. Such two-body states are related to each-other by relation \eqref{eq::2BS_pToJ}. By omitting some passive coefficients for brevity, the following expression is obtained,
\begin{equation}
\begin{aligned}
&\ket{JM\mu ; w_1 w_2 w_3 ; \lambda_1 \lambda_2 \lambda_3} \\
&\hspace{0.cm}= \sum_{\lambda_1',\lambda_2'}\; [...]\int \diff\alpha \, \diff\!\cos\!\beta \, \diff\gamma \;(...)
\sum_{j_{12}=|\lambda_1'-\lambda_2'|}^{\infty}\,\sum_{\lambda_{12}=-j_{12}}^{j_{12}} \sqrt{\frac{2j_{12}+1}{4\pi}} D^{j_{12}}_{\lambda_{12}\,\lambda_1'-\lambda_2'}(3\pi/2,\pi/2-\phi_{12},0)
\;U\left(R\left(\alpha,\beta,\gamma\right)\right) \\ 
&\hspace{4.4cm}\Big(U(R(5\pi/2,\pi/2,0) 
L_{z}(m_{12},p_3)) \ket{p_{12};j_{12}\lambda_{12};\lambda_1'\lambda_2'}\,\otimes\,\ket{p_3(\pi/2)(3\pi/2);s_3\lambda_3}_{\pi}\Big).
\end{aligned} \label{eq::BtoW_intermediary}
\end{equation}
The structure of the state inside the large parentheses almost follows the pattern of Wick's $p$-helicity states \eqref{eq::WD_3bdef_1}. The main difference lays in the rotation that acts on the two-body state. It involves a $5\pi/2$ rotation around the $z$-axis where a rotation in between $0$ and $2\pi$ is expected. For bosonic states, $2\pi$ rotations are identified to the identity and can freely be ignored. For fermionic states, these rotations give rise to minus signs which have to be taken into account. Both cases can be taken into account at once by adding a $(-1)^{2\lambda_1'+2\lambda_2'}$ factor. A $(-1)^{\lambda_3-s_3}$ phase factor has also to be added to truely correspond to Wick's $p$-helicity state definition,
\begin{equation}
\begin{aligned}
&\ket{JM\mu ; w_1 w_2 w_3 ; \lambda_1 \lambda_2 \lambda_3} \\
&\hspace{0.5cm} =
\sum_{\lambda_1',\lambda_2'}\; [...]\int \diff\alpha \, \diff\!\cos\!\beta \, \diff\gamma \;(...)\; 
\sum_{j_{12}=|\lambda_1'-\lambda_2'|}^{\infty}\,\sum_{\lambda_{12}=-j_{12}}^{j_{12}} \sqrt{\frac{2j_{12}+1}{4\pi}} D^{j_{12}}_{\lambda_{12}\,\lambda_1'-\lambda_2'}(3\pi/2,\pi/2-\phi_{12},0)\\ 
&\hspace{3.4cm} U\left(R\left(\alpha,\beta,\gamma\right)\right)(-1)^{2\lambda_1'+2\lambda_2'+s_3-\lambda_3} \ket{p_3(\pi/2)(\pi/2);j_{12}\lambda_{12}s_3\lambda_3;p_{12}s_1\lambda_1's_2\lambda_2'}.
\end{aligned}
\end{equation}
A second use of relation \eqref{eq::2BS_pToJ} allows to replace this state with defined momenta by a sum of states with defined total angular momentum,
\begin{equation}
\begin{aligned}
\ket{JM\mu ; w_1 w_2 w_3 ; \lambda_1 \lambda_2 \lambda_3}\hspace{12.4cm}\\
=\sum_{\lambda_1',\lambda_2'}\; [...]\int \diff\alpha \, \diff\!\cos\!\beta \, \diff\gamma \;(...)\;
\sum_{j_{12}=|\lambda_1'-\lambda_2'|}^{\infty}\,\sum_{\lambda_{12}=-j_{12}}^{j_{12}} \{...\} \sum_{\bar J=|\lambda_{12}-\lambda_3|}^{\infty}\,\sum_{\bar M=-\bar J}^{\bar J} \sqrt{\frac{2\bar J +1}{4\pi}}\, D^{\bar J}_{\bar M\,\lambda_{12}-\lambda_3}(\pi/2,\pi/2,0)&\\
\phantom{\sum_{J'=|\lambda_{12}-\lambda_3|}^{\infty}}U\left(R\left(\alpha,\beta,\gamma\right)\right) \ket{p_3;\bar J\bar M;j_{12}\lambda_{12}s_3\lambda_3;p_{12}s_1\lambda_1's_2\lambda_2'}&. 
\end{aligned}\label{eq::BtoW_JH2}
\end{equation}
Dots have again been used to replace passive coefficients in the expression. The relation can still be simplified. To start with, the transformation rule of angular momentum eigenstates under rotations \cite{wein95,mart70} is used to eliminate the $U\left(R(\alpha,\beta,\gamma)\right)$ operator,
\begin{equation}
\begin{aligned}
\ket{JM\mu ; w_1 w_2 w_3 ; \lambda_1 \lambda_2 \lambda_3}\hspace{12.4cm}\\
=\sum_{\lambda_1',\lambda_2'}\; [...]\int \diff\alpha \, \diff\!\cos\!\beta \, \diff\gamma \;(...)\;
\sum_{j_{12}=|\lambda_1'-\lambda_2'|}^{\infty}\,\sum_{\lambda_{12}=-j_{12}}^{j_{12}} \{...\} \sum_{\bar J=|\lambda_{12}-\lambda_3|}^{\infty}\,\sum_{\bar M=-\bar J}^{\bar J} \sqrt{\frac{2\bar J +1}{4\pi}}\, D^{\bar J}_{\bar M\,\lambda_{12}-\lambda_3}(\pi/2,\pi/2,0)&\\
\sum_{M'=-\bar J}^{\bar J} D^{\bar J}_{M'\bar M}\left(\alpha,\beta,\gamma\right) \ket{p_3;\bar J M';j_{12}\lambda_{12}s_3\lambda_3;p_{12}s_1\lambda_1's_2\lambda_2'}&. 
\end{aligned}
\end{equation}
Aforementioned passive coefficients will be reintroduced one after the others. Brackets, curly brackets and parentheses are used consistently to trace their origin. The $(\alpha,\beta,\gamma)$ dependence is confined in two Wigner $D$ matrices,
\begin{equation}
\begin{aligned}
\ket{JM\mu ; w_1 w_2 w_3 ; \lambda_1 \lambda_2 \lambda_3}\hspace{11.5cm}\\
=\sum_{\lambda_1',\lambda_2'}\; [...]
\sum_{j_{12}=|\lambda_1'-\lambda_2'|}^{\infty}\,\sum_{\lambda_{12}=-j_{12}}^{j_{12}} \{...\} \sum_{\bar J=|\lambda_{12}-\lambda_3|}^{\infty}\,\sum_{\bar M=-\bar J}^{\bar J} \sqrt{\frac{2\bar J +1}{4\pi}}\, D^{\bar J}_{\bar M\,\lambda_{12}-\lambda_3}(\pi/2,\pi/2,0)\hspace{1.7cm}&\\
\sum_{M'=-\bar J}^{\bar J} \left(\int \diff\alpha \, \diff\!\cos\!\beta \, \diff\gamma \left(D^{J*}_{M\mu}\left(\alpha,\beta,\gamma\right)\right)D^{\bar J}_{M'\bar M}\left(\alpha,\beta,\gamma\right)\right) \ket{p_3;\bar J M';j_{12}\lambda_{12}s_3\lambda_3;p_{12}s_1\lambda_1's_2\lambda_2'}&. 
\end{aligned}
\end{equation}
The orthogonality of Wigner $D$ matrices \cite{vars88} can now be used. It produces three Kronecker deltas that eliminate three sums among the six,
\begin{equation}
\begin{aligned}
\ket{JM\mu ; w_1 w_2 w_3 ; \lambda_1 \lambda_2 \lambda_3} =\left[\sqrt{\frac{2J+1}{8\pi^2}}\,\right]\sum_{\lambda_1',\lambda_2'}\; [...]
\sum_{j_{12}=|\lambda_1'-\lambda_2'|}^{\infty}\,\sum_{\lambda_{12}=-j_{12}}^{j_{12}} \{...\}\, \sqrt{\frac{2 J +1}{4\pi}}\, D^{J}_{\mu\,\lambda_{12}-\lambda_3}(\pi/2,\pi/2,0)&\\
\frac{8\pi^2}{2J+1}\ket{p_3;J M;j_{12}\lambda_{12}s_3\lambda_3;p_{12}s_1\lambda_1's_2\lambda_2'}&.
\end{aligned}
\end{equation}
At this stage, reintroducing all the passive coefficients, the following expression has been obtained,
\begin{equation}
\begin{aligned}
\ket{JM\mu ; w_1 w_2 w_3 ; \lambda_1 \lambda_2 \lambda_3} =\sqrt{2\pi}\sum_{\lambda_1',\lambda_2'}\left[D^{s_1}_{\lambda_1^\prime\,\lambda_1}(R_W^1) D^{s_2}_{\lambda_2^\prime\,\lambda_2}(R_W^2) e^{i(2s_2+\lambda_2'-\lambda_1'+2\lambda_3)\pi/2}\right]\hspace{3cm}&\\
\sum_{j_{12}=|\lambda_1'-\lambda_2'|}^{\infty}\,\sum_{\lambda_{12}=-j_{12}}^{j_{12}}\left\{ (-1)^{2\lambda_1'+2\lambda_2'+s_3-\lambda_3} \sqrt{\frac{2j_{12}+1}{4\pi}} D^{j_{12}}_{\lambda_{12}\,\lambda_1'-\lambda_2'}(3\pi/2,\pi/2-\phi_{12},0)\right\}&\\
\, D^{J}_{\mu\,\lambda_{12}-\lambda_3}(\pi/2,\pi/2,0) \ket{p_3;J M;j_{12}\lambda_{12}s_3\lambda_3;p_{12}s_1\lambda_1's_2\lambda_2'}&.
\end{aligned}
\end{equation}
Complex numbers inside the summation can be made explicit by turning complex Wigner $D$ matrices into real $d$ matrices and phases \cite[section 4.3]{vars88},
\begin{equation}
    \begin{aligned}
    D^{j_{12}}_{\lambda_{12}\,\lambda_1'-\lambda_2'}(3\pi/2,\pi/2-\phi_{12},0)& D^{J}_{\mu\,\lambda_{12}-\lambda_3}(\pi/2,\pi/2,0)= e^{-i(3\lambda_{12}+\mu)\pi/2} d^{j_{12}}_{\lambda_{12}\,\lambda_1'-\lambda_2'}\left(\pi/2-\phi_{12}\right)d^{J}_{\mu\,\lambda_{12}-\lambda_3}(\pi/2).\\
    \end{aligned}
\end{equation}
Compiling phase factors and noticing that $\lambda_{12}$ and $\lambda_1'-\lambda_2'$ have the same integer or half-integer nature, one gets the final expression,
\begin{equation}
\begin{aligned}
\ket{JM\mu ; w_1 w_2 w_3 ; \lambda_1 \lambda_2 \lambda_3} &=\sum_{\lambda_1',\lambda_2'}D^{s_1}_{\lambda_1^\prime\,\lambda_1}(R_W^1) D^{s_2}_{\lambda_2^\prime\,\lambda_2}(R_W^2) e^{i(2s_2+2s_3+\lambda_2'-\lambda_1'-\mu)\pi/2}\\
&\sum_{j_{12}=|\lambda_1'-\lambda_2'|}^{\infty}\,\sum_{\lambda_{12}=-j_{12}}^{j_{12}} e^{i\lambda_{12}\pi/2} \sqrt{\frac{2j_{12}+1}{2}} d^{j_{12}}_{\lambda_{12}\,\lambda_1'-\lambda_2'}(\pi/2-\phi_{12})d^{J}_{\mu\,\lambda_{12}-\lambda_3}(\pi/2) \\
&\hspace{6.22cm}\ket{p_3;J M;j_{12}\lambda_{12}s_3\lambda_3;p_{12}s_1\lambda_1's_2\lambda_2'}.
\end{aligned}
\end{equation}

\subsubsection*{Normalisation consistence}
\label{ssec::WtoB_NormCons}

To confirm relations \eqref{eq::BtoW_fin} and \eqref{eq::BtoW_fin_massless}, one may try to deduce the orthonormalisation of Berman's $J$-helicity states from the one of Wick's $J$-helicity states. This subsection is devoted to this consistency check supposing three massless particles. The scalar product of two arbitrary Berman's $J$-helicity states is evaluated by making use of relation \eqref{eq::BtoW_fin_massless} while Wick's states orthonormalisation \eqref{eq::WD_norm_1} is assumed. After a few algebra, one gets
\begin{equation}
\begin{aligned}
&\braket{\bar J \bar M \bar \mu ; \bar w_1 \bar w_2 \bar w_3; \bar \lambda_1 \bar \lambda_2 \bar \lambda_3|JM\mu ; w_1w_2w_3 ; \lambda_1 \lambda_2 \lambda_3} \\
& \hspace{0.5cm} = \frac{4\sqrt{p_3^2+4p_{12}^2}}{p_3p_{12}}\,\delta(\bar p_3-p_3)
\delta(\bar{p}_{12}-p_{12})\,\delta_{\bar{J}J}\delta_{\bar{M}M}\delta_{\bar\lambda_1\,\lambda_1}\delta_{\bar\lambda_2\,\lambda_2}\delta_{\bar\lambda_3\,\lambda_3} e^{i(\bar \mu-\mu) \pi/2} \\
& \hspace{1cm}  \sum_{j_{12}=|\lambda_1-\lambda_2|}^{\infty} \,\sum_{\lambda_{12}=-j_{12}}^{j_{12}} \frac{2j_{12}+1}{2}\, d^{j_{12}}_{\lambda_{12}\,\lambda_1-\lambda_2}(\pi/2-\bar\phi_{12})\, d^{j_{12}}_{\lambda_{12}\,\lambda_1-\lambda_2}(\pi/2-\phi_{12}) d^{J}_{\bar \mu\,\lambda_{12}-\lambda_3}(\pi/2)d^{J}_{\mu\,\lambda_{12}-\lambda_3}(\pi/2).\label{eq::BtoW_norm_int_1}  
\end{aligned}
\end{equation}
Let us remind that a dependence on the energy variable $w_1$ ($\bar w_1$) is hidden inside the $\phi_{12}$ ($\bar \phi_{12}$) angle. Comparing with the announced orthonormalisation of Berman's $J$-helicity states, the remaining summations on $j_{12}$ and $\lambda_{12}$ are expected to produce a Kronecker delta in $\mu$ and a Dirac delta in $w_1$. Using properties of $d$ matrices \cite{vars88}, these summations can be reduced analytically,
\begin{equation}
\begin{aligned}
 \sum_{j_{12}=|\lambda_1-\lambda_2|}^{\infty} \,\sum_{\lambda_{12}=-j_{12}}^{j_{12}}
\frac{2j_{12}+1}{2}\,
d^{j_{12}}_{\lambda_{12} \,\lambda_1-\lambda_2}\left(\bar u\right) d^{j_{12}}_{\lambda_{12}\,\lambda_1-\lambda_2}\left(u\right) d^{J}_{\bar{\mu}\,\lambda_{12}-\lambda_3}(\pi/2)  d^{J}_{\mu\,\lambda_{12}-\lambda_3}(\pi/2) = \delta\left(u-\bar u \right) \delta_{\mu\bar\mu}.
\end{aligned}
\end{equation}
where the shorter notation $u$ for the variable $\cos(\pi/2-\phi_1')$ has been used. Dirac deltas on $p_3$, $p_{12}$ and $u$ can also be turned into Dirac deltas on $w_1$, $w_2$ and $w_3$ using \eqref{eq::BtoW_w1w2_um},
\begin{equation}
\begin{aligned}
\delta(\bar u- u)\delta(\bar p_{12}- p_{12})\delta(\bar p_3- p_3) &= \frac{2p_3p_{12}}{\sqrt{4p_{12}^2+p_3^2}}\, \delta(\bar w_1- w_1)\delta(\bar w_2- w_2)\delta(\bar w_3- w_3).\label{eq::BtoW_um_w1w2_Jaco}
\end{aligned}
\end{equation}
As a result, one get the expected orthonormalisation of Berman's $J$-helicity states,
\begin{equation}
\begin{aligned}
\braket{\bar J \bar M \bar \mu ; \bar w_1 \bar w_2 \bar w_3; \bar \lambda_1 \bar \lambda_2 \bar \lambda_3|JM\mu ; w_1w_2w_3 ; \lambda_1 \lambda_2 \lambda_3} = 8\, \delta(\bar w_1- w_1)\delta(\bar w_2-w_2)\delta(\bar w_3-w_3)&\\
\delta_{\bar{J}J}\delta_{\bar{M}M}\delta_{\bar\mu\mu}\delta_{\lambda_1\,\bar\lambda_1}\delta_{\lambda_2\,\bar\lambda_2}\delta_{\lambda_3\,\bar\lambda_3}&.
\end{aligned}
\end{equation}


\section{Symmetry considerations about equation (\ref{eq::3GB_ME_Vunsym_fin})}
\label{app::simpl}
First of all, let us have a quick look at coefficients \eqref{eq::3GB_ME_coefficients},
\begin{equation}
\left(\mathcal{C}_{J\bar\mu;\bar\lambda_1\bar\lambda_2\lambda_3}^{ j_{12}\lambda_{12}}\right)^*\mathcal{C}_{J\mu;\lambda_1\lambda_2\lambda_3}^{j_{12}\lambda_{12}} = (-1)^{(\lambda_2-\bar\lambda_2)/2 + (\bar\lambda_1-\lambda_1)/2 + (\bar\mu-\mu)/2}\,\frac{2j_{12}+1}{2}\, d_{\bar\mu\,\lambda_{12}-\lambda_3}^{J}(\pi/2)d_{\mu\,\lambda_{12}-\lambda_3}^{J}(\pi/2).
\end{equation}
These coefficients are easily shown to stay real, independently of the spin of the particles. If $\mu$ or $\bar \mu$ is null, properties of Wigner $d$ matrices allows to show that, as long as $J-\lambda_{12}+\lambda_3$ is an odd number, the whole coefficient cancels \cite{vars88}. Because $\lambda_3$ belongs to $\{+1,-1\}$ and because the total angular momentum of a symmetric state with $\mu=0$ must be odd, whenever $\mu$ or $\bar\mu$ is null, all terms with odd $\lambda_{12}$ cancel, thereby avoiding evaluating them.

The dependency on helicity quantum numbers of the $\tilde \Psi$ function \eqref{eq::3GB_ME_modPsi} can be investigated. The symmetry properties of Wigner $d$ matrices allows to equate different evaluations of $\tilde \Psi$. Following equalities immediately comes out of these symmetry properties,
\begin{equation}
\begin{aligned}
\tilde\Psi(...;j_{12},\lambda_{12},\Delta\lambda;...)=\tilde\Psi(...;j_{12},-\Delta\lambda,-\lambda_{12};...)
=(-1)^{\lambda_{12}}\tilde\Psi(...;j_{12},-\lambda_{12},-\Delta\lambda;...)&\\
=(-1)^{\lambda_{12}} \tilde\Psi(...;j_{12},\Delta\lambda,\lambda_{12};...)&.
\end{aligned} \label{eq::3GB_ME_unsym_tildepsisym_1}
\end{equation}
Other symmetries require $\Psi$ to be an even function of $u$. In that case, performing the change of variables $u \rightarrow-u$ allows to show that
\begin{equation}
\begin{aligned}
\tilde\Psi(...;j_{12},\lambda_{12},\Delta\lambda;...)=(-1)^{j_{12}}\tilde\Psi(...;j_{12},-\lambda_{12},\Delta\lambda;...)=(-1)^{j_{12}}\tilde\Psi(...;j_{12},-\Delta\lambda,\lambda_{12};...)&\\
=(-1)^{j_{12}+\lambda_{12}}\tilde\Psi(...;j_{12},\Delta\lambda,-\lambda_{12};...)=(-1)^{j_{12}+\lambda_{12}} \tilde\Psi(...;j_{12},\lambda_{12},-\Delta\lambda;...)&.
\end{aligned} \label{eq::3GB_ME_unsym_tildepsisym_2}
\end{equation}
The assumption that requires $\Psi$ to be even is consistent with the trial shape \eqref{eq::3GB_ME_trialFunc_ump}. However, it may be invalidated if kinematic factors such as those from relations \eqref{eq::3GB_spinJ_1e13e23} and \eqref{eq::3GB_spinJ_e} need to be included in order to ensure the symmetry of the state. Regardless of this precaution, relations \eqref{eq::3GB_ME_unsym_tildepsisym_1} and/or \eqref{eq::3GB_ME_unsym_tildepsisym_2} allow for the recycling of most evaluations of $\tilde{\Psi}$, thereby reducing computational cost. These eight symmetry relations can even show that $\tilde\Psi$ cancels for some combinations of quantum numbers, such as for $\Delta \lambda = 0$, $\lambda_{12}$ even and $j_{12}$ odd.

Finally, the ME on two-body $J$-helicity states appearing in the integrand from equation \eqref{eq::3GB_ME_Vunsym_fin} is also to be analysed. Its dependence on helicity quantum numbers can be investigated by means of formula \eqref{eq::2BS_posMatElEv_2} from Section \ref{ssec:2BS}. Using properties of Clebsh-Gordan coefficients, it can be shown that 
\begin{equation}
\begin{aligned}
\mathcal{C}^{j_{12};11}_{l_{12}s_{12};\lambda_1\lambda_2} = (-1)^{s_{12}}\,\mathcal{C}^{j_{12};11}_{l_{12}s_{12};-\lambda_2-\lambda_1} = (-1)^{l_{12}-j_{12}}\,\mathcal{C}^{j_{12};11}_{l_{12}s_{12};-\lambda_1-\lambda_2}
\end{aligned}
\end{equation}
where $\mathcal{C}$ coefficients are those defined in Section \ref{ssec:2BS}. Applying this result to equation \eqref{eq::2BS_posMatElEv_2} reveals that certain helicity quadruplets  $(\lambda_1,\lambda_2,\bar\lambda_1,\bar\lambda_2)$ yield identical values. Specifically, these quadruplets can be grouped into five families, within which ME on two-body $J$-helicity states are equal each-other. These families are displayed in Table~\ref{tab::3GB_ME_quadruplet}. Let us also mention that, again in view of \eqref{eq::2BS_posMatElEv_2}, these matrix elements are real-valued. This allows to freely exchange the bra and the ket,
\begin{equation}
\begin{aligned}
\bra{\bar p_{12}; j_{12}\lambda_{12};\bar\lambda_1\bar\lambda_2}\mathcal{O}(r_{12})\ket{p_{12};j_{12}\lambda_{12};\lambda_1\lambda_2} = \bra{p_{12}; j_{12}\lambda_{12};\lambda_1\lambda_2}\mathcal{O}(r_{12})\ket{\bar p_{12};j_{12}\lambda_{12};\bar\lambda_1\bar\lambda_2}&.
\end{aligned}\label{eq::3GB_ME_unsym_TBME_sym}
\end{equation}
Observations about $\tilde\Psi$ and about the two-body ME are to be combined to deduce symmetry properties of the whole integral from \eqref{eq::3GB_ME_Vunsym_fin}. Simplifications occur depending on the assumptions made on $\Psi$ and $\bar \Psi$.

A special case of prime importance in the following concerns $\Psi$ and $\bar \Psi$ being two equal real functions which are even for $u$. These assumptions imply two symmetry properties at the level of the integrals. First, the equality and reality hypothesis allows to freely exchange $\Delta\lambda$ and $\Delta\bar\lambda$ by exchanging both integration variables. On the other hand, the hypothesis about the parity in $u$ implies that the sign of each $\Delta\lambda$ can be flipped, at worst, at the cost of a phase factor $(-1)^{j_{12}+\lambda_{12}}$. These two properties implies that integrals line up with the five families of helicity quadruplets from Table~\ref{tab::3GB_ME_quadruplet}, except for the $5^{\text{th}}$ family in which quadruplets above and below the dashed line differ by a $(-1)^{j_{12}+\lambda_{12}}$ phase factor. Each evaluation of an integral from \eqref{eq::3GB_ME_Vunsym_fin} for a given helicity quadruplet provides a value valid for any other quadruplet in the same family, potentially up to a minus sign. 

Apart from this special case, a second simplification turns out relevant for the upcoming developments. In the first family from Table~\ref{tab::3GB_ME_quadruplet}, both $\Delta \lambda$ and $\Delta \bar \lambda$ are always zero. As a result and without further conditions, calculations for both helicity quadruplet in this family will yield identical values for integrals. This result will prove valuable, as the two helicity quadruplets in question are precisely those encountered during calculations with $A''_2$ states.

\begin{table}
    \centering
    \begin{tblr}{|l| c c c c|}
    \hline
    $1^{\text{st}}$ family & & $(+,+,+,+)$ & $(-,-,-,-)$ &  \\
    \hline \hline
    $2^{\text{nd}}$ family & & $(+,+,-,-)$ & $(-,-,+,+)$ & \\
    \hline \hline
    $3^{\text{rd}}$ family & & $(-,+,-,+)$ & $(+,-,+,-)$ & \\
    \hline \hline
    $4^{\text{th}}$ family & & $(+,-,-,+)$ & $(-,+,+,-)$ & \\
    \hline \hline
    $5^{\text{th}}$ family & $(+,-,+,+)$ & $(+,+,+,-)$ & $(+,-,-,-)$ & $(-,-,+,-)$ \\
    \cline[dashed]{2-5}
    & $(-,+,+,+)$ & $(+,+,-,+)$ & $(-,+,-,-)$ & $(-,-,-,+)$ \\
    \hline 
    \end{tblr}
    \caption{Table of the different helicities quadruplets $(\lambda_1,\lambda_2,\bar\lambda_1,\bar\lambda_2)$ that give rise to the same ME on two-body $J$-helicity states $\bra{\bar p_{12}; j_{12}\lambda_{12};\bar\lambda_1\bar\lambda_2}\mathcal{O}(r_{12})\ket{p_{12};j_{12}\lambda_{12};\lambda_1\lambda_2}$.}
    \label{tab::3GB_ME_quadruplet}
\end{table}

\section{Potential for three-gluon glueballs}
\label{app:pot}

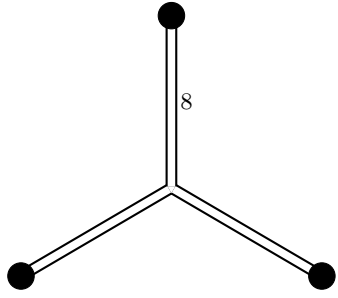
\begin{figure}
\centering
\begin{tikzpicture}
    \usetikzlibrary{calc}
    \usetikzlibrary{decorations.pathmorphing} 
    \coordinate (A) at (0, 0);
    \coordinate (B) at (4, 0);
    \coordinate (C) at (2, {2*sqrt(3)});
    \coordinate (Dh) at (2, {2*sqrt(3)/3+0.04});
    \coordinate (Dl) at ({2-0.03}, {2*sqrt(3)/3});
    \coordinate (Dr) at ({2+0.03}, {2*sqrt(3)/3});

    \draw[-, double distance=1mm, thick] (A) -- (Dl);
    \draw[-, double distance=1mm, thick] (B) -- (Dr);
    \draw[-, double distance=1mm, thick] (C) -- (Dh) node[midway, right] {$8$};
    \filldraw[black] (A) circle (5pt);
    \filldraw[black] (B) circle (5pt);
    \filldraw[black] (C) circle (5pt);
\end{tikzpicture}
\caption{Schematic representation of a $Y$ junction modeling three-gluon glueballs. The three gluons bound at once by pooling their three $8$ flux tubes.}
\label{fig:app_pot_Y}
\end{figure}

The interactions between three gluons can be described in two main geometries. From the perspective discussed in the main text, each gluon generates two flux tubes that transform with the fundamental and the anti-fundamental representation of $SU(3)$. These flux tubes merge pairwise, in a so-called $\Delta$ junction. Alternatively, each gluon can be viewed as producing a single flux tube in the adjoint representation of $SU(3)$. In that case, the three flux tubes merge in a so-called $Y$ junction \cite{math08b}, as illustrated in Figure \ref{fig:app_pot_Y}. Differences between $Y$ and $\Delta$ junctions are discussed in \cite{buis09,card08}, with the conclusions favoring the $\Delta$ junction.

For comparison, let us consider here a $Y$ junction. In this configuration, the interaction potential used in constituent approaches writes down as follows,
\begin{equation}
    V(\bm{r_1},\bm{r_2},\bm{r_3}) = \min_{\bm{Y}} \sum_{i=1}^3 \beta \braket{F_i^2} |\bm{r_i}-\bm{Y}| + \sum_{i<j}^3 \frac{\alpha_s\braket{F_i\cdot F_j}}{|\bm{r_i} - \bm{r_j}|}.
\end{equation}
Here, $\beta$ is the fundamental string tension, and $\alpha_s$ is the strong coupling constant. The factor $F_i^2$ corrresponds to the $SU(3)$ Casimir operator acting on the $i^{\text{th}}$ particle. Including this factor in the confining potential is known as the Casimir scaling hypothesis \cite{bali00,sema04,bicu08}. For flux tubes transforming under the fundamental or the anti-fundamental representation of $SU(3)$, this factor is shown to equal $4/3$, while for flux tubes transforming with the adjoint representation of $SU(3)$, it is shown to equal $3$. The one-gluon exchange term includes a factor $F_i\cdot F_j$ which combines the $SU(3)$ generators of particles $i$ and $j$. Its value can be deduced from those of the Casimir because
\begin{equation}
    \braket{F_i\cdot F_j} = \frac{\braket{(F_i+F_j)^2}-\braket{F_i^2}-\braket{F_j^2}}{2},
\end{equation}
where $(F_i+F_j)^2$ is the $SU(3)$ Casimir operator for the pair of particles $i$ and $j$. Substituting these factors for a system of three gluons leads to the following expression,
\begin{equation}
    V(\bm{r_1},\bm{r_2},\bm{r_3}) = 3 \beta \min_{\bm{Y}} \sum_{i=1}^3  |\bm{r_i}-\bm{Y}| - \sum_{i<j}^3 \frac{3\alpha_s}{2|\bm{r_i} - \bm{r_j}|}.
\end{equation}
The $Y$ junction introduces a genuine three-body interaction, which is incompatible with the formalism developed in the current work. As proposed in reference \cite{silv04}, this three-body interaction can be approximated by three effective two-body interactions with a rescaling parameter close to $1/2$. The interaction then becomes 
\begin{equation}
    V(\bm{r_1},\bm{r_2},\bm{r_3}) = \frac{3}{2} \beta \sum_{i<j}^3  |\bm{r_i}-\bm{r_j}| - \sum_{i<j}^3 \frac{3\alpha_s}{2|\bm{r_i} - \bm{r_j}|}.
\end{equation}
To compare this with the potential \eqref{eq::3GB_spec_ham_pot} used in the $\Delta$ junction, one must relate the mesonic and the fundamental string tensions. For mesons, the Casimir scaling factor is $4/3$ for both the quark and the anti-quark flux tubes, 
\begin{equation}
    \sigma = \frac{4}{3} \beta.
\end{equation}
Expressing the interaction potential in terms of the mesonic string tension gives
\begin{equation}
    V(\bm{r_1},\bm{r_2},\bm{r_3}) = \frac{9}{8} \sigma \sum_{i<j}^3  |\bm{r_i}-\bm{r_j}| - \sum_{i<j}^3 \frac{3\alpha_s}{2|\bm{r_i} - \bm{r_j}|}.
\end{equation}
This potential differs from the $\Delta$ junction potential \eqref{eq::3GB_spec_ham_pot} by a factor of $9/8$. This factor, a little higher than $1$, slightly increases the overall energy of the glueball if $Y$ junction is considered. This conclusion aligns with the LQCD arguments from reference \cite{card08}. For this reason, a $\Delta$ junction is preferred in the present work.


\bibliographystyle{elsarticle-num-names.bst}
\bibliography{ref}


\end{document}